\documentclass[11pt,a4paper]{article}
\pdfoutput=1
\usepackage{jheppub}
\usepackage{amsmath}
\usepackage{epsfig}
\usepackage{amssymb}
\usepackage{graphics}
\usepackage[active]{srcltx}
\usepackage{epstopdf}
\usepackage{pdfsync}
\usepackage{shuffle}
\usepackage{tikz}
\newcommand{\bea}{\begin{eqnarray}}
\newcommand{\eea}{\end{eqnarray}}
\newcommand{\bean}{\begin{eqnarray*}}
\newcommand{\eean}{\end{eqnarray*}}
\newcommand{\nn}{\nonumber \\}

\def\W #1{\widetilde{#1}}
\def\WH #1{\widehat{#1}}

\def\eref#1{(\ref{#1})}

\def\d{{\rm d}}

\def\a{{\alpha}}

\def\b{{\beta}}

\def\d{\partial}

\def\eps{\epsilon}

\def\Label#1{\label{#1}%
  \smash{\hbox to0pt{\raise1ex\hbox{\tiny[#1]}\hss}}}

\def\YM{{\tiny\mbox{YM}}}

\def\YMs{{\tiny\mbox{YMs}}}

\def\EYM{{\tiny \mbox{EYM}}}

\title{Expansion of EYM Amplitudes in Gauge Invariant Vector Space}

\author[a,c]{Bo Feng,}
\author[a]{Xiao-Di Li,}
\author[b]{Rijun Huang}%\footnote{The email addresses are: fengbo@zju.edu.cn, lixiaodi@zju.edu.cn, huang@njnu.edu.cn}
\affiliation[a]{Zhejiang Institute of Modern Physics, Department of Physics,
 Zhejiang University,\\
 No.38 Zheda Road, Hangzhou 310027, P.R. China.}
\affiliation[b]{Institute of Theoretical Physics, School of Physics and Technology, Nanjing Normal University,\\
 No.1 Wenyuan Road, Nanjing 210046, P.R. China.}
\affiliation[c]{Center of Mathematical Science,
  Zhejiang University,\\
  No.38 Zheda Road, Hangzhou 310027, P.R. China.}

\emailAdd{fengbo@zju.edu.cn}
\emailAdd{lixiaodi@zju.edu.cn}
\emailAdd{huang@njnu.edu.cn}

\date{\today}
\abstract{
Motivated by the problem of expanding  single-trace tree-level amplitude of Einstein-Yang-Mills theory to the BCJ basis of Yang-Mills amplitudes, we present an alternative  expansion formula in the gauge invariant vector space. Starting from a generic vector space consisting of polynomials of momenta and polarization vectors, we define a new sub-space as gauge invariant vector space by imposing constraints of gauge invariant conditions. To characterize this sub-space, we compute its dimension and construct an explicit gauge invariant basis from it. We propose an expansion formula in the gauge invariant basis with expansion coefficients being linear combinations of Yang-Mills amplitude, manifesting the gauge invariance of both expansion basis and coefficients. With help of quivers, we compute the expansion coefficients via differential operators and demonstrate the general expansion algorithm by several examples. }

\begin{document}
\maketitle \flushbottom

\newpage

%%%%%%%%%%%%%%%%%%
\section{Introduction}

In recent decades there have been rapid developments in the field of scattering amplitudes. For instance, complicated multi-loop amplitudes are being computed by new computational techniques \cite{Bern:1994cg,Bern:1994zx,Britto:2004nc,Goncharov:2010jf,Henn:2013pwa}, while new formalisms are being constructed encoding inspiring mathematical structures \cite{Arkani-Hamed:2013jha,Arkani-Hamed:2013kca,Arkani-Hamed:2016rak,Arkani-Hamed:2017tmz,Cachazo:2013gna,Cachazo:2013hca,Cachazo:2013iea,Cachazo:2014nsa,Cachazo:2014xea}. Among these advances, the study of scattering amplitudes of gravity and gauge theories as well as the intimate relationships between them attract heavy attentions. It is already well-known that there are non-trivial relations between tree level color-ordered Yang-Mills amplitudes such as $U(1)$-relations, Kleiss-Kuijf (KK) relations \cite{Kleiss:1988ne, DelDuca:1999rs} and Bern-Carrasco-Johansson (BCJ) relations \cite{Bern:2008qj, Chen:2011jxa}, which reduce the minimal number of independent color-ordered Yang-Mills amplitudes to $(n-3)!$. For gravity amplitude, the Kawai-Lewellen-Tye (KLT) relations \cite{Kawai:1985xq}, which originally state that a closed string amplitude is a combination of products of two open string amplitudes, degenerate to similar relations between gravity and Yang-Mills amplitudes in the field theory limit. Besides, the BCJ double copy conjecture reveals another new way of constructing gravity amplitude from Yang-Mills amplitudes based on the exciting idea of color-kinematic duality \cite{Bern:2008qj, Bern:2010ue, Bern:2010yg}.

In addition to these relations, amplitudes of Einstein-Yang-Mills (EYM) theories where gravitons are allowed to interact with gauge bosons are also investigated from many aspects \cite{Bern:1999bx, Chiodaroli:2014xia, Cachazo:2014nsa, Cachazo:2014xea}. Especially in \cite{Cachazo:2014xea}, a generalized KLT relation is proposed from the study of Cachazo-He-Yuan(CHY) formalism \cite{Cachazo:2013gna,Cachazo:2013hca,Cachazo:2013iea,Cachazo:2014nsa,Cachazo:2014xea}, schematically formulated for the tree-level single-trace EYM amplitude\footnote{Hereafter we will always abbreviate tree-level single-trace EYM amplitude as EYM amplitude for simplicity.} as
\begin{align}
A^{\EYM}_{r,s}(\alpha)
=\sum_{\sigma,\tilde{\sigma}\in S_{n-3}} A^{\YM}_{n}(n-1,n,\sigma,1)\mathcal{S}[\sigma|\tilde{\sigma}]A^{\YMs}_{r,s}(\alpha|1,\tilde{\sigma},n-1,n)~,~~~\label{generalized-KLT}
\end{align}
with $A^{{\YMs}}$ being amplitudes of Yang-Mills-scalar theory and $\mathcal{S}$ the momentum kernel defined in \cite{Bern:1998ug,BjerrumBohr:2010ta,BjerrumBohr:2010zb}. Parallel to the study of monodromy relations of string theory, in \cite{Stieberger:2016lng} the authors present a new relation formulating the EYM amplitude with $n$ gluons and one graviton as a linear combination of $(n+1)$-point Yang-Mills amplitudes in a compact expression. Shortly, this result is generalized to the situations with more than one gravitons \cite{Nandan:2016pya, Chiodaroli:2017ngp} and double color traces \cite{Nandan:2016pya} in the framework of CHY formalism. Furthermore, in paper \cite{Fu:2017uzt}, by studying the constraints of gauge invariance, a compact recursive formula is presented for the expansion of EYM amplitudes with $m$ gravitons in terms of KK basis of color-ordered Yang-Mills amplitudes, and the result is also proven in the CHY formalism \cite{Teng:2017tbo} and generalized to multi-trace amplitudes \cite{Du:2017gnh}. Upon the purpose of current paper, we recall the expansion of EYM amplitudes to color-ordered Yang-Mills amplitudes in KK basis as in the paper \cite{Fu:2017uzt, Du:2017gnh},
\begin{align}
A^{\EYM}_{n,m}(1,2,\ldots,n;\mathbf{H})
=\sum_{\shuffle} \sum_{\mathbf{h}|\tilde{h}=\mathbf{H}\backslash h_a}
C_{h_a}(\mathbf{h}) A^{\EYM}_{n+m-|\tilde{h}|,|\tilde{h}|} (1,\{2,\ldots,n-1\}\shuffle\{\mathbf{h},h_a\},n;\tilde{h})~,~~~\label{EYM2KK}
\end{align}
where $\mathbf{H}=\{h_1,h_2,\ldots,h_m\}$ is a set of $m$ gravitons, and $\alpha\shuffle \beta$ stands for the shuffle permutations between two ordered sets $\alpha,\beta$, {\sl {\sl i.e.}}, permutations of $\alpha\cup\beta$ keeping the orderings of $\alpha$ and $\beta$ respectively. In this expansion legs $1$ and $n$ are always fixed in the first and last positions in the color-ordering. Hence by the recursive formula, at the end the EYM amplitude would be expanded to the basis of Yang-Mills amplitudes with legs $1$ and $n$ being fixed. Coefficient of each Yang-Mills amplitude is a linear combination of $C_{h_a}(\mathbf{h})$'s, which are polynomial functions of polarization vectors and momenta whose precise definition can be found in \cite{Fu:2017uzt}.

While the expansion of EYM amplitude in KK basis of Yang-Mills amplitudes has been solved completely, since KK basis is not the minimal basis of color-ordered Yang-Mills amplitudes, a question naturally arises: what would happen when expanding an EYM amplitude to the minimal basis, {\sl i.e.}, the BCJ basis of Yang-Mills amplitudes? In a first thought, it seems that this question has already been solved by the generalized KLT relation (\ref{generalized-KLT}). However in (\ref{generalized-KLT}) the momentum kernel $\mathcal{S}[\sigma|\tilde{\sigma}]$ and $A_{R}$ are difficult to compute and we also need to sum over all $S_{n-3}$ permutations. Hence the generalized KLT relation dose not work well in practical computation. One could also start with expression \eref{EYM2KK} and reformulate KK basis to BCJ basis by BCJ relations. However, computation of several examples is suffice to suggest that the algebraic manipulations are rather complicated. The resulting expansion coefficients are rather cumbersome without any hints of systematic and compact reorganization, because there are too many equivalent expressions. In paper \cite{Feng:2019tvb}, a new method is proposed by introducing the differential operators into this problem. The differential operator is originally applied to the research of the relationships of amplitudes of different theories \cite{Cheung:2017ems}, and later a series of work show how to apply  differential operators to the expansion of EYM amplitude to KK basis \cite{Feng:2019tvb,Hu:2019qdq,Zhou:2019mbe}. Then naturally  differential operators are applied into the expansion of EYM amplitude into BCJ basis being limited to some simple cases where EYM amplitudes contain one, two or three gravitons. However a systematic method for generic EYM amplitude with $n$ gluons and $m$ gravitons is still in demand.

In this paper, we are trying to fulfill this request by providing a systematic method for computing the expansion coefficients of EYM amplitude with $m$ gravitons in the BCJ basis. Besides the use of differential operators, we would also need the principle of gauge invariance. Since Yang-Mills amplitudes of BCJ basis are linearly independent, if we can write an EYM amplitude as linear combination of Yang-Mills amplitudes of BCJ basis, the gauge invariance of polarization tensors of gravitons would be transformed partially into the gauge invariance of expansion coefficients, which contain one half polarization vectors of the polarization tensors. Hence the gauge invariance
put strong constraints on the form of the expansion coefficients. In fact, the gauge invariance principle has already played important roles in the study of scattering amplitude. It is expected that the gauge invariance could completely determine the amplitudes of certain field theories \cite{Boels:2016xhc, Arkani-Hamed:2016rak, Rodina:2016jyz}, and further exploration can be found in various aspects \cite{Barreiro:2013dpa, Boels:2017gyc, Boels:2018nrr, Fu:2017uzt, Cheung:2017ems, Barreiro:2019ncv}. Especially demonstrated in \cite{Fu:2017uzt}, it is the constraints of gauge invariance that make a compact formula available for expansion of EYM amplitude in KK basis. However the potential applications of gauge invariance are still not fully exploited. In this paper, we would like to take a different understanding of gauge invariance. Just as what we have been done for the symmetries in amplitudes of $\mathcal{N}=4$ super-Yang-Mills theory, since the principle of gauge invariance is a strong constraint for gauge theory, we prefer to make it manifest in the level of scattering amplitudes.

With the new understanding of gauge invariance, in this paper we will show how to expand general EYM amplitude into BCJ basis of Yang-Mills amplitudes systematically. Organization for this paper is as follows. In \S\ref{section-EYM}, we review some backgrounds. In \S\ref{sec-basis}, we introduce the gauge invariant vector space living in a general vector space consisting polynomials of Lorentz contractions of momenta and polarization vectors. We compute the dimension of gauge invariant space, characterize the explicit form of vectors, and finally construct the gauge invariant basis. In \S\ref{sec-coefficient}, we define gauge invariant vectors and differential operators in quiver representations, which is the description of mathematical structures of these vectors and operators. With help of quivers, we implement a systematic algorithm to compute expansion coefficients. In \S\ref{section-examples}, we illustrate our method by several explicit examples, the EYM amplitudes with up to four gravitons in the purpose of clarifying some subtleties. In \S\ref{section-conclusion}, we conclude our discussion and point out some problems to be solved in future. Detailed proofs of some propositions as well as some explicit BCJ coefficients in BCJ relations are presented in appendices.

%%%%%%%%%%%%%%%%%%
\section{The expansion of EYM amplitudes to Yang-Mills amplitudes in BCJ basis} \label{section-EYM}
%%%%%%%%%%%%%%%%%

In this section, we review some background knowledge which is useful in the later discussion of expanding EYM amplitude to BCJ basis of Yang-Mills amplitudes. Firstly, as reviewed in \cite{Feng:2019tvb}, an arbitrary color-ordered Yang-Mills amplitude can be expanded to BCJ basis with three particles being fixed in certain positions relating to the color-ordering, as
\bea A_n (1,\b_1,...,\b_r,2,\a_1,...,\a_{n-r-3},n) & = &
\sum_{\{\xi\} \in \{\b\} \shuffle_{\cal P} \{\a\}} {\cal
C}_{\{\a\},\{\b\};\{\xi\}}
A_n(1,2,\{\xi\},n)~.~~~\label{BCJ-exp-1}\eea
The expansion coefficients, namely BCJ coefficients, are firstly conjectured in \cite{Bern:2008qj} and later proven in \cite{Chen:2011jxa}, with the expression
\bea {\cal C}_{\{\a\},\{\b\};\{\xi\}} & = & \prod_{k=1}^r {{\cal
F}_{\b_k}(\{\a\},\{\b\};\{\xi\}) \over {\cal
K}_{1\b_1...\b_k}}~.~~~\label{BCJ-exp-2} \eea
Notations in above expression and explicit examples are presented in Appendix \ref{sec-appen-BCJ}.

Secondly, we review the differential operators which are originally introduced in \cite{Cheung:2017ems}. An important differential operator is the {\sl insertion operator}  defined by
\bea {\cal
T}_{ik(i+1)}:=\partial_{k_i\cdot \epsilon_k}-\partial_{k_{i+1}\cdot \epsilon_k}~.~~~\label{operator-insertion}
\eea
Physically it stands for changing a graviton $k$ into a gluon and inserting it
between $i$ and $i+1$ in the color ordering of gluons. If two gluons are
not adjacent, for instance $i, i+2$, we will have
\bea {\cal T}_{ik(i+2)}={\cal T}_{ik(i+1)}+{\cal
T}_{(i+1)k(i+2)}~,~~~\label{Inser-not-nearby}\eea
and its physical meaning is also clear\footnote{If $i,j$ are not in the same trace,
it has no clear physical meaning.}. Another important operator is the {\sl gauge invariance differential operator}, defined as
\begin{equation}
{\cal G}_{a}:=\sum_{i\neq a} (k_a\cdot k_i) \frac{\partial}{\partial (\epsilon_a\cdot k_i)}+ \sum_{j\neq a} (k_a\cdot \epsilon_j) \frac{\partial}{\partial (\epsilon_a\cdot \epsilon_j)}~.~~~\label{operator-Wald}
\end{equation}
It has a physical meaning of imposing gauge invariance, {\sl i.e.}, changing $\eps_a\to k_a$. For an arbitrary polynomial of polarization vectors and momenta, if it vanishes under operator ${\cal G}_a$, we can conclude it is gauge invariant for polarization vector $\epsilon_a$. Gauge invariance operators are commutative, {\sl i.e.}, $[{\cal G}_{a},{\cal G}_{b}]=0$, so the result of a multiplication of a sequential operators does not depend on the ordering, and we can denote a sequential gauge invariance operator as
\bea {\cal G}_{i_1 i_2...i_s}:= {\cal G}_{i_1}{\cal G}_{i_2}\cdots {\cal G}_{i_s}~~~,~~~i_1<i_2<\cdots <i_s~.~~~\label{G-more}\eea
The insertion operator and gauge invariance operator satisfy the following commutative relation,
\begin{align}
[\mathcal{T}_{ijk},\mathcal{G}_{l}]=\delta_{li}\mathcal{T}_{ij}-\delta_{lk}\mathcal{T}_{jk}~,~~~ \label{operator-equation}
\end{align}
with $T_{ij}:=\partial_{(\epsilon_i\cdot \epsilon_j)}$, and it is valid after applying to any functions of polarization vectors and momenta\footnote{For detailed description of these differential operators and their relations please refer to paper \cite{Cheung:2017ems}.}.

Finally let us present a general discussion on the expansion of EYM amplitude to BCJ basis. For particles with spin, the corresponding Lorentz representations are carried out by polarizations, {\sl e.g.}, polarization vector $\tilde{\epsilon}_i^{~\mu}$ for gluon and polarization tensor $\epsilon_{h_i}^{\mu\nu}$ for graviton. When expanding EYM amplitude to BCJ basis, the polarization tensor of graviton is factorized into two parts $\epsilon_{h_i}^{\mu\nu}=\tilde{\epsilon}_{h_i}^{~\mu}\otimes \epsilon_{h_i}^{\nu}$. The part $\tilde{\epsilon}_{h_i}^{~\mu}$ is inherent by the polarization vector of gluon in Yang-Mills basis, while the other part $\epsilon_{h_i}^{\nu}$ is absorbed into expansion coefficients. More explicitly, the expansion coefficients are rational function of momenta $k^\mu_\kappa, \kappa=1,\ldots,n,h_1,\ldots,h_m$ and polarization vectors $\epsilon^\mu_{h_\kappa}, \kappa=1,\ldots,m$. A crucial difference between expanding to KK basis and BCJ basis is that, the BCJ basis is truly an algebraic independent basis and the corresponding expansion coefficients must be {\sl gauge invariant}, {\sl i.e.},
\begin{equation} A^{\EYM}=\sum c_{{\tiny \mbox{gauge-inv}}}\times (A^{\YM}~ \mbox{in BCJ basis})~.~~~\label{rearrangeEYM-0}\end{equation}
This observation inspires us to consider another form of expansion
\begin{equation}A^{\EYM}=\sum (\mbox{linear sum of}~A^{\YM})\times b_{{\tiny\mbox{gauge-inv}}} ~.~~~\label{rearrangeEYM}\end{equation}
In the former formulation \eref{rearrangeEYM-0}, independent Yang-Mills amplitudes are taken to be expansion basis, and each coefficient as a function of momenta and polarization vectors $\epsilon_{h_\kappa}$ should satisfy conditions of gauge invariance for \textit{all} $\epsilon_{h_\kappa}$ with $\kappa=h_1, h_2,\ldots,h_m$. In the latter formulation \eref{rearrangeEYM}, $b_{{\tiny\mbox{gauge-inv}}}$'s are the expansion basis and the expansion coefficients become a linear combination of $A^{\YM}$'s with coefficients being rational functions of momenta. The later form has already appeared in \cite{Feng:2019tvb}, and in order to distinguish the two different kinds of basis we call the later ones $b_{{\tiny\mbox{gauge-inv}}}$ as {\sl gauge invariant building blocks}\footnote{Although we already know the formulation \eref{rearrangeEYM} is more suitable for applying differential operators, in paper \cite{Feng:2019tvb} we are not able to push the discussion further since the discussion of building blocks are too difficult at that time.}.

%%%%%%%%%%%%%%%%%%
\section{Building up expansion basis in gauge invariant vector space}
\label{sec-basis}
%%%%%%%%%%%%%%%%%

As mentioned earlier, in the expansion of EYM amplitudes, the gauge invariant coefficients $c_{{\tiny \mbox{gauge-inv}}}$ as well as expansion basis $b_{{\tiny\mbox{gauge-inv}}}$ are crucial. They are polynomial functions of polarization vectors which vanish under conditions of gauge invariance. In this section we would like to start from a most general vector space and localize a gauge invariant subspace of it. The expansion basis we are looking for is living in this subspace exactly.

%%%%%%%%%%%%%%%%%%%
\subsection{Gauge invariant vector space and its dimension}
%%%%%%%%%%%%%%%%%%%%%%

Let us start from the most general polynomial $\mathfrak{h}$, constructed by Lorentz contractions of $n$ momenta $k_1,k_2,\ldots,k_n$ and $m$ polarizations $\epsilon_1,\ldots,\epsilon_m$ with $m\leq n$. By Lorentz invariance and {\sl multi-linearity} of $\eps_i$, this polynomial must be the form schematically as
\begin{equation}
 \mathfrak{h}_{n,m}(k_1,\ldots,k_n,\epsilon_1,\ldots,\epsilon_m)=\alpha_0(\epsilon\cdot k)^m+\alpha_1(\epsilon\cdot \epsilon)(\epsilon\cdot k)^{m-2}+\cdots+\alpha_{\lfloor\frac{m}{2}\rfloor}(\epsilon\cdot \epsilon)^{\lfloor\frac{m}{2}\rfloor}(\epsilon\cdot k)^{m-\lfloor\frac{m}{2}\rfloor}~,~~~\label{gen-c}
\end{equation}
where for each monomial the degree of $\epsilon$ is $m$ and each $\epsilon_i, i=1,\ldots,m$ appears once and only once, while the coefficients $\a$'s are rational functions of  Mandelstam variables of momenta. If we take all monomials $\mathbb{B}[V]:=\{ (\epsilon\cdot \epsilon)^j(\epsilon\cdot k)^{m-2j}~,~0\leq j\leq \lfloor\frac{m}{2}\rfloor \}$ as a generating set,\footnote{These monomials are not linearly independent. There are relations between them generating by momentum conservation $\sum_i k_i=0$ and transverse condition $\eps_i \cdot k_i=0$. Furthermore, we consider only the parity even case, {\sl i.e.}, without total antisymmetric tensor $\eps_{\mu_1...\mu_D}$.}
 then we can build up a vector space $\mathcal{V}_{n,\{\eps_1,...,\eps_m\}}$ over the filed of rational functions of Mandelstam variables, where any polynomial $\mathfrak{h}_{n,m}$ belongs to this vector space.

In order to carve out the gauge invariant vector space from $\mathcal{V}_{n,\{\eps_1,...,\eps_m\}}$, let us impose gauge invariant conditions on $\mathfrak{h}_{n,m}$. This can be achieved by applying differential operators ${\cal G}_i$'s  to \eref{gen-c}, {\sl i.e.},
\begin{equation*}
  {\cal G}_i~\mathfrak{h}_{n,m}:=\mathfrak{h}_{n,m-1}(\epsilon_i\rightarrow k_i)~~~\mbox{for~each} ~~~ i=1,...,m~.
\end{equation*}
Such operator establishes a linear mapping between different vector spaces as
\bea \mathcal{V}_{n,\{\eps_1,...,\eps_m\}}\xrightarrow{\mathcal{G}_{t}} \mathcal{V}_{n,\{\eps_1,...,\eps_{t-1},\WH \eps_t,\eps_{t+1},...,\eps_m\}}~,~~~\label{map-1}\eea
where in the resulting vector space the polarization $\epsilon_t$ does appear and is replaced by $k_t$, denoted by $\WH \epsilon_t$. This linear map is
{\it surjective}\footnote{The property of surjectivity is the cornerstone in our discussion. For the vector space of polynomials without term $(\epsilon\cdot k)^m$ surjectivity of the map no longer holds.} by noticing the reduction of $\mathbb{B}[V]$, {\sl i.e.},
\bea \mathrm{Im}~\mathcal{G}_t[\mathcal{V}_{n,\{\eps_1,...,\eps_m\}}]= \mathcal{V}_{n,\{\eps_1,...,\eps_{t-1},\WH \eps_t,\eps_{t+1},...,\eps_m\}}~.~~~\label{map-2}\eea
We can successively apply different gauge invariant operators ${\mathcal{G}}_i$'s, $i=1,\ldots,m$ and establish a mapping chain of vector spaces. Since all ${\mathcal{G}}_i$'s are commutative, the result dose not depend on the ordering of successive applying, and we can denote the mapping chain as
\bea \mathcal{V}_{n,m}\xrightarrow{{\cal G}_{i_1 i_2...i_s}} \mathcal{V}^{(i_1 i_2...i_s)}_{n,m-s}~.~~~\label{map-3}\eea
The superscripts label the removed polarization vectors $\epsilon_i$ in the vector space. Note that different orderings of applying $\mathcal{G}_i$'s produce different mapping chains which at the end lead to the same vector space, so (\ref{map-3}) in fact represents a collection of mapping chains.

The kernel of linear map $\mathcal{G}_i:~\mathcal{V}_{n,s}\to \mathcal{V}_{n,s-1}^{(i)}$ is defined by
\begin{equation}
\mathrm{Ker} ~\mathcal{G}_i[\mathcal{V}_{n,s}]=\{~v\in \mathcal{V}_{n,s}~|~~\mathcal{G}_i[\mathcal{V}_{n,s}]=0~~\}~.~~~\label{kernel-G}
\end{equation}
Physically it means that the vectors of kernel are gauge invariant for $i$-th particle. Using the fact that the linear map is surjective \eref{map-3}, by {\sl fundamental theorem of linear map} \cite{axler2015linear}, we get
\begin{equation}
\dim \mathcal{V}_{n,s+1}=\dim \mathrm{Ker}~\mathcal{G}_i[\mathcal{V}_{n,s+1}]+\dim \mathrm{Im}~\mathcal{G}_i[\mathcal{V}_{n,s+1}]=\dim \mathrm{Ker}~\mathcal{G}_i[\mathcal{V}_{n,s+1}]+\dim \mathcal{V}_{n,s}~.~~~\label{dim-compute}
\end{equation}
Then the dimension of kernel can be computed by the difference of dimensions of vector space as
\begin{equation} \dim \mathrm{Ker}~\mathcal{G}_i[\mathcal{V}_{n,s+1}]=
\dim \mathcal{V}_{n,s+1}-\dim \mathcal{V}_{n,s}~.~~~\label{dim-compute-1}
\end{equation}
When applying more than one $\mathcal{G}_i$'s, this relation can be generalized to
\begin{equation} \dim \mathrm{Ker}~\mathcal{G}_{i_1 i_2.. i_t}[\mathcal{V}_{n,s}]=
\dim \mathcal{V}_{n,s}-\dim \mathcal{V}_{n,s-t}~.~~~\label{dim-compute-2}
\end{equation}
For example let us consider the simplest case $s=1$,
\begin{equation}
\dim \mathrm{Ker}~\mathcal{G}_1[\mathcal{V}_{n,1}]=\dim \mathcal{V}_{n,1}-\dim \mathcal{V}_{n,0}~.~~~\label{dim-compute-s=0}
\end{equation}
Vector space $\mathcal{V}_{n,0}$ is the field of rational functions of Mandelstam variables, so the basis is just $1$ and $\dim \mathcal{V}_{n,0}=1$. For vector space with only one polarization, the kernel $\mathrm{Ker}~\mathcal{G}_1[\mathcal{V}_{n,1}]$ consists of all vectors vanishing under gauge invariant operator. This is the gauge invariant vector sub-space $\mathcal{W}_{n,1}$ in a vector space $\mathcal{V}_{n,1}$. Thus we get
\begin{equation}
\dim \mathcal{W}_{n,1}:=\dim \mathrm{Ker}~\mathcal{G}_1[\mathcal{V}_{n,1}]=\dim \mathcal{V}_{n,1}-1~.~~~\label{dim-W-m=1}
\end{equation}
For a general vector space $\mathcal{V}_{n,m}$ with $m$ polarizations, we can define {\sl the gauge invariant vector sub-space} as the intersection of kernels of all possible linear maps
$\mathcal{G}_i$'s as,
\begin{equation}\mathcal{W}_{n,m}:=\bigcap_{i=1}^m \mathrm{Ker}~\mathcal{G}_i[\mathcal{V}_{n,m}]=\{~v\in \mathcal{V}_{n,m}~|~~\mathcal{G}_i(v)=0~~\forall i=1,2,\ldots,m~\}~.~~~\label{space-W}\end{equation}
This means that a vector in $\mathcal{W}_{n,m}$ would vanish under any linear map $\mathcal{G}_i$. This is exactly the sub-space where all gauge invariant coefficients $c_{{\tiny \mbox{gauge-inv}}}$ of \eref{rearrangeEYM-0} and the expansion basis $b_{{\tiny\mbox{gauge-inv}}}$ of \eref{rearrangeEYM} live.

Let us try to compute the dimension of $\mathcal{W}_{n,m}$ and start with the case $m=2$. Generally, for any two linear spaces $U_1, U_2$, we have the following relation for the dimension\footnote{ Suppose $U_1,\cdots, U_m$ are subspaces of V, then the {\sl sum} of $U_1,\cdots, U_m$ is defined as the set of all possible sums of elements of $U_1,\cdots, U_m$, explicitly $U_1+\cdots+U_m=\{u_1+\cdots+u_m:u_1\in U_1,\cdots, u_m\in U_m\}$. We should note that the definition of sum is different of direct sum, a sum $U_1+U_2$ is a {\sl direct sum} $U_1\oplus U_2$ if and only if $U_1\cap U_2=\{0\}$, and for direct sum $\dim(U_1\oplus U_2)=\dim U_1+\dim U_2$. The definitions are following \cite{axler2015linear}. },
\bea \dim U_1+ \dim U_2= \dim (U_1+U_2)-\dim (U_1\bigcap U_2)~.~~~\label{V-rel-1}\eea
Apply this relation to the vector spaces of kernels, {\sl i.e.}, $U_i=\mathrm{Ker}~\mathcal{G}_i[\mathcal{V}_{n,m}]$, we get
\bea \dim \mathcal{W}_{n,2}&:= & \dim (\mathrm{Ker}~\mathcal{G}_1\cap \mathrm{Ker}~\mathcal{G}_2)=\dim \mathrm{Ker}~\mathcal{G}_1+\dim \mathrm{Ker}~\mathcal{G}_2-\dim (\mathrm{Ker}~\mathcal{G}_1+\mathrm{Ker}~\mathcal{G}_2)~.~~~\label{m=2-1}\eea
The first two terms in the RHS can be computed by \eref{dim-compute-1}, and in order to compute the third term, we need to use the following proposition\footnote{Proof of proposition 1 and proposition 2 can be found in Appendix \ref{appendix-proof}.},
\begin{quote}
{\bf P{\footnotesize ROPOSITION 1 }}: {\sl any two kernels of linear maps $\mathcal{G}_{i}$'s satisfy} {\sl  the splitting formula},
\begin{equation}
\mathrm{Ker}~\mathcal{G}_1+\mathrm{Ker}~\mathcal{G}_2=\mathrm{Ker}~\mathcal{G}_{12}
~,~~~\label{proposition1}
\end{equation}
\end{quote}
and its generalization,
\begin{quote}
{\bf P{\footnotesize ROPOSITION 1 EXTENDED }}: {\sl the kernels of linear maps $\mathcal{G}_{i}$'s satisfy the generalized splitting formula},
\begin{equation}
\mathrm{Ker}~\mathcal{G}_1+\mathrm{Ker}~\mathcal{G}_2+\cdots +\mathrm{Ker}~\mathcal{G}_m=\mathrm{Ker}~\mathcal{G}_{12...m}~.~~~\label{proposition1ex}
\end{equation}
\end{quote}
Together with \eref{dim-compute-2}, we can rewrite \eref{m=2-1} as
\bea \dim \mathcal{W}_{n,2}=2 (\dim \mathcal{V}_{n,2}-\dim \mathcal{V}_{n,1})-(\dim \mathcal{V}_{n,2}-\dim \mathcal{V}_{n,0})=\dim \mathcal{V}_{n,2}-2\dim \mathcal{V}_{n,1}+\dim \mathcal{V}_{n,0}~.~~~\label{dim-W-n=2}\eea

Recursively using \eref{V-rel-1}, we want to generalize above result to arbitrary $m$. For simplicity let us denote $U_i:=\mathrm{Ker}~\mathcal{G}_i$, and when $m=3$ we get
\begin{eqnarray}
\dim (U_1+U_2+U_3)&=&\dim (U_1+U_2)+\dim U_3-\dim ((U_1+U_2)\cap U_3)\nonumber\\
&=&\dim U_1+\dim U_2+\dim U_3-\dim (U_1\cap U_2)-\dim ((U_1+U_2)\cap U_3)~.~~~\label{gen-m-1}
\end{eqnarray}
In the second line, the first three terms have already been computed, while in order to compute the fourth term we need to use the following proposition\footnote{In general $(U_1+U_2)\cap U_3=U_1\cap U_3+U_2\cap U_3$ is not true. For example, in a two-dimension space $U$, let us choose $U_1, U_2, U_3$ to be line $y=0$, $x=0$ and $x=y$ respectively. Then $U_1+U_2$ is the whole XY-plane, and $(U_1+U_2)\cap U_3$ is the line $x=y$. While in the RHS, $U_1\cap U_3$ and $U_2\cap U_3$ are just the origin $(0,0)$. So the RHS is a point.}
\begin{quote}
{\bf P{\footnotesize ROPOSITION 2 }}: {\sl three kernels of linear maps $\mathcal{G}_{i}$'s satisfy} {\sl  the distribution formula},
\begin{equation}
(\mathrm{Ker}~\mathcal{G}_1+\mathrm{Ker}~\mathcal{G}_2)\cap \mathrm{Ker}~\mathcal{G}_3=\mathrm{Ker}~\mathcal{G}_1\cap \mathrm{Ker}~\mathcal{G}_3+\mathrm{Ker}~\mathcal{G}_2\cap \mathrm{Ker}~\mathcal{G}_3~,~~~\label{proposition2-0}
\end{equation}
\end{quote}
and its generalization
\begin{quote}
{\bf P{\footnotesize PROPOSITION 2 EXTENDED }}: {\sl the kernels of linear maps $\mathcal{G}_{i}$'s satisfy the generalized distribution formula},
\begin{equation}
\left(\sum_{i=1}^{m-1}\mathrm{Ker}~\mathcal{G}_i\right)\cap \mathrm{Ker}~\mathcal{G}_m=\sum_{i=1}^{m-1}\mathrm{Ker}~\mathcal{G}_i\cap\mathrm{Ker}~\mathcal{G}_m~.~~~\label{proposition2ex}
\end{equation}
\end{quote}
Together with \eref{V-rel-1}, we can rewrite \eref{gen-m-1} as
\begin{eqnarray}
&&\dim (\mathrm{Ker}~\mathcal{G}_1+\mathrm{Ker}~\mathcal{G}_2+\mathrm{Ker}~\mathcal{G}_3)=\dim \mathrm{Ker}~\mathcal{G}_1+\dim \mathrm{Ker}~\mathcal{G}_2+\dim \mathrm{Ker}~\mathcal{G}_3-\dim (\mathrm{Ker}~\mathcal{G}_1\cap \mathrm{Ker}~\mathcal{G}_2)\nonumber\\
&&~~~~~~~-\dim (\mathrm{Ker}~\mathcal{G}_1\cap \mathrm{Ker}~\mathcal{G}_3)-\dim (\mathrm{Ker}~\mathcal{G}_2\cap \mathrm{Ker}~\mathcal{G}_3)+\dim (\mathrm{Ker}~\mathcal{G}_1\cap \mathrm{Ker}~\mathcal{G}_2\cap \mathrm{Ker}~\mathcal{G}_3)~.~~~\label{dim-three-kernel}
\end{eqnarray}
In equation \eref{dim-three-kernel}, in order to compute the dimension $\dim \mathcal{W}_{n,3}:=\dim (\mathrm{Ker}~\mathcal{G}_1\cap \mathrm{Ker}~\mathcal{G}_2\cap \mathrm{Ker}~\mathcal{G}_3)$, we need the result of $\dim (\mathrm{Ker}~\mathcal{G}_1+\mathrm{Ker}~\mathcal{G}_2+\mathrm{Ker}~\mathcal{G}_3)$, which by  proposition 1 extended (\ref{proposition1ex}) it equals to $\dim \mathrm{Ker}~\mathcal{G}_{123}$. Using \eref{dim-compute-2}, we get
\begin{equation}
\dim \mathrm{Ker}~\mathcal{G}_i=\mathcal{V}_{n,3}-\mathcal{V}_{n,2}~~~,~~~ \dim \mathrm{Ker}~\mathcal{G}_{ij}=\mathcal{V}_{n,3}-\mathcal{V}_{n,1}~~~,~~~\dim \mathrm{Ker}~\mathcal{G}_{ijk}=\mathcal{V}_{n,3}-\mathcal{V}_{n,0}~.~~~
\end{equation}
Then
\begin{equation}
  \dim \mathcal{W}_{n,3}=\dim \mathcal{V}_{n,3}-3\dim \mathcal{V}_{n,2}+3\dim \mathcal{V}_{n,1}-\dim \mathcal{V}_{n,0}~.~~~
\end{equation}
Notice that the numerical factors $1,3,3,1$ are nothing but $\binom{3}{i}$ for $i=0,1,2,3$.

Let us proceed further to arbitrary $m$. With proposition 1 extended and proposition 2 extended, equations (\ref{m=2-1}) and (\ref{dim-three-kernel}) are exactly the same as {\sl the principle of inclusion-exclusion}. By the well-known principle of inclusion-exclusion, we get
\begin{equation}
\dim \left(\sum_{i=1}^{m} \mathrm{Ker}~ \mathcal{G}_i\right)=\sum_{s=1}^{m}(-)^{s-1}
\sum_{{\tiny\mbox{all}}~s-{\tiny \mbox{subsets}}} \dim \left(\bigcap_{j=1}^{s} \mathrm{Ker}~\mathcal{G}_{i_j}\right)~,~~~\label{gen-m-I}
\end{equation}
where the second summation is over all subsets with $s$ indices.
It is also well-known that starting from the principle of inclusion-exclusion we can arrive at
\begin{equation}
\dim \left(\bigcap_{i=1}^{m} \mathrm{Ker}~\mathcal{G}_i\right)=\sum_{s=1}^{m}(-)^{s-1}\sum_{{\tiny\mbox{all}}~s-{\tiny \mbox{subsets}}} \dim \left(\sum_{j=1}^{s} \mathrm{Ker}~\mathcal{G}_{i_j}\right)~.~~~\label{gen-m-II}
\end{equation}
By proposition 1 extended, we can write
\begin{equation}
\dim \left(\sum_{j=1}^{s} \mathrm{Ker}~\mathcal{G}_{i_j}\right)=\dim \mathrm{Ker}~\mathcal{G}_{i_1i_2\cdots i_s}=\dim \mathcal{V}_{n,m}-\dim \mathcal{V}_{n,m-s}~.~~~\label{gen-m-III}
\end{equation}
Substituting (\ref{gen-m-III}) back to (\ref{gen-m-II}), we get
\begin{align}
 \dim \mathcal{W}_{n,m}:= \text{dim}\ (\bigcap_{i=1}^m \text{Ker}\ \mathcal{G}_i)
 =& \sum_{s=1}^m \sum_{i_1<\cdots<i_s}(-1)^{s-1} ( \text{dim}\ \mathcal{V}_{n,m}-\text{dim}\ \mathcal{V}_{n,m-s}^{(i_1\cdots i_s)}) \notag\\
 =& \sum_{s=1}^m (-1)^{s-1}\binom{m}{s}\text{dim}\ \mathcal{V}_{n,m}+\sum_{s=1}^m (-1)^{s}\binom{m}{s}\text{dim}\ \mathcal{V}_{n,m-s} \notag\\
 =&  \sum_{s=0}^m (-1)^{s}\binom{m}{s}\text{dim}\ \mathcal{V}_{n,m-s}~,~~~ \label{dimension-W}
\end{align}
%
\iffalse
Using \eref{gen-m-I} and \eref{gen-m-II} alternatively to reduce down, and the result \eref{gen-m-III},
we finally arrive at the formula of the dimension of  generic gauge invariant vector space as
we get
\begin{align}
\text{dim}\ (\bigcap_{i=1}^m \text{Ker}\ \mathcal{G}_i)
 =& \sum_{s=1}^m \sum_{i_1<\cdots<i_s}(-1)^{s-1} \text{dim}(\text{Ker}\ \mathcal{G}_{i_1}+\cdots+\text{Ker}\ \mathcal{G}_{i_s}) \notag\\
 =& \sum_{s=1}^m \sum_{i_1<\cdots<i_s}(-1)^{s-1} \text{dim}\ \text{Ker}\ (\mathcal{G}_{i_1}\cdots\mathcal{G}_{i_s}) \notag\\
 =& \sum_{s=1}^m \sum_{i_1<\cdots<i_s}(-1)^{s-1} ( \text{dim}\ \mathcal{V}_{n,m}-\text{dim}\ \mathcal{V}_{n,m-s}^{(i_1\cdots i_s)}) \notag\\
 =& \sum_{s=1}^m \sum_{i_1<\cdots<i_s}(-1)^{s-1}\text{dim}\ \mathcal{V}_{n,m}-\sum_{s=1}^m \sum_{i_1<\cdots<i_s}(-1)^{s-1}\text{dim}\ \mathcal{V}_{n,m-s}^{(i_1\cdots i_s)} \notag\\
 =& \sum_{s=1}^m (-1)^{s-1}\binom{m}{s}\text{dim}\ \mathcal{V}_{n,m}+\sum_{s=1}^m (-1)^{s}\binom{m}{s}\text{dim}\ \mathcal{V}_{n,m-s} \notag\\
 =& \text{dim}\ \mathcal{V}_{n,m}+\sum_{s=1}^m (-1)^{s}\binom{m}{s}\text{dim}\ \mathcal{V}_{n,m-s} = \sum_{s=0}^m (-1)^{s}\binom{m}{s}\text{dim}\ \mathcal{V}_{n,m-s}. \Label{dimension-W}
\end{align}
%
or
%
\begin{equation}
\dim \mathcal{W}_{n,m}=\sum_{s=0}^{m}(-)^s\binom{m}{s}\dim \mathcal{V}_{n,m-s}~,~~~\Label{dim-Wnm}
\end{equation}
%
\fi
where the dimension of vector space $\mathcal{V}_{n,m}$ can be computed via\footnote{The counting of \eref{dim-Vnm} can be carried out as follows. Firstly we select $i$ pairs of $\eps$, and there are $\binom{m}{2i}$ choices, while each left $\eps$ can be contracted with $(n-2)$ momenta after $(\eps\cdot k_n)$ by momentum conservation. For $2i$ $\eps$'s, the number of different contractions is $\frac{(2i)!}{2^i~(i!)}$.}
\begin{equation}
\dim \mathcal{V}_{n,m}=\sum_{i=0}^{\lfloor \frac{m}{2}\rfloor} \binom{m}{2i} \frac{(2i)!}{2^i~(i!)}(n-2)^{m-2i}~.~~~\label{dim-Vnm}
\end{equation}
Hence the dimension of arbitrary gauge invariant vector space $\mathcal{W}_{n,m}$ can be computed by formula (\ref{dimension-W}) and (\ref{dim-Vnm}).

Let us present a few examples demonstrating the computation of dimensions. For the special case $m=n$, $\dim \mathcal{W}_{n,n}$ with first few $n$'s are listed as,
\begin{center}
\begin{tabular}{|c|c|c|c|c|c|c|c|}
\hline
~~~~~$n$~~~~~ & ~~~~~4~~~~~ & ~~~~~5~~~~~ & ~~~~~6~~~~~ & ~~~~~7~~~~~ & ~~~~~8~~~~~ & ~~~~~9~~~~~ & ~~~~~10~~~~~ \\
\hline
$\dim \mathcal{W}_{n,n}$  & 10 & 142 & 2364 & 45028 &  969980 &  23372550 &  623805784  \\
\hline
\end{tabular}
\end{center}
In paper \cite{Boels:2016xhc} the same result has  been provided up to $n=7$.\footnote{In paper \cite{Boels:2016xhc}, there are two types of spaces being considered. The another one is the space with at least one contraction between polarization vectors in polynomials, {\sl i.e.}, polynomials without monomial $(\epsilon\cdot k)^m$, which is exactly the vector space that Yang-Mills amplitudes live in. Its dimension is $(n-3)!$.}. Comparing with that result, our calculation shows more efficiency than that of solving linear equations of gauge invariance directly. Furthermore, several examples of $\dim \mathcal{W}_{n,m}$ and $\dim \mathcal{W}_{n+m,m}$ with arbitrary $n$ but definite value of $m$ are listed below as
\begin{center}
\begin{tabular}{|c|c|c|c|c|}
\hline
~~~~~~~~~$m$~~~~~~~~~ & ~~~~~~~~~1~~~~~~~~~ & ~~~~~~~~~2~~~~~~~~~ & ~~~~~~~~~3~~~~~~~~~ & ~~~~~~~~~4~~~~~~~~~ \\
\hline
$\dim \mathcal{W}_{n,m}$ & $n-3$  &  $(n-3)^2+1$  & $(n-3)^3+3(n-3)$  & $(n-3)^4+6(n-3)^2+3$ \\
\hline
$\dim \mathcal{W}_{n+m,m}$ &  $n-2$   & $(n-1)^2+1$ &  $n^3+3n$ &  $(n+1)^4+6(n+1)^2+3$ \\
\hline
\end{tabular}
\end{center}
%

%%%%%%%%%%%%%%%%%%%
\subsection{Gauge invariant vectors}
%%%%%%%%%%%%%%%%%%%%%%

The dimension of gauge invariant vector space characterizes the minimal number of vectors to expand an arbitrary vector, while the explicit form of vector is not constrained. From the working experiences of EYM amplitude expansion with one, two and three gravitons \cite{Feng:2019tvb}, we get the insight that the coefficients appearing therein could be recast in a manifestly gauge invariant form as linear combinations of multiplications of fundamental $f$-terms. Here the fundamental $f$-terms stand for two types of Lorentz contractions of field strength $f_i^{\mu\nu}=k_i^{\mu}\epsilon_i^{\nu}-\epsilon_i^{\mu}k_i^{\nu}$  and external momenta, with at most two $f_i$'s,
\begin{equation} \mbox{Fundamental}~f\mbox{-terms:}~~~~~~~~~ k_i\cdot f_a \cdot k_j~~~\mbox{and}~~~k_i\cdot f_a\cdot f_b\cdot k_j~.~~~ \label{fundamental-f-term}\end{equation}
This observation can be generalized beyond $m=3$, and it can be stated as follows. For any vector in gauge invariant vector space $\mathcal{W}_{n,m}$ with $m<n$\footnote{We should emphasize the condition $m<n$, which is different from previous discussion where $m$ could equal to $n$. Proof of the statement in this subsection can not be trivially generalized to the $m=n$ case, so if results in this subsection could be applied to the case $m=n$ is still a question for us. },
\begin{quote}
{\sl Every vector in $\mathcal{W}_{n,m}$ can be recast in a manifestly gauge invariant form, which is a linear combination of the multiplications of fundamental $f$-terms with the total number of field strength $f$ in every monomial being $m$.}
\end{quote}
We shall prove this statement by induction. The cases with $m=1,2,3$ have already been shown to be true in \cite{Feng:2019tvb}. Following the idea of induction, we assume that this statement is true for all $s<m$, and prove that it must be true for $m$.

A polynomial $\mathfrak{h}_{n,m}\in \mathcal{W}_{n,m}$ with $m$ polarizations $\epsilon_1,\epsilon_2,\ldots, \epsilon_m$ can be generally written as
\begin{equation}
\mathfrak{h}_{n,m}=\sum_{i=2}^{m}(\epsilon_1\cdot \epsilon_i) T_{1i}+\sum_{i=2}^{m}(\epsilon_1\cdot k_i)(\epsilon_i\cdot T'_{1i})+\sum_{i=m+1}^{n-1}(\epsilon_1\cdot k_i)T''_{1i}~,~~~\label{vector-hnm}
\end{equation}
where momentum conservation has been applied to eliminate $\epsilon_1\cdot k_n$, so that all $(\epsilon_1\cdot \epsilon_i), (\epsilon_1\cdot k_i)$ appearing in $\mathfrak{h}_{n,m}$ are linearly independent. Polynomials $T_{1i}\in \mathcal{V}_{n,m-2}$ and $\epsilon_i\cdot T'_{1i}~,~ T''_{1i}\in \mathcal{V}_{n,m-1}$. Since $\mathfrak{h}_{n,m}\in \mathcal{W}_{n,m}$, by definition we have
\begin{equation}
\mathcal{G}_a ~\mathfrak{h}_{n,m}=0~~~,~~~\forall(1\leq a\leq m)~.~~~
\end{equation}
From the operator equation (\ref{operator-equation}), we explicitly have $[\mathcal{T}_{a1n},\mathcal{G}_a]=\mathcal{T}_{a1}$ with $a=2,\cdots,m$. Applying them to $\mathfrak{h}_{n,m}$ generates a set of equations as
\begin{align}
  [\mathcal{T}_{a1n},\mathcal{G}_a]\mathfrak{h}_{n,m} = \mathcal{T}_{a1}\mathfrak{h}_{n,m}
~\to~ & -\mathcal{G}_a(\partial_{\epsilon_1\cdot k_a}-\partial_{\epsilon_1\cdot k_n})\mathfrak{h}_{n,m} = \partial_{\epsilon_1\cdot \epsilon_a}\mathfrak{h}_{n,m}
~\to~  -(k_a\cdot T'_{1a})= T_{1a}~,~~~
\end{align}
where we have considered the fact that $\mathfrak{h}_{n,m}$ does not contain $(\epsilon_1\cdot k_n)$. With above result we can rewrite $\mathfrak{h}_{n,m}$  as
\begin{align}
 \mathfrak{h}_{n,m}
=&\sum_{i_1=2}^{m}(\epsilon_1\cdot f_{i_1} \cdot T'_{1i_1})
+\sum_{i_1=m+1}^{n-1}(\epsilon_1\cdot k_{i_1})T''_{1i_1}~.~~~
\end{align}
We also need to consider the gauge invariance of $\mathfrak{h}_{n,m}$ with respect to polarization vector $\epsilon_1$,
\begin{align}
 \mathfrak{h}_{n,m}(\epsilon_1\rightarrow k_1)
=\sum_{i_1=2}^{m}(k_1\cdot f_{i_1} \cdot T'_{1i_1})
+\sum_{i_1=m+1}^{n-1}(k_1\cdot k_{i_1})T''_{1i_1}
=0~.~~~
\end{align}
Then we get
\begin{align}
 T''_{1(n-1)}=-\sum_{i_1=2}^{m}\frac{(k_1\cdot f_{i_1} \cdot T'_{1i_1})}{(k_1\cdot k_{n-1})}
-\sum_{i_1=m+1}^{n-2} \frac{(k_1\cdot k_{i_1})}{(k_1\cdot k_{n-1})} T''_{1i_1}~.~~~ \label{equ:coeE1}
\end{align}
After substituting above results back to $\mathfrak{h}_{n,m}$, we get
\begin{align}
 \mathfrak{h}_{n,m}
=& \sum_{i_1=2}^{m}\frac{(k_{n-1}\cdot f_1\cdot f_{i_1} \cdot T'_{1i_1})}{(k_1\cdot k_{n-1})}
+\sum_{i_1=m+1}^{n-2} \frac{(k_{n-1}\cdot f_1\cdot k_{i_1})}{(k_1\cdot k_{n-1})} T''_{1i_1}~.~~~
\end{align}
So $\mathfrak{h}_{n,m}$ is already manifestly gauge invariant for polarization vector $\epsilon_1$. In fact, we can also choose to eliminate other coefficients in (\ref{equ:coeE1}) and introduce different poles in denominator of $\mathfrak{h}_{n,m}$.

We can also generate another set of equations by considering the operator relations $[\mathcal{T}_{i1n},\mathcal{G}_a]=0$ with $i=m+1,\cdots,n-2$ and $a=2,\cdots,m$. Applying them to $\mathfrak{h}_{n,m}$ produces
\begin{align}
 [\mathcal{T}_{i1n},\mathcal{G}_a]\mathfrak{h}_{n,m}=0 \
 ~\to~ \quad  -\mathcal{G}_a(\partial_{\epsilon_1\cdot k_i}-\partial_{\epsilon_1\cdot k_n})\mathfrak{h}_{n,m}=0 \
 ~\to~ \quad  \mathcal{G}_aT''_{1i}=0~,~~~
\end{align}
which means $T''_{1i_1}$ is gauge invariant for $\epsilon_{2},\epsilon_3,\cdots,\epsilon_m$. By assumption of induction, $T''_{1i_1}$ can be written as a linear combination of multiplication of fundamental $f$-terms. Because $\mathfrak{h}_{n,m}$ and $T''_{1i_1}$ are gauge invariant for $\epsilon_a$ with $a=2,\cdots,m$, and $(k_{n-1}f_1f_{i_1}T'_{1i_1})$'s are linearly independent, $T'_{1i_1}$ is also gauge invariant for all its own polarization vectors. Again by assumption of induction, any $(A f_{i_1} T'_{1i_1})$ can also be written in a manifest gauge invariant form with only $f$ appears. Thus as a linear function of $(k_{n-1}\cdot f_1\cdot f_{i_1} \cdot T'_{1i_1})$ and $T''_{1i_1}$, the polynomial $\mathfrak{h}_{n,m}$ can also be written in a manifest gauge invariant form, and we have proven the first part of our statement.

To complete our proof, we need to apply above procedure to $(\epsilon_{i_1}\cdot T'_{1i_1})$ in \eref{vector-hnm} and rewrite it as
\begin{align}
 (\epsilon_{i_1}\cdot T'_{1i_1})
 =\sum_{i_2=2,i_2\ne i_1}^{m}(\epsilon_{i_1}\cdot \epsilon_{i_2})T_{1i_1i_2}
   +\sum_{i_2=2,i_2\ne i_1}^{m} (\epsilon_{i_1}\cdot k_{i_2})(\epsilon_{i_2}\cdot T'_{1i_1i_2})
   +\sum_{i_2=m+1}^{n-1,1}(\epsilon_{i_1}\cdot k_{i_2})T''_{1i_1i_2}~,~~~
\end{align}
where in the last summation $i_2$ can equal to $1$. Let us again apply operator equations $[\mathcal{T}_{ai_1 n},\mathcal{G}_a]=\mathcal{T}_{ai_1}$, with $a=2,\cdots,m$ and $a\ne i_1$, which generates a set of equations,
\begin{align}
(\epsilon_{i_1}\cdot T'_{1i_1})
 =&\sum_{i_2=2,i_2\ne i_1}^{m} (\epsilon_{i_1}\cdot f_{i_2}\cdot T'_{1i_1i_2})
   +\sum_{i_2=m+1}^{n-1,1}(\epsilon_{i_1}\cdot k_{i_2})T''_{1i_1i_2}~.~~~
\end{align}
So $\mathfrak{h}_{n,m}$ becomes
\begin{align}
\mathfrak{h}_{n,m}
=&\sum_{i_1=2}^{m}\sum_{i_2=2,i_2\ne i_1}^{m} \frac{(k_{n-1}f_1f_{i_1}f_{i_2}T'_{1i_1i_2})}{(k_1k_{n-1})}
   +\sum_{i_1=2}^{m}\sum_{i_2=m+1}^{n-1,1}  \frac{(k_{n-1}f_1 f_{i_1} k_{i_2})}{(k_1k_{n-1})} T''_{1i_1i_2}
+\sum_{i_1=m+1}^{n-2} \frac{(k_{n-1}f_1 k_{i_1})}{(k_1k_{n-1})}T''_{1i_1}~.~~~
\end{align}
Then we apply $[\mathcal{T}_{ji_1n},\mathcal{G}_a]=0$ with $j=m+1,m+2,\cdots,n-1,1$ and $a=2,\cdots,i_1-1,i_1+1,\cdots,m$ to $(\epsilon_{i_1}\cdot T'_{1i_1})$, which leads to $\mathcal{G}_a T''_{1i_1j}=0$. It says that $T''_{1i_1i_2}$ is gauge invariant for its own polarization vectors, and it can be written as linear combination of multiplication of fundamental $f$-terms. For the same reason as before, we conclude that $T'_{1i_1i_2}$ is also gauge invariant for its own polarization vectors. Continuously applying the same procedure to $T'$ until to the last polarization vector, we would arrive at
\begin{equation}
\mathfrak{h}_{n,m}=\sum_{s=2}^{m}\widetilde{\mathfrak{h}}_{n,s}+\sum_{i=m+1}^{n-2}\frac{k_{n-1}\cdot f_1\cdot k_i}{k_1\cdot k_{n-1}}T''_{1i}~,~~~
\end{equation}
where
\begin{equation}
\widetilde{\mathfrak{h}}_{n,s}=\sum_{i_1=2}^{m} \sum_{i_2=2\atop i_2\neq i_1}^{m}\cdots \sum_{i_{s-1}=2\atop i_{s-1}\neq i_1,i_2,\ldots,i_{s-2}}^{m} \sum_{i_s=m+1\atop i_s=1,i_1,i_2,\ldots,i_{s-2}}^{n-1}\frac{k_{n-1}\cdot f_1\cdot f_{i_1}\cdots f_{i_{s-1}}\cdot k_{i_s}}{k_1\cdot k_{n-1}}T''_{(1i_1\cdots i_{s-1})i_s}~,~~~\label{vector-hnmtilde}
\end{equation}
with polynomial $T''_{(1i_1\cdots i_{s-1})i_s}\in \mathcal{W}_{n,m-s}$.

To further reduce the expression $(k\cdot f\cdots f\cdot k)$ to the fundamental $f$-terms,  we should get help from the following identities,
\bea (B\cdot f_p \cdot A)(C\cdot k_p)=(B\cdot f_p \cdot
C)(A\cdot k_p)+(C\cdot f_p \cdot A)(B\cdot
k_p)~,~~~\label{use-identity-1} \eea
where $A,B,C$ could be any strings. More explicitly, applying above identity to expression with three $f$'s, we get
\begin{equation}
(k_i\cdot f_{a_1}\cdot f_{a_2}\cdot f_{a_3}\cdot k_j)(k_l\cdot k_{a_2})=(k_i\cdot f_{a_1}\cdot k_{a_2})(k_l\cdot f_{a_2}\cdot f_{a_3}\cdot k_j)+(k_i\cdot f_{a_1}\cdot f_{a_2}\cdot k_l)(k_{a_2}\cdot f_{a_3}\cdot k_j)~.~~~
\end{equation}
So any $f$-term with any number of $f_i$'s can be reduced to fundamental $f$-terms, while at the same time $T''_{(1i_1\cdots i_{s-1})i_s}$ has been reorganized as a linear combination of multiplication of fundamental $f$-terms. This ends the proof of statement by induction method.

Before ending this subsection, let us take a look on another gauge invariant $f$-term that mentioned in \cite{Feng:2019tvb}, {\sl i.e.}, the trace ${\rm Tr}(f_{a_1} f_{a_2} \cdots f_{a_k})=f_{a_1}^{\mu\nu}f_{a_2,\nu\rho}\cdots f_{a_k, \mu}^{\sigma}$. It can be expanded as
\bea  {\rm Tr}(f_{a_1} f_{a_2}\cdots f_{a_s}\cdots  f_{a_k})&=& [(\eps_{a_1}\cdot f_{a_2}\cdots f_{a_s}\cdots f_{a_k}\cdot k_{a_1})
-(k_{a_1}\cdot f_{a_2}\cdots f_{a_s}\cdots  f_{a_k} \eps_{a_1})]{(A\cdot k_{a_s})\over (A\cdot k_{a_s})}\nn
& = & { (\eps_{a_1} f_{a_2}\cdots  f_{a_s}\cdot A) (k_{a_s}\cdot f_{a_{s+1}}\cdots f_{a_k}\cdot k_{a_1}) + (\eps_{a_1}\cdot f_{a_2}\cdots  f_{a_{s-1}}\cdot k_{a_s})(A\cdot f_{a_s}\cdots  f_{a_k}\cdot k_{a_1})\over (A\cdot k_{a_s})}\nn
& - &{ (k_{a_1}\cdot f_{a_2}\cdots  f_{a_s}\cdot A) (k_{a_s}\cdot f_{a_{s+1}}\cdots f_{a_k}\cdot \eps_{a_1}) + (k_{a_1}\cdot f_{a_2}\cdots  f_{a_{s-1}}\cdot k_{a_s})(A\cdot f_{a_s}\cdots  f_{a_k}\cdot \eps_{a_1})\over (A\cdot k_{a_s})}~,~~~\nonumber\eea
where identity \eref{use-identity-1} has been used in the derivation. Combining the first and third term as well as the second and fourth term, we can get
\bea
{\rm Tr}(f_{a_1} f_{a_2}\cdots f_{a_s}\cdots f_{a_k})
={  (k_{a_s}\cdot f_{a_{s+1}}\cdots  f_{a_k} f_{a_1} f_{a_2}\cdots  f_{a_s}\cdot A)\over (A\cdot k_{a_s})}+{ (A\cdot f_{a_s}\cdots f_{a_k} f_{a_1} f_{a_2}\cdots  f_{a_{s-1}}\cdot k_{a_s})\over (A\cdot k_{a_s})}~.~~~\label{trace-f}
\eea
A simple example is ${\rm Tr}(f_{a_1} f_{a_2})=2(k_{a_2}\cdot f_{a_1}\cdot f_{a_2}\cdot A)/ (A\cdot k_{a_2})$. So this type of gauge invariant $f$-terms, which is originally viewed as a new type different from $(kf\cdots fk)$, are also composed by fundamental $f$-term.

%%%%%%%%%%%%%%%%%%%
\subsection{Gauge invariant basis}
%%%%%%%%%%%%%%%%%%%%%%

Any gauge invariant vector in  $\mathcal{W}_{n,m}$ could be an element to form a gauge invariant basis $b_{{\tiny \mbox{gauge-inv}}}$ in the EYM amplitude expansion \eref{rearrangeEYM}. However, in order to turn a subset of $\mathcal{W}_{n,m}$ to a complete basis,
we should choose a set of vectors satisfying the following two properties,
\begin{enumerate}
\item all vectors in the set are linearly independent,
\item the number of vectors in the set equals to the dimension of gauge invariant vector space.
\end{enumerate}
Note that the fundamental $f$-terms are not completely independent from each other. For instance, using \eref{use-identity-1} it is easy to see that
\bea (k_i\cdot  f_a \cdot f_b \cdot k_j)  (k_1\cdot  k_a) =(k_i\cdot  f_a\cdot  k_1)(k_a\cdot  f_b\cdot  k_j)+(k_1\cdot  f_a\cdot  f_b\cdot  k_j)(k_i\cdot  k_a)~.~~~\label{f-term-red-1} \eea
So one can always reduce any fundamental $f$-terms to the following form,
\begin{equation}
k_1\cdot f_a\cdot f_b\cdot k_1~~~\mbox{and}~~~k_1\cdot f_a\cdot k_i~.~~~
\end{equation}
From the definition of $f_i^{\mu\nu}$, it's easy to get
\begin{equation}
k_1\cdot f_a\cdot f_b\cdot k_1=k_1\cdot f_b\cdot f_a\cdot k_1~~~,~~~k_1\cdot f_a\cdot k_1=0~~~,~~~k_1\cdot f_a\cdot k_a=0~.~~~
\end{equation}
In the case of $A^{\EYM}_{n,m}$, the momentum list is $\{k_1,\ldots,k_n,k_{h_1},\ldots,k_{h_m}\}$ while the polarization vector list is $\{\epsilon_{h_1},\ldots,\epsilon_{h_m}\}$, so by default the above subscripts $a,b\in \{h_1,\ldots,h_m\}$ and $i\in \{1,\ldots,n,h_1,\ldots,h_m\}$. After using momentum conservation to eliminate $k_n$, we can restrict the fundamental $f$-terms to be
\begin{eqnarray}
&& k_1\cdot f_{h_i}\cdot f_{h_j}\cdot k_1~~~,~~~1\leq i<j\leq m~,~~~\label{fundamental-f-term-kffk}\\
&& k_1\cdot f_{h_i}\cdot k_j~~~,~~~i\in \{1,\ldots,m\}~~~,~~~j\in\{2,\ldots,n-1,h_1,\ldots,h_m\}/\{h_i\}~.~~~\label{fundamental-f-term-kfk}
\end{eqnarray}
Using above fundamental $f$-terms,  we can construct a set of vectors as
\begin{equation}
\left(\prod_{i=1}^{s}  k_1\cdot f_{h_{\alpha_{2i-1}}}\cdot f_{h_{\alpha_{2i}}}\cdot k_1\right)\left(\prod_{i=2s+1}^{m}  k_1\cdot f_{h_{\beta_i}}\cdot k_j\right)~~~,~~~s=0,1,\ldots,\lfloor\frac{m}{2}\rfloor~,~~~\label{basis-original}
\end{equation}
with the convention
\begin{equation}
\alpha_{2i-1}<\alpha_{2i+1}~~\forall(1\leq i\leq s-1)~~,~~\alpha_{2i-1}<\alpha_{2i}~~\forall(1\leq i\leq s)~~,~~\beta_i<\beta_{i+1}~~~\forall(2s+1\leq i\leq m-1)~.~~~
\end{equation}
The linear independence of these vectors (\ref{basis-original}) is obvious. In order to demonstrate that they form a real basis of $\mathcal{W}_{n+m,m}$, we should show  the total number of these vectors equals to $\dim \mathcal{W}_{n+m,m}$ according to property 2. We can count the total number of independent vectors with respect to specific $s$ as
\begin{equation}
  \frac{m!}{s!~ 2^s~(m-2s)!} (n+m-3)^{m-2s}~~~\to~~~\#(\mbox{vectors})=\sum_{s=0}^{\lfloor\frac{m}{2} \rfloor} \frac{m!}{s!~ 2^s~(m-2s)!} (n+m-3)^{m-2s}~.~~~\label{dim-vector}
\end{equation}
According to (\ref{dimension-W}) and (\ref{dim-Vnm}), the dimension of $\mathcal{W}_{n+m,m}$ is
\begin{eqnarray}
\dim~\mathcal{W}_{n+m,m}&=&\sum_{s=0}^{m}\sum_{i=0}^{\lfloor\frac{m-s}{2} \rfloor}(-)^{s}\binom{m}{s}\binom{m-s}{2i}\frac{(2i)!}{2^i~(i!)}(n+m-2)^{m-s-2i}\nonumber\\
&=& \sum_{i=0}^{\lfloor \frac{m}{2}\rfloor}\sum_{s=0}^{m-2i}(-)^s\frac{m!}{(s!)((m-s-2i)!)~2^i~(i!)}(n+m-2)^{m-s-2i}~.~~~
\end{eqnarray}
Noticing the relation
\begin{equation*}
\sum_{s=0}^{m-2i}(-)^s \frac{(n+m-2)^{m-s-2i}}{(s!)((m-s-2i)!)}=\frac{(n+m-2-1)^{m-2i}}{(m-2i)!}~,~~~
\end{equation*}
we immediately get $\dim\mathcal{W}_{n+m,m}=\#(\mbox{vectors})$ defined in (\ref{basis-original}). Hence the set of vectors defined in (\ref{basis-original}) satisfies the required two conditions and could be chosen as an expansion basis for $A^{\EYM}_{n,m}$ in \eref{rearrangeEYM}. In practice we would prefer a basis with minimal dimension, then we define the fundamental $f$-terms as
\begin{eqnarray}
&& \mathsf{F}_{h_ih_j}:=\frac{k_1\cdot f_{h_i}\cdot f_{h_j}\cdot k_1}{(k_1\cdot k_{h_i})(k_1\cdot k_{h_j})}~~~,~~~1\leq i<j\leq m~,~~~\label{fundamental-f-term-Fhh}\\
&& \mathsf{F}_{h_i}^{h_j}:=\frac{k_1\cdot f_{h_i}\cdot k_{h_j}}{k_1\cdot k_{h_i}}~~~,~~~i\in\{1,\ldots,m\}~,~j\in\{1,\ldots,m\}/\{i\}~,~~~\label{fundamental-f-term-Fh}\\
&& \mathsf{F}_{h_i}^{a}:=\frac{k_1\cdot f_{h_i}\cdot K_a}{k_1\cdot k_{h_i}}~~~,~~~i\in \{1,\ldots,m\}~,~a\in\{2,\ldots,n-1\}~,~~~\label{fundamental-f-term-Fa}
\end{eqnarray}
where $K_a:= \sum_{i=2}^{a} k_i$. The vectors in the expansion basis can be constructed from above fundamental $f$-terms as
\begin{equation}
\prod_{i=1}^{p} {\sf{F}}_{h_{\alpha_{2i-1}}h_{\alpha_{2i}}}\prod_{i=1}^{q}{\sf{F}}_{h_{\beta_i}}^{h_{\beta'_i}}\prod_{i=1}^{r} {\sf{F}}_{h_{\gamma_i}}^{a_{\gamma_i}}~~~,~~~p,q,r\in\mathbb{N}~~~\mbox{and}~~~2p+q+r=m~,~~~\label{basis-def-2}
\end{equation}
with the convention
\begin{equation}
\begin{array}{l}
 \alpha_{2i-1}<\alpha_{2i+1}~~\forall(1\leq i\leq p-1)~~~,~~~\alpha_{2i-1}<\alpha_{2i}~~\forall(1\leq i\leq p) \\
 \beta_i<\beta_{i+1}~~\forall(1\leq i\leq q-1)~~~,~~~\gamma_i<\gamma_{i+1}~~\forall(1\leq i\leq r-1)
\end{array}
~.~~~\label{basis-def-indices}
\end{equation}
They contribute to a complete set of expansion basis, and a general EYM amplitude can be expanded into this basis as
\begin{align}
A^{{\tiny\mbox{EYM}}}_{n;m}(k_1,k_2,\ldots,k_n;\bold{H})
= &\sum_{h_{\beta'_{1}}\in \bold{H}/\{h_{\beta_1}\}}\cdots \sum_{h_{\beta'_{q}}\in \bold{H}/\{h_{\beta_q}\}}  \sum_{a_{\gamma_1}=2}^{n-1}\cdots \sum_{a_{\gamma_r}=2}^{n-1}      \nonumber\\
& \sum_{\mathfrak{a}\cup{\mathfrak{b}}\cup{\mathfrak{c}}=\bold{H}}{\mspace{-24mu}~'}~~~\mathcal{C}[{\sf{F}}_{h_{\alpha_1}h_{\alpha_2}}\cdots {\sf{F}}_{h_{\alpha_{2p-1}}h_{\alpha_{2p}}}{\sf{F}}_{h_{\beta_1}}^{h_{\beta'_1}}\cdots {\sf{F}}_{h_{\beta_q}}^{h_{\beta'_q}}{\sf{F}}_{h_{\gamma_1}}^{a_{\gamma_1}}\cdots{\sf{F}}_{h_{\gamma_r}}^{a_{\gamma_r}}]     \nonumber\\
& ~~~~~~~~~~~~\times \mathcal{B}[{\sf{F}}_{h_{\alpha_1}h_{\alpha_2}}\cdots {\sf{F}}_{h_{\alpha_{2p-1}}h_{\alpha_{2p}}}{\sf{F}}_{h_{\beta_1}}^{h_{\beta'_1}}\cdots {\sf{F}}_{h_{\beta_q}}^{h_{\beta'_q}}{\sf{F}}_{h_{\gamma_1}}^{a_{\gamma_1}}\cdots{\sf{F}}_{h_{\gamma_r}}^{a_{\gamma_r}}]~,~~~ \label{expansionGI}
\end{align}
where $\bold{H}/{h_i}$ is the set of gravitons excluding $h_i$, and the three sets $\mathfrak{a}=\{h_{\alpha_1},\ldots,h_{\alpha_{2p}}\}$, $\mathfrak{b}=\{h_{\beta_1},\ldots,h_{\beta_q}\}$, $\mathfrak{c}=\{h_{\gamma_1},\ldots,h_{\gamma_r}\}$ with $2p+q+r=m$ are a splitting of all gravitons. $\mathcal{B}[\cdots]$ represents a particular vector in the expansion basis $\mathcal{B}$, $\mathcal{C}[\cdots]$ represents the coefficient of the corresponding vector, and the reduced summation $\sum'$ runs over all possible splittings $\mathfrak{a}\cup\mathfrak{b}\cup\mathfrak{c}=\bold{H}$ with the prime meaning that terms with index circle should be excluded\footnote{Discussion of index circle can be found in \cite{Feng:2019tvb}, and we will return back to it later.}.  We can see that all the information of polarization vectors $\epsilon_{h_{\kappa}}$ is encoded in $\mathcal{B}$ as expected.

%%%%%%%%%%%%%%%%%%
\section{Determining expansion coefficients via differential operators} \label{sec-coefficient}
%%%%%%%%%%%%%%%%%

We have defined the gauge invariant expansion basis, and the next step is to determine the expansion coefficients. As earlier mentioned, the EYM amplitude can be expanded schematically in the form,
\begin{equation}
A^{\EYM}_{n,m}(1,2,\ldots,n;\mathbf{H})=(~~\mbox{Coefficients}~~) \otimes (~~\mbox{Gauge~Invariant~Basis}~~)~,~~~\label{expansionGI-abbr}
\end{equation}
or more explicitly see \eref{expansionGI}. The expansion coefficients are linear combinations of Yang-Mills amplitudes $A^{\YM}_{n+m}$. To use \eref{expansionGI-abbr} efficiently, a crucial point is to find a way to distinguish vectors in the gauge invariant basis from each other. Inspired from the explicit form of vectors in \eref{gen-c}, we notice that the signature of vectors is the structure $(\eps\cdot \eps)^p(\eps\cdot k)^q$, where $(\epsilon\cdot \epsilon)$'s and $(\epsilon\cdot k)$'s could be linearly independent. This motivates us to consider two kinds of differential operators as
\begin{equation}{\mathcal{T}}_{ah_ib}:=\frac{\partial}{\partial(\epsilon_{h_i}\cdot k_a)}-\frac{\partial}{\partial(\epsilon_{h_i}\cdot k_b)}~~~,~~~ {\mathcal{T}}_{ab}:=\frac{\partial}{\partial(\epsilon_a\cdot \epsilon_b)}~.~~~\label{operator-insertion-h}\end{equation}
Applying these operators to the RHS of \eref{expansionGI-abbr} all terms will vanish except those containing corresponding $(\epsilon\cdot \epsilon)$ and $(\epsilon\cdot k)$. While applying these operators to the LHS of \eref{expansionGI-abbr}, the physical meaning will be different. Applying ${\mathcal{T}}_{ab}$ to single-trace EYM amplitudes produces multi-trace EYM amplitudes which would complicate the amplitude expansion, however applying $\mathcal{T}_{ah_i b}$ to a single-trace EYM amplitude produces another single-trace EYM amplitude but with one less graviton. $\mathcal{T}_{ah_i b}$ will transform the graviton $h_i$ to a gluon $h_i$ and insert the gluon in the positions between gluons $a, b$ respecting the color-ordering. So each time applying an insertion operator to (\ref{expansionGI-abbr}), the number of gravitons is reduced by one, then a multiplication of $m$ insertion operators would transform the LHS of \eref{expansionGI-abbr} to Yang-Mills amplitudes completely, as expected\footnote{Alternatively, we could also apply less insertion operators to generate a set of linear equations of single-trace EYM amplitudes, and recursively use the expansion of single-trace EYM amplitude with less number of gravitons into Yang-Mills amplitudes.}.

In fact, we can take one step further and define a {\sl differential operator} as a multiplication of $m$ properly chosen insertion operators. When applying the differential operator to \eref{expansionGI}, besides some vectors with already known coefficients, there would be one and only one vector with unknown coefficient in the RHS of \eref{expansionGI} remains, and all other vectors vanish,
\begin{equation} \mbox{Differential~Operator~on}~A^{\EYM}_{n,m}=~\mbox{Coefficient}\times (~~\mbox{Differential~Operator~on}~\mathcal{B}~~)~.~~~\end{equation}
As a consequence, we get a linear equation with only one unknown variable, and the corresponding expansion coefficient can be computed directly as a function of $A^{\YM}_{n+m}$ that generated by differential operator applying on the RHS of \eref{expansionGI}\footnote{The idea of selecting only one unknown variable at each step is similar with that of the OPP reduction method \cite{Ossola:2006us} for one-loop amplitude.}. The problem of EYM amplitude expansion is then translated to the construction of properly defined differential operators, which would be the major purpose of this section. Surprisingly, we find it very helpful to use quivers to represent the gauge invariant basis and differential operators for our purpose.

%%%%%%%%%%%%%%%
\subsection{The gauge invariant basis and its quiver representation}
%%%%%%%%%%%%%%%%%%%

The definition of insertion operator (\ref{operator-insertion-h}) indicates that a differential operator would only affect the Lorentz contraction $(\epsilon_{h}\cdot k)$, so all other types of Lorentz contractions $(k\cdot k)$ and $(\epsilon\cdot \epsilon)$ can be treated as unrelated factors. In order to characterize the structure of $(\epsilon_h\cdot k)$ in a gauge invariant vector, we can assign a quiver, {\sl i.e.}, directed graph, to it\footnote{The idea of using arrows to represent Lorentz contractions has already been applied in literatures \cite{Hou:2018bwm,Du:2019vzf}, where all types of Lorentz contractions are considered. However, we are only interested in Lorentz contraction of the type $\epsilon\cdot k$ in this paper.}. So in this subsection first we define the quiver representation of atomic factors like $(\epsilon_i\cdot k_j)$, secondly give the quiver representations of fundamental $f$-terms, and finally consider the quiver representation of gauge invariant vectors and talk about the properties of their quiver representation.
\\

We call a directed graph representing all $(\eps\cdot k)$'s of a vector as {\bf $(\eps k)$-quiver} of the vector. In a quiver, we use a directed solid line to represent $(\epsilon_{h_i}\cdot k_{h_j})$ with an arrow pointing to a graviton momentum $k_{h_i}$, and a directed dashed line to represent $(\epsilon_{h_i}\cdot k_j)$ with an arrow pointing to a gluon momentum $k_j$ as\footnote{From now on, we will identify an directed line with its corresponding $(\eps\cdot k)$ term, and sometimes when we refer to a specific directed line connecting two nodes from $a$ to $b$, we will use the label $(ab)$ and $b$ is called the head, $a$ called the tail. }
\begin{center}
\begin{tikzpicture}
  \draw [fill] (0,0) circle [radius=0.05] (2,0) circle [radius=0.05] (6,0) circle [radius=0.05] (8,0) circle [radius=0.05];
  \draw [thick, ->] (0,0)--(1,0);
  \draw [thick] (1,0)--(2,0);
  \draw [thick, dashed, ->] (6,0)--(7,0);
  \draw [thick, dashed] (7,0)--(8,0);

  \node [above] at (0,0) {$h_i$};
  \node [above] at (2,0) {$h_j$};
  \node [above] at (6,0) {$h_i$};
  \node [above] at (8,0) {$j$};
  \node [right] at (8,0) {.};

  \node [] at (1,-0.5) {$\epsilon_{h_i}\cdot k_{h_j}$};
  \node [] at (7,-0.5) {$\epsilon_{h_i}\cdot k_j$};

\end{tikzpicture}
\end{center}
As for the fundamental $f$-term (\ref{fundamental-f-term-kffk}), which can be expanded as
\begin{eqnarray}
k_1\cdot f_{h_i}\cdot f_{h_j}\cdot k_1&=& (k_1\cdot k_{h_i})(\epsilon_{h_i}\cdot k_{h_j})(\epsilon_{h_j}\cdot k_1) -(k_1\cdot k_{h_i})(\epsilon_{h_i}\cdot \epsilon_{h_j})(k_{h_j}\cdot k_1) \nonumber\\
 &&~~~~~~~~~~~~-(k_1\cdot \epsilon_{h_i})(k_{h_i}\cdot k_{h_j})(\epsilon_{h_j}\cdot k_1)+(k_1\cdot \epsilon_{h_i})(k_{h_i}\cdot \epsilon_{h_j})(k_{h_j}\cdot k_1)~,~~~
\end{eqnarray}
then its $(\eps k)$-quiver representation consists of three $(\eps k)$-directed graphs as\footnote{Notice that there are four terms in the expansion of $k_1\cdot f_{h_i}\cdot f_{h_j}\cdot k_1$, while the $\epsilon_{h_i}\cdot \epsilon_{h_j}$ term is the most crucial signature to distinguish it from other fundamental $f$-terms. However in this paper we only consider insertion operators so that $\epsilon \cdot \epsilon$ is out of our sight.}
\begin{center}
\bea
\begin{tikzpicture}
\draw [fill] (0,0) circle [radius=0.05]  (2,0) circle [radius=0.05]  (4,0) circle [radius=0.05]  (6,0) circle [radius=0.05]  (8,0) circle [radius=0.05]  (10,0) circle [radius=0.05]  (1,1.5) circle [radius=0.05]  (4,1.5) circle [radius=0.05]  (10,1.5) circle [radius=0.05];
\draw [thick, ->] (6,0)--(5,0);
\draw [thick] (5,0)--(4,0);
\draw [thick, ->] (8,0)--(9,0);
\draw [thick] (9,0)--(10,0);

\draw [thick, dashed, ->] (4,0)--(4,0.75);
\draw [thick, dashed] (4,0.75)--(4,1.5);
\draw [thick, dashed, ->] (10,0)--(10,0.75);
\draw [thick, dashed] (10,0.75)--(10,1.5);
\draw [thick, dashed, ->] (0,0)--(0.5,0.75);
\draw [thick, dashed] (0.5,0.75)--(1,1.5);
\draw [thick, dashed, ->] (2,0)--(1.5,0.75);
\draw [thick, dashed] (1.5,0.75)--(1,1.5);

\node [] at (3,0.75) {$+$};
\node [] at (7,0.75) {$+$};

\node [below] at (0,0) {$h_i$};
\node [below] at (2,0) {$h_j$};
\node [below] at (4,0) {$h_i$};
\node [below] at (6,0) {$h_j$};
\node [below] at (8,0) {$h_i$};
\node [below] at (10,0) {$h_j$};

\node [above] at (1,1.5) {$1$};
\node [above] at (4,1.5) {$1$};
\node [above] at (10,1.5) {$1$};

\node [] at (-3,0.75) {$k_1\cdot f_{h_i}\cdot f_{h_j}\cdot k_1$:};
\node [right] at (10.3,0.75) {.};

\end{tikzpicture}\label{kffk-fig}\eea
\end{center}
Since each graph denotes a multiplication of $(\epsilon\cdot k)$ terms, hence when applying the following derivatives
\begin{equation}
\frac{\partial}{\partial (\epsilon_{h_i}\cdot k_1)}\frac{\partial}{\partial (\epsilon_{h_j}\cdot k_1)}~~~,~~~\frac{\partial}{\partial (\epsilon_{h_j}\cdot k_{h_i})}\frac{\partial}{\partial (\epsilon_{h_i}\cdot k_1)}~~~,~~~\frac{\partial}{\partial (\epsilon_{h_i}\cdot k_{h_j})}\frac{\partial}{\partial (\epsilon_{h_j}\cdot k_1)}
\end{equation}
to $(k_1\cdot f_{h_i}\cdot f_{h_j}\cdot k_1)$, we will get non-vanishing results. Similarly, for $(k_1\cdot f_{h_i}\cdot k)$  their $(\eps k)$-quivers are
\begin{center}
\bea
  \begin{tikzpicture}
    \draw [fill] (0,0) circle [radius=0.05]  (0,2) circle [radius=0.05] (2,0) circle [radius=0.05] (2,2) circle [radius=0.05] (8,0) circle [radius=0.05] (10,0) circle [radius=0.05] (8,2) circle [radius=0.05] (10,2) circle [radius=0.05];
    \draw [thick, ->] (0,0)--(1,0);
    \draw [thick] (1,0)--(2,0);
    \draw [thick, dashed, ->] (0,2)--(1,2);
    \draw [thick, dashed] (1,2)--(2,2);
    \draw [thick, dashed, ->] (8,0)--(9,0);
    \draw [thick, dashed] (9,0)--(10,0);
    \draw [thick, dashed, ->] (8,2)--(9,2);
    \draw [thick, dashed] (9,2)--(10,2);

    \node [] at (1,1) {$+$};
    \node [] at (9,1) {$+$};

    \node [left] at (0,0) {$h_i$};
    \node [right] at (2,0) {$h_j$};
    \node [left] at (0,2) {$h_i$};
    \node [right] at (2,2) {$1$};

    \node [left] at (8,0) {$h_i$};
    \node [right] at (10,0) {$j$};
    \node [left] at (8,2) {$h_i$};
    \node [right] at (10,2) {$1$};

    \node [] at (-2.5,1) {$k_1\cdot f_{h_i}\cdot k_{h_j}$:};
    \node [] at (5.5,1) {$k_1\cdot f_{h_i}\cdot k_{j}$:};
  \end{tikzpicture}\label{kfk-fig} \eea
\end{center}
where we have distinguished two cases, $k_{h_j}$ being the momentum of a graviton $h_j$ and $k_j$ being a momentum of a gluon.

\begin{figure}
\begin{center}
\begin{tikzpicture}
  \draw [thick, cyan] (0,0) to [out=45, in=180] (1,0.3) to [out=0, in=135] (2,0);
  \draw [thick, brown] (0,0) to [out=315, in=180] (1,-0.3) to [out=0, in=225] (2,0);
  \draw [thick, cyan, ->] (0.9,0.3)--(1.1,0.3);
  \draw [thick, brown, ->] (1.1,-0.3)--(0.9,-0.3);
  \draw [fill] (0,0) circle [radius=0.05]  (2,0) circle [radius=0.05] (5,0) circle [radius=0.05] (7,0) circle [radius=0.05] (10,0) circle [radius=0.05] (12,0) circle [radius=0.05];
  \draw [thick, ->] (5,0)--(6,0);
  \draw [thick] (6,0)--(7,0);
  \draw [thick, dashed, ->] (10,0)--(11,0);
  \draw [thick, dashed] (11,0)--(12,0);

  \node [left] at (0,0) {$h_i$};
  \node [right] at (2,0) {$h_j$};
  \node [left] at (5,0) {$h_i$};
  \node [right] at (7,0) {$h_j$};
  \node [left] at (10,0) {$h_i$};
  \node [right] at (12,0) {$K_a$};

  \node [] at (1,-1) {$\mathsf{F}_{h_ih_j}$};
  \node [] at (6,-1) {$\mathsf{F}_{h_i}^{h_j}$};
  \node [] at (11,-1) {$\mathsf{F}_{h_i}^a$};

\end{tikzpicture}
\caption{The quiver representation of fundamental $f$-terms.}  \label{fundamental-quivers}
\end{center}
\end{figure}
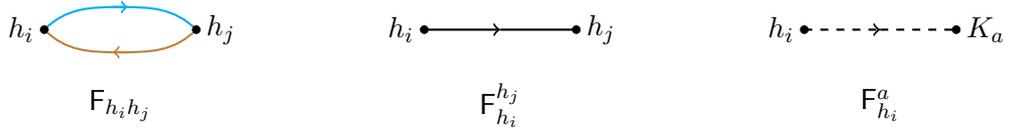

Note that the factor $(\epsilon_{h_i}\cdot k_1)$ exists in both $(k_1\cdot f_{h_i}\cdot f_{h_j}\cdot k_1)$ and $(k_1\cdot f_{h_i}\cdot k)$, so the action of derivative $\partial_{\epsilon_{h_i}\cdot k_1}$ on them both are non-zero. Consequently, we prefer to eliminate the dashed lines representing $\epsilon_{h_i}\cdot k_1$ in the graphs of $(\eps k)$-quivers to obtain a simple presentation. Furthermore, to represent one fundamental $f$-term by only one graph and distinguish $\mathsf{F}_{h_ih_j}$ from $\mathsf{F}_{h_i}^{h_j}$, we combine the two solid arrows in \eref{kffk-fig} to a loop. Finally, the fundamental $f$-terms $\mathsf{F}_{h_ih_j}$, $\mathsf{F}_{h_i}^{h_j}$, $\mathsf{F}_{h_i}^a$ defined in (\ref{fundamental-f-term-Fhh}), (\ref{fundamental-f-term-Fh}) and (\ref{fundamental-f-term-Fa}) are represented by quivers in Fig.\ref{fundamental-quivers}. To distinguish these quivers from $(\eps k)$-quivers, we will call them {\bf basis quivers} or just quivers.

We should emphasize that from a basis quiver it is easy to recover all corresponding $(\eps k)$-quivers by replacing any one solid or dashed arrow $(\epsilon_{h_i}\cdot k)$ in the graph by a dashed arrow $(\epsilon_{h_i}\cdot k_1)$, {\sl i.e.}, from Fig.\ref{fundamental-quivers} to \eref{kffk-fig},\eref{kfk-fig}. However, given a $(\eps k)$-quiver, it is hard to tell which basis quiver it comes from, especially when there are many $(\eps\cdot k_1)$ lines. The fact that there is no one-to-one correspondence between basis quivers and $(\eps k)$-quivers causes some technical difficulties in the construction of differential operators. Fortunately, for a gauge invariant vector, its basis quiver and $(\eps k)$-quivers do possess a common property: they all contains $m$ and only $m$ lines (counting both dashed line and solid line), since each line carries one $\eps_{h_i}$.

Note that the basis quiver for $\mathsf{F}_{h_ih_j}$ is a colored loop, where colors are to remind us that it is an overlapping of three $(\eps k)$-quivers after eliminating dashed lines. We call such a colored loop as {\bf pseudo-loop}. In general there are also real loops. For example, the $\mathsf{F}_{h_1}^{h_2}\mathsf{F}_{h_2}^{h_1}$ containing a monomial $(\epsilon_{h_1}k_{h_2})(\epsilon_{h_2}k_{h_1})$  and  $\mathsf{F}_{h_1}^{h_2}\mathsf{F}_{h_2}^{h_3}\mathsf{F}_{h_3}^{h_1}\mathsf{F}_{h_4}^{h_3}$
containing a monomial $(\epsilon_{h_1}k_{h_2})(\epsilon_{h_2}k_{h_3})(\epsilon_{h_3}k_{h_1})(\epsilon_{h_4}k_{h_3})$ can be represented as
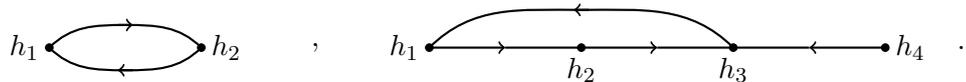
\begin{center}
\begin{tikzpicture}
  \draw [fill] (0,0) circle [radius=0.05] (2,0) circle [radius=0.05] (5,0) circle [radius=0.05] (7,0) circle [radius=0.05] (9,0) circle [radius=0.05] (11,0) circle [radius=0.05];
  \draw [thick] (0,0) to [out=45, in=180] (1,0.3) to [out=0, in=135] (2,0);
  \draw [thick] (0,0) to [out=315, in=180] (1,-0.3) to [out=0, in=225] (2,0);
  \draw [thick, ->] (0.9,0.3)--(1.1,0.3);
  \draw [thick, ->] (1.1,-0.3)--(0.9,-0.3);

  \draw [thick, ->] (5,0)--(6,0);
  \draw [thick] (6,0)--(7,0);
  \draw [thick, ->] (7,0)--(8,0);
  \draw [thick] (8,0)--(9,0);
  \draw [thick, ->] (11,0)--(10,0);
  \draw [thick] (10,0)--(9,0);
  \draw [thick] (5,0) to [out=45, in=180] (7,0.5) to [out=0, in=135] (9,0);
  \draw [thick, ->] (7.1,0.5)--(6.9,0.5);
  \node [] at (3.5,0) {$,$};

  \node [left] at (0,0) {$h_1$};
  \node [right] at (2,0) {$h_2$};
  \node [left] at (5,0) {$h_1$};
  \node [below] at (7,0) {$h_2$};
  \node [below] at (9,0) {$h_3$};
  \node [right] at (11,0) {$h_4$};
  \node [] at (12,0) {.};

\end{tikzpicture}
\end{center}
However as explained in \cite{Feng:2019tvb}, the terms with indices or part of indices forming a closed circle will not present in the expansion of EYM amplitude, although such terms do appear in the gauge invariant basis. So we will exclude basis quivers with real loops in practical computation.
\\

Next let us consider the quiver representation of a vector in the gauge invariant basis. As shown in (\ref{basis-def-2}), such a vector is a multiplication of fundamental $f$-terms as
\begin{equation}
\left( \prod_{i=1}^{p} {\sf{F}}_{h_{\alpha_{2i-1}}h_{\alpha_{2i}}}\right) \left(\prod_{i=1}^{q}{\sf{F}}_{h_{\beta_i}}^{h_{\beta'_i}}\right) \left( \prod_{i=1}^{r} {\sf{F}}_{h_{\gamma_i}}^{a_{\gamma_i}}\right)~.~~~\label{basis-def-example}
\end{equation}
Since each $\eps_{h_i}$ appears only once in a vector, then only one $(\eps_{h_i}\cdot k)$, so we can conclude that each point labelled by $h_i$ in the basis quiver of a gauge invariant vector has at most one out-going line, but possibly several in-coming lines. Consequently, all pseudo-loops are topological disconnected from each other. The point labelled by $K_a$ is connected by only in-coming lines but not out-going lines, hence all such points are also topological disconnected from each other. Furthermore, pseudo-loops can not be connected with points labelled by $K_a$ either. So a quiver graph could have many disconnected components, whose number is at least $p$ and at most $p+r$, since several dashed lines can be connected to the same node $K_{a_i}$. While a solid line for $\mathsf{F}_{h_i}^{h_j}$ can be connected to one and only one disconnect component.

With above analysis, let us discuss the possible structures appearing in a quiver representation for a vector in gauge invariant basis (\ref{basis-def-example}). Firstly, since each $\mathsf{F}_{h_{\gamma_i}}^{a_{\gamma_i}}$ is represented by a dashed directed line with arrow pointing to $K_{a_{\gamma_i}}$, its head can never be connected with a pseudo-loop or a solid line. Secondly, each $\mathsf{F}_{h_{\beta_i}}^{h_{\beta'_i}}$ is represented by a solid line with arrow pointing to $h_{\beta'_i}\neq h_{\beta_i}$, so if $h_{\beta'_i}\in \{h_{\gamma_1},\ldots, h_{\gamma_r}\}$ its head is linked with a dashed line, while if $h_{\beta'_i}\in \{h_{\alpha_1},\ldots,h_{\alpha_{2p}}\}$ its head is linked with a pseudo-loop, and if $h_{\beta'_i}\in \{h_{\beta_1},\ldots, h_{\beta_r}\}/\{h_{\beta_{i}}\}$, for instance $h_{\beta'_i}=h_{\beta_j}$ its head is linked with another solid line, and the latter's head is further linked with a pseudo-loop, a dashed line or a solid line. A succession of solid lines should stops at a dashed line or a pseudo loop finally, otherwise it would form a real loop which should be excluded.

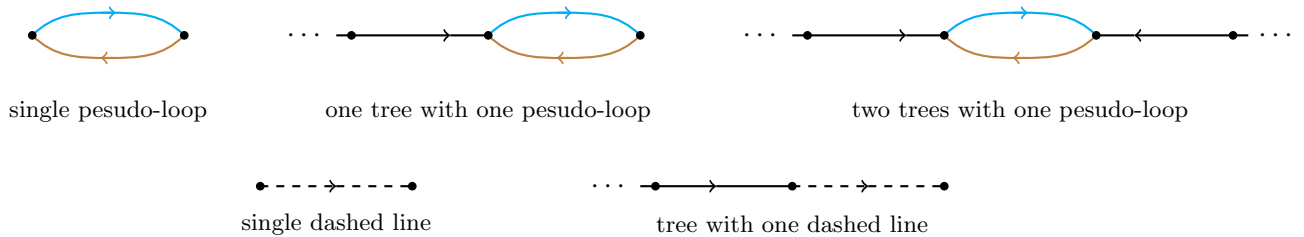
\begin{figure}
  \centering
  \begin{tikzpicture}
    \draw [thick, cyan] (-1,0) to [out=45, in=180] (0,0.3) to [out=0, in=135] (1,0);
    \draw [thick, brown] (-1,0) to [out=315, in=180] (0,-0.3) to [out=0, in=225] (1,0);
    \draw [thick, ->, brown] (0.1,-0.3)--(-0.1,-0.3);
    \draw [thick, ->, cyan] (-0.1,0.3)--(0.1,0.3);

    \draw [thick, cyan] (5,0) to [out=45, in=180] (6,0.3) to [out=0, in=135] (7,0);
    \draw [thick, brown] (5,0) to [out=315, in=180] (6,-0.3) to [out=0, in=225] (7,0);
    \draw [thick, ->, brown] (6.1,-0.3)--(5.9,-0.3);
    \draw [thick, ->, cyan] (5.9,0.3)--(6.1,0.3);
    \draw [thick, ->] (3,0)--(4.5,0);
    \draw [thick] (4.5,0)--(5,0);
    \draw [fill] (3.2,0) circle [radius=0.05];
    \node [left] at (3,0) {$\cdots$};

    \draw [thick, cyan] (11,0) to [out=45, in=180] (12,0.3) to [out=0, in=135] (13,0);
    \draw [thick, brown] (11,0) to [out=315, in=180] (12,-0.3) to [out=0, in=225] (13,0);
    \draw [thick, ->, brown] (12.1,-0.3)--(11.9,-0.3);
    \draw [thick, ->, cyan] (11.9,0.3)--(12.1,0.3);
    \draw [thick, ->] (9,0)--(10.5,0);
    \draw [thick] (10.5,0)--(11,0);
    \draw [thick, ->] (15,0)--(13.5,0);
    \draw [thick] (13.5,0)--(13,0);
    \draw [fill] (9.2,0) circle [radius=0.05];
    \node [left] at (9,0) {$\cdots$};
    \draw [fill] (14.8,0) circle [radius=0.05];
    \node [right] at (15,0) {$\cdots$};

    \draw [dashed, thick, ->] (2,-2)--(3,-2);
    \draw [dashed, thick] (3,-2)--(4,-2);

    \draw [dashed, thick, ->] (9,-2)--(10,-2);
    \draw [dashed, thick] (10,-2)--(11,-2);
    \draw [thick, ->] (7,-2)--(8,-2);
    \draw [thick] (8,-2)--(9,-2);
    \draw [fill] (7.2,-2) circle [radius=0.05];
    \node [left] at (7,-2) {$\cdots$};

    \draw [fill] (-1,0) circle [radius=0.05]  (1,0) circle [radius=0.05] (5,0) circle [radius=0.05] (7,0) circle [radius=0.05] (11,0) circle [radius=0.05] (13,0) circle [radius=0.05] (2,-2) circle [radius=0.05] (4,-2) circle [radius=0.05] (9,-2) circle [radius=0.05] (11,-2) circle [radius=0.05];

    \node [] at (0,-1) {{\footnotesize single pesudo-loop}};
    \node [] at (5,-1) {{\footnotesize one tree with one pesudo-loop}};
    \node [] at (12,-1) {{\footnotesize two trees with one pesudo-loop}};
    \node [] at (3,-2.5) {{\footnotesize single dashed line}};
    \node [] at (9,-2.5) {{\footnotesize tree with one dashed line}};
  \end{tikzpicture}
  \caption{Possible structures that can appear in the quiver representation of gauge invariant basis. The solid line without starting point denotes possible tree or line segments. The dashed lines could be connected at the same point $K_a$.}\label{Fig-quiver-structure}
\end{figure}

To summarize, the quiver representation of a vector in gauge invariant basis could contain the following sub-structures,
\begin{enumerate}
\item only a single dashed line,
\item a dashed line linked with a tree consisting of solid lines,
\item only a single pseudo-loop,
\item a pseudo-loop connected with a tree consisting of solid lines in one side,
\item a pseudo-loop connected with two trees consisting of solid lines in both sides,
\end{enumerate}
as shown in Fig.\ref{Fig-quiver-structure}.
Two examples of quiver representations for two vectors in the gauge invariant basis of $A^{\EYM}_{n,6}$ are shown as,
\begin{align}
\begin{tikzpicture}
  \draw [thick, cyan] (0,0) to [out=45, in=180] (0.5,0.15) to [out=0, in=135] (1,0);
  \draw [thick, brown] (0,0) to [out=315, in=180] (0.5,-0.15) to [out=0, in=225] (1,0);
  \draw [thick, ->, brown] (0.55,-0.15)--(0.45,-0.15);
  \draw [thick, ->, cyan] (0.45,0.15)--(0.55,0.15);
  \draw [thick] (0,0) to [out=45, in=180] (1,0.3) to [out=0, in=135] (2,0);
  \draw [thick, ->] (1.1,0.3)--(0.9,0.3);
  \draw [thick, ->] (3,0)--(3.5,0);
  \draw [thick] (3.5,0)--(4,0);
  \draw [dashed, thick, ->] (4,0)--(4,0.5);
  \draw [dashed, thick] (4,0.5)--(4,1);
  \draw [dashed, thick, ->] (5,0)--(5,0.5);
  \draw [dashed, thick] (5,0.5)--(5,1);
  \draw [thick] (8,0) to [out=45, in=180] (9.5,0.5) to [out=0, in=135] (11,0);
  \draw [thick] (9,0) to [out=45, in=180] (10,0.3) to [out=0, in=135] (11,0);
  \draw [thick, ->] (9.4,0.5)--(9.6,0.5);
  \draw [thick, ->] (10.2,0.3)--(10.3,0.3);
  \draw [thick, ->] (11,0)--(10.5,0);
  \draw [thick] (10.5,0)--(10,0);
  \draw [dashed, thick, ->] (10,0)--(10,0.8);
  \draw [dashed, thick] (10,0.8)--(10,1);
  \draw [dashed, thick, ->] (12,0)--(12,0.8);
  \draw [dashed, thick] (12,0.8)--(12,1);
  \draw [dashed, thick, ->] (13,0)--(11.5,0.5);
  \draw [dashed, thick] (11.5,0.5)--(10,1);
  \draw [fill] (0,0) circle [radius=0.05]  (1,0) circle [radius=0.05]  (2,0) circle [radius=0.05]  (3,0) circle [radius=0.05]  (4,0) circle [radius=0.05]  (5,0) circle [radius=0.05]  (4,1) circle [radius=0.05]  (5,1) circle [radius=0.05];
  \draw [fill] (8,0) circle [radius=0.05]  (9,0) circle [radius=0.05]  (10,0) circle [radius=0.05]  (11,0) circle [radius=0.05]  (12,0) circle [radius=0.05]  (13,0) circle [radius=0.05]  (12,1) circle [radius=0.05]  (10,1) circle [radius=0.05];
  \node [below] at (0,0) {$h_1$};
  \node [below] at (1,0) {$h_2$};
  \node [below] at (2,0) {$h_3$};
  \node [below] at (3,0) {$h_4$};
  \node [below] at (4,0) {$h_5$};
  \node [below] at (5,0) {$h_6$};
  \node [above] at (4,1) {$K_2$};
  \node [above] at (5,1) {$K_7$};
  \node [below] at (8,0) {$h_1$};
  \node [below] at (9,0) {$h_2$};
  \node [below] at (10,0) {$h_3$};
  \node [below] at (11,0) {$h_4$};
  \node [below] at (12,0) {$h_5$};
  \node [below] at (13,0) {$h_6$};
  \node [above] at (10,1) {$K_4$};
  \node [above] at (12,1) {$K_6$};
  \node [] at (2.5,-1) {$\mathsf{F}_{h_1h_2}\mathsf{F}_{h_3}^{h_1}\mathsf{F}_{h_4}^{h_5}\mathsf{F}_{h_5}^{2}\mathsf{F}_{h_6}^{7}$};
  \node [] at (10.5,-1) {$\mathsf{F}_{h_1}^{h_4}\mathsf{F}_{h_2}^{h_4}\mathsf{F}_{h_4}^{h_3}\mathsf{F}_{h_3}^{4}\mathsf{F}_{h_5}^{6}\mathsf{F}_{h_6}^{4}$};
\end{tikzpicture} \label{quiver-examples}
 \end{align}
The two examples illustrate our previous discussions very well. There are three disconnected components for the first one, and two for the second one. In the second graph, two dashed lines is connected to one node representing the fundamental $f$-terms  $\mathsf{F}_{h_3}^{4}$, $\mathsf{F}_{h_6}^{4}$. All directed solid lines stop at pseudo-loops or dashed lines.

In fact, we can give a more precise description of the structures of basis quivers by using the concept of {\bf rooted tree} \cite{diestel10}. The quiver of a vector in gauge invariant basis consists of some disconnected components and each component contains only one pseudo-loop or a node $K_{a_i}$. If we focus on a disconnected component with node $K_{a_i}$, it is exactly a rooted tree with the root being the node $K_{a_i}$. More precisely, it is a directed rooted tree with an orientation towards the root, {\sl i.e.}, the direction of all lines in the tree directs to the root from leaves, as illustrated in the previous two examples. For the disconnected component with a pseudo-loop, we could split the pseudo-loop into two colored lines resulting in two sub-graphs. For each sub-graph, we take the node with only in-coming lines as the root, thus we obtain two rooted trees from a disconnected component with a pseudo-loop. The picture of rooted trees will help us to construct the differential operators and understand many properties of our algorithm later.

%%%%%%%%%%%%%%%
\subsection{Constructing differential operators}
\label{operators}
%%%%%%%%%%%%%%%%

Since a vector in the gauge invariant basis is a polynomial of $(\epsilon\cdot k)$'s, it will be non-vanishing under the action of a derivative $\partial_{\epsilon_{h_i}\cdot k}$ only if its  $(\eps k)$-quiver representation contains a solid or dashed line corresponding to $\epsilon_{h_i}\cdot k$. Hence by constructing a differential operator as a proper combination of some derivatives $\partial_{\epsilon_{h_i}\cdot k}$'s, we expect ideally under its action only one vector is non-vanishing, so it can select a particular non-vanishing vector in gauge invariant basis. Although in fact we can not do this, we succeed in dividing the computation of coefficients of gauge invariant basis into many steps, and in each step by applying an appropriate differential operator, only one new vector is non-vanishing except some vectors whose coefficients are already known. The goal in this subsection is to construct such differential operators.

The expected differential operators can be constructed by three types of insertion operators (\ref{operator-insertion-h}). The first type of insertion operator takes the form,
\begin{equation}
\mathcal{T}_{ah_i(a+1)}=\partial_{\epsilon_{h_i}\cdot k_a}-\partial_{\epsilon_{h_i}\cdot k_{a+1}}~~~,~~~a=2,3,\ldots, n-1~,~~~
\end{equation}
where $k_a$ is the momentum of a gluon. A vector is non-zero under $\mathcal{T}_{ah_i(a+1)}$ if its $(\eps k)$-quiver contains a dashed line corresponding to $\epsilon_{h_i}\cdot k_a$ or $\epsilon_{h_i}\cdot k_{a+1}$. Applying this insertion operator to the fundamental $f$-terms, we get
\begin{equation}
\mathcal{T}_{ah_i(a+1)}~\mathsf{F}_{h_{\alpha_{2j-1}}h_{\alpha_{2j}}}=0~~~,~~~\mathcal{T}_{ah_i(a+1)}~\mathsf{F}_{h_{\beta_j}}^{h_{\beta'_j}}=0~,~~~\label{Tkhk-relation}
\end{equation}
and
\begin{equation}
\mathcal{T}_{a h_i (a+1)}~ \mathsf{F}_{h_j}^b=\left[\partial_{(\epsilon_{h_i}\cdot k_a)}-\partial_{(\epsilon_{h_i}\cdot k_{a+1})}\right]\sum_{l=2}^{b}\frac{(k_1\cdot k_{h_j})(\epsilon_{h_j}\cdot k_l)-(k_1\cdot \epsilon_{h_j})(k_{h_j}\cdot k_l)}{k_1\cdot k_{h_j}}=\delta_{ij}\delta_{ab}~.~~~ \label{Tff-relation}
\end{equation}
The above results tell us that if the basis quiver of a vector in gauge invariant basis contains a dashed line representing $\mathsf{F}_{h_{\gamma_i}}^a$, then a differential operator containing the insertion operator $\mathcal{T}_{a h_i (a+1)}$ will select out this vector and other vectors containing the same dashed line. The relation \eref{Tff-relation} can be graphically represented as,
\begin{align}
\begin{tikzpicture}
\draw [fill] (0,0) circle [radius=0.05]  (2,0) circle [radius=0.05];
\node [left] at (0,0) {$h_i$};
\node [right] at (2,0) {$K_a$};
\draw [thick, dashed, ->] (0,0)--(1,0);
\draw [thick, dashed] (1,0)--(2,0);
\node [] at (0.6,0) {$\mathcal{T}_{a h_i (a+1)}~\Big(~~~~~~~~~~~~~~~~~~~~~~~~~~~~\Big)=1~.$};
\end{tikzpicture}\label{Tkhk-relation2}
\end{align}
%
%So this insertion operator is ideal for distinguishing $\mathsf{F}_{h_{\gamma_i}}^a$ from each other, or $\mathsf{F}_{h_{\alpha_{2j-1}}h_{\alpha_{2j}}}$ and $\mathsf{F}_{h_{\beta_j}}^{h_{\beta'_j}}$.
The second type of insertion operators takes the form  $\mathcal{T}_{h_jh_in}=\partial_{\epsilon_{h_i}\cdot k_{h_j}}-\partial_{\epsilon_{h_i}\cdot k_n}$, where the Lorentz contraction of a polarization vector with a graviton momentum has been included. Since by definition the momentum $k_n$ does not appear in fundamental $f$-terms, when applying $\mathcal{T}_{h_jh_in}$ to them only the derivative $\partial_{\epsilon_{h_i}\cdot k_{h_j}}$ works. Explicitly, we get
\begin{equation}
\mathcal{T}_{h_jh_in}~\mathsf{F}_{h_i'}^{a_{i'}}=0~~~,~~~
\mathcal{T}_{h_jh_in}~\mathsf{F}_{h_{i'}}^{h_{j'}}=\delta_{ii'}\delta_{jj'}~~~,~~~
\mathcal{T}_{h_jh_in}~\mathsf{F}_{h_{i'}h_{j'}}=\frac{\epsilon_{h_j}\cdot k_1}{k_1\cdot k_{h_j}} \left( \delta_{ii'}\delta_{jj'}+\delta_{ij'}\delta_{ji'} \right)~,~~~
\label{Thhk-relation}\end{equation}
represented in quivers as
\begin{center}
\begin{tikzpicture}
  \draw [thick, ->] (-1,0)--(0,0);
  \draw [thick] (0,0)--(1,0);

  \draw [thick, cyan] (6,0) to [out=45, in=180] (7,0.3) to [out=0, in=135] (8,0);
  \draw [thick, brown] (6,0) to [out=315, in=180] (7,-0.3) to [out=0, in=225] (8,0);
  \draw [thick, ->, brown] (7.1,-0.3)--(6.9,-0.3);
  \draw [thick, ->, cyan] (6.9,0.3)--(7.1,0.3);

  \node [left] at (-1,0) {$h_i$};
  \node [right] at (1,0) {$h_j$};
  \node [left] at (6,0) {$h_i$};
  \node [right] at (8,0) {$h_j$};

  \node [] at (-2.3,0) {$\mathcal{T}_{h_jh_in}$};
  \node [] at (0,0) {$\Big(~~~~~~~~~~~~~~~~~~~~~~~~\Big)$};
  \node [] at (2.2,0) {$=1$};
  \node [] at (3.5,0) {$,$};

  \node [] at (4.7,0) {$\mathcal{T}_{h_jh_in}$};
  \node [] at (7,0) {$\Big(~~~~~~~~~~~~~~~~~~~~~~~~\Big)$};
  \node [] at (9.3,0) {$=$};

  \draw [fill] (-1,0) circle [radius=0.05]  (1,0) circle [radius=0.05] (6,0) circle [radius=0.05] (8,0) circle [radius=0.05];

  \draw [fill] (10,0) circle[radius=0.05] (12,0) circle [radius=0.05];
  \draw [thick,->, dashed] (10,0)--(11,0);
  \draw [thick, dashed] (11,0)--(12,0);
  \node [left] at (10,0) {$h_j$};
  \node [right] at (12,0) {$1~.$};

\end{tikzpicture}
\end{center}
Since both $\mathsf{F}_{h_ih_j}$ and $\mathsf{F}_{h_i}^{h_j}$ are non-vanishing under $\mathcal{T}_{h_jh_in}$, we may conclude this insertion operator is not sufficient to distinguish these two terms. However, we shall note that the insertion operator is actually a differential operator which works through the more smaller pieces, {\sl i.e.}, Lorentz contractions $(\eps_{h_i}\cdot k_j)$, rather than fundamental $f$-terms $\mathsf{F}_{h_ih_j}$ and $\mathsf{F}_{h_i}^{h_j}$. According to this view, it is easy to accept that $\mathsf{F}_{h_ih_j}$ and $\mathsf{F}_{h_i}^{h_j}$ are non-vanishing under the action of $\mathsf{F}_{h_ih_j}$, since the quivers of them both contain the solid line from $h_i$ to $h_j$.

In order to construct a differential operator that can distinguish $\mathsf{F}_{h_ih_j}$ from $\mathsf{F}_{h_i}^{h_j}$, we need to consider a third type of composite insertion operators. The key difference of these two terms is that $\mathsf{F}_{h_ih_j}$ has two polarization vectors, while $\mathsf{F}_{h_i}^{h_j}$ has only one. In other words, in the $(\eps k)$-quiver of $\mathsf{F}_{h_ih_j}$, there are always two lines linked together, a solid line $(h_i h_j)$ and a dashed line $(1 h_i)$ or $(1 h_j)$, so we can multiply $\mathcal{T}_{h_jh_in}$ by an additional insertion operator containing the derivative $\d_{\eps_{h_j}\cdot k_1}$, and under such operators $\mathsf{F}_{h_i}^{h_j}$ always vanishes. Then choosing the operator  $\mathcal{T}_{1h_j2}\mathcal{T}_{h_jh_in}$, and applying it to $\mathsf{F}_{h_ih_j}$ we have
\begin{equation}
(k_1\cdot k_{h_j})\mathcal{T}_{1h_j2}\mathcal{T}_{h_jh_in}~\mathsf{F}_{h_{i'}h_{j'}}  =  \delta_{ii'}\delta_{jj'}~.~~~\label{ThhkT1h2-relation}
\end{equation}
It is easy to see that the operator satisfy our requirement of distinguishing $\mathsf{F}_{h_ih_j}$ and $\mathsf{F}_{h_i}^{h_j}$, and it also distinguishes the pseudo-loop of $\mathsf{F}_{h_{i}h_{j}}$ from all other pseudo-loops. However $\mathcal{T}_{1h_j2}$ causes some additional troubles, since there will be some multiplications of fundamental $f$-terms that do ont vanish, such as
\footnote{Note that in the $(\eps k)$-quiver of $\mathsf{F}_{h_k}^{a_t}$ there is also the contraction $\eps\cdot k_2$, which would produce non-vanishing result under operator $\mathcal{T}_{1h_j2}$.}
\begin{equation}
  (k_1\cdot k_{h_j})\mathcal{T}_{1h_j2}\mathcal{T}_{h_jh_in}\mathsf{F}_{h_i}^{h_j}\mathsf{F}_{h_j}^{a_t}=-k_{h_j}\cdot (k_1+K_{a_t})~~~,~~~(k_1\cdot k_{h_j})\mathcal{T}_{1h_j2}\mathcal{T}_{h_jh_in}\mathsf{F}_{h_i}^{h_j}\mathsf{F}_{h_j}^{h_p}
 =-k_{h_j}\cdot k_{h_p}~.~~~\label{FF-relation}
\end{equation}
%
%
%\begin{equation}
% (k_1\cdot k_{h_j})\mathcal{T}_{1h_j2}\mathcal{T}_{h_jh_in}\mathsf{F}_{h_i}^{h_j}\mathsf{F}_{h_j}^{a_t}=-k_{h_j}\cdot (k_1+K_{a_t}). \Label{FF-relation}
%\end{equation}
%
%as well as
%
%\begin{equation}
% (k_1\cdot k_{h_j})\mathcal{T}_{1h_j2}\mathcal{T}_{h_jh_in}\mathsf{F}_{h_i}^{h_j}\mathsf{F}_{h_j}^{h_p}
% =-k_{h_j}\cdot k_{h_p}. \Label{FF-relation-1}
%\end{equation}
%
It means that although $\mathcal{T}_{1h_j2}\mathcal{T}_{h_jh_in}$ is able to distinguish one pseudo-loop from the others, it would mix contributions from vectors without pseudo-loop. However, it is not a problem at all, if we try to solve the coefficients of basis in multiple steps. We can firstly compute the coefficients of $\mathsf{F}_{h_i}^{h_j}\mathsf{F}_{h_j}^{a_j}$ and $\mathsf{F}_{h_i}^{h_j}\mathsf{F}_{h_j}^{h_p}$ by differential operator $\mathcal{T}_{a_jh_j(a_j+1)}\mathcal{T}_{h_jh_in}$ and $\mathcal{T}_{h_p h_j n}\mathcal{T}_{h_jh_in}$ respectively, under which $\mathsf{F}_{h_ih_j}$ has no contribution at all. Then apply $\mathcal{T}_{h_jh_in}\mathcal{T}_{1h_j2}$ to compute the coefficient of $\mathsf{F}_{h_ih_j}$, and treat the coefficients of $\mathsf{F}_{h_i}^{h_j}\mathsf{F}_{h_j}^{a_j}$, $\mathsf{F}_{h_i}^{h_j}\mathsf{F}_{h_j}^{h_p}$ as known input.
\\

After above discussions, we can roughly give a general picture of constructing a differential operator to select a particular vector in the gauge invariant basis through the quiver representation. The major idea is to construct a new special $(\eps k)$-quiver from a vector's basis quiver, which can be used to construct the expected differential operators. The reasonable method of giving these new $(\eps k)$-quivers is following: a dashed line in the basis quiver of a vector suggests us that there is also a dashed line in the new $(\eps k)$-quiver but representing $(\eps_{h_i}\cdot K_a)$, and a solid line $(h_i h_j)$ in the basis quiver also suggests that there is a solid line in the new $(\eps k)$-quiver representing $(\eps_{h_i}\cdot k_{h_j})$, while for a pseudo-loop in the basis quiver we can choose to construct either a solid line $(\eps_{h_i}\cdot k_{h_j})$ connected with a dashed line $(\eps_{h_j}\cdot k_1)$ or a solid line $(\eps_{h_j}\cdot k_{h_i})$ connected with a dashed line $(\eps_{h_i}\cdot k_1)$ in the new $(\eps k)$-quiver. We are free to take any one of the two choices when meeting a pseudo-loop. Finally we get a new $(\eps k)$-quiver which are used to construct differential operators.

Just as we have discussed in the last of the previous subsection, {\sl the $(\eps k)$-quiver is a collection of rooted trees}. The disconnected component of a pseudo-loop in the basis quiver of a vector have been split into two branches, each branch is a rooted tree with root being $k_1$ and is good according to our choice, and the components without pseudo-loops directly give us rooted trees.
Further more, a collection of rooted trees can be algebraically represented as the embedded structure where at each level we write as
$\{ {\rm root}: {\rm leaf~1}; ...; {\rm leaf~m}\}$.\footnote{For example, the second quiver in (\ref{quiver-examples}) can be represented as
\bea \{K_6: h_5\}~~~,~~~\{K_4:~h_6;~\{h_3, h_4: h_1; h_2\}\}~.~~~\label{one-rep}\eea
}

Secondly, having obtained the desired $(\eps k)$-quivers, we can construct the corresponding differential operators by the following rules:
\begin{enumerate}
\item assign an operator $\mathcal{T}_{a h_i(a+1)}$ to each dashed line $(h_i K_a)$ in the new $(\eps k)$-quiver, which uniquely picks up the corresponding dashed line in a vector's basis quiver;

\item assign an operator $\mathcal{T}_{h_jh_in}$ to each solid line $(h_i h_j)$ in the new $(\eps k)$-quiver, which uniquely picks up the corresponding solid line in a vector's basis quiver;

\item assign an operator $(k_1\cdot k_{h_i})\mathcal{T}_{1 h_i 2}$ to each dashed line $(h_ik_1)$ in the new $(\eps k)$-quiver.
\end{enumerate}
Above rules can be represented graphically as
\begin{align}
\begin{tikzpicture}
%%% solid line
\draw [fill] (-6,0) circle [radius=0.05] (-4,0) circle [radius=0.05];
\draw [thick,dashed,->] (-6,0)--(-5,0);
\draw [thick,dashed] (-5,0)--(-4,0);
\node [left] at (-6,0) {$h_a$};
\node [right] at (-4,0) {$K_b$};
\node [right] at (-3.5,0) {$:=\mathcal{T}_{bh_a(b+1)}$~,};
%%% solid line
\draw [fill] (0,0) circle [radius=0.05] (2,0) circle [radius=0.05];
\draw [thick,->] (0,0)--(1,0);
\draw [thick] (1,0)--(2,0);
\node [left] at (0,0) {$h_a$};
\node [right] at (2,0) {$h_b$};
\node [right] at (2.5,0) {$:=\mathcal{T}_{h_bh_an}$~,};
%%% dashed line
\draw [fill] (6,0) circle [radius=0.05];
\draw [thick, dashed,->] (6,0)--(7,0);
\node [left] at (6,0) {$h_{a}$};
\node [right] at (7,0) {$k_1$};
\node [right] at (7.5,0) {$:=(k_1k_a)\mathcal{T}_{1h_a2}$~.~~};
\end{tikzpicture}\label{map-D-rule}
\end{align}
So the corresponding differential operator for a vector in gauge invariant basis is defined by multiplying all assigned operators in the new $(\eps k)$-quiver together, then we call the $(\eps k)$-quivers constructed according to the above rules as
{\bf $D$-quivers}. We want to emphasize that: (1) there is a one-to-one map between $D$-quivers and differential operators, then one quiver defines an unique differential operator, (2) $D$-quiver is a special $(\eps k)$-quiver, which can be associated to a given basis quiver.

Finally, above discussions can be summarized as the following map stating  from a given vector to a corresponding differential operator,
\bea
& & B_i=\left( \prod_{i=1}^{p} {\sf{F}}_{h_{\alpha_{2i-1}}h_{\alpha_{2i}}}\right) \left(\prod_{i=1}^{q}{\sf{F}}_{h_{\beta_i}}^{h_{\beta'_i}}\right) \left( \prod_{i=1}^{r} {\sf{F}}_{h_{\gamma_i}}^{a_{\gamma_i}}\right)\nonumber\\
&& ~~~~~\to D_i=\left( \prod_{i=1}^{p} (k_1\cdot k_{h_{\alpha_{2i}}}) \mathcal{T}_{h_{\alpha_{2i}}h_{\alpha_{2i-1}}n}\mathcal{T}_{1h_{\alpha_{2i}}2}\right) \left(\prod_{i=1}^{q} \mathcal{T}_{h_{\beta'_{i}}h_{\beta_{i}}n}\right) \left( \prod_{i=1}^{r} \mathcal{T}_{a_{\gamma_i}h_{\gamma_{i}}(a_{\gamma_i+1})}\right)~,~~~\label{differential-operator}
\end{eqnarray}
where $B_i\in \mathcal{B}$.
There are several technical points we want to explain. First, the mapping rule is defined such that
\bea D_i[B_i]=1~.~~~\label{Di-def-1}\eea
Second, although insertion operators are commutative, when acting on EYM amplitudes we need to choose a proper ordering to make the physical meaning clear. We shall apply insertion operators of the type $\mathcal{T}_{ah_{\gamma}a'}, \mathcal{T}_{1h_\alpha 2}$ first, then the types $\mathcal{T}_{h_\alpha h_{\alpha}' n}$ and  $\mathcal{T}_{h_{\beta}h_\beta' n}$. More explicitly, the ordering of applying insertion operators is from the roots to the leaves in the $D$-quiver opposite to the direction of arrows.
\\

In fact, we can make the result more concrete when acting $D_i$ on $A^{\rm EYM}_{n,m}$. As mentioned, each $D_i$ can be represented by a $D$-quiver as the collection of rooted trees. For example, the $D$-quiver for a differential operator is
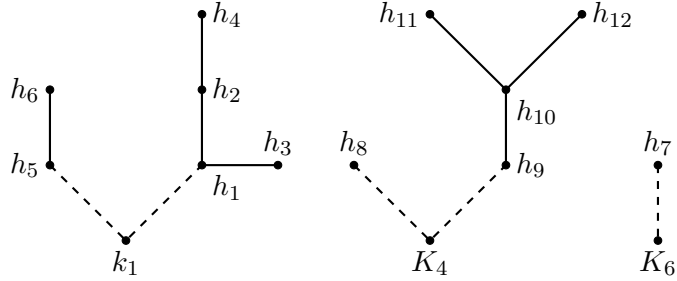
\begin{center}
  \begin{tikzpicture}
    \draw [fill] (0,0) circle [radius=0.05] (-1,1) circle [radius=0.05] (-1,2) circle [radius=0.05] (1,1) circle [radius=0.05] (1,2) circle [radius=0.05] (1,3) circle [radius=0.05] (2,1) circle [radius=0.05] (4,0) circle [radius=0.05] (3,1) circle [radius=0.05] (5,1) circle [radius=0.05] (5,2) circle [radius=0.05] (4,3) circle [radius=0.05] (6,3) circle [radius=0.05] (7,0) circle [radius=0.05] (7,1) circle [radius=0.05];
    \draw [thick, dashed] (-1,1)--(0,0)--(1,1) (3,1)--(4,0)--(5,1) (7,1)--(7,0);
    \draw [thick] (-1,2)--(-1,1) (1,3)--(1,1)--(2,1) (4,3)--(5,2)--(6,3) (5,2)--(5,1);
    \node [below] at (0,0) {$k_1$};
    \node [below] at (4,0) {$K_4$};
    \node [below] at (7,0) {$K_6$};
    \node [left] at (-1,1) {$h_5$};
    \node [left] at (-1,2) {$h_6$};
    \node [below right] at (1,1) {$h_1$};
    \node [above] at (2,1) {$h_3$};
    \node [right] at (1,2) {$h_2$};
    \node [right] at (1,3) {$h_4$};
    \node [above] at (3,1) {$h_8$};
    \node [right] at (5,1) {$h_9$};
    \node [below right] at (5,2) {$h_{10}$};
    \node [left] at (4,3) {$h_{11}$};
    \node [right] at (6,3) {$h_{12}$};
    \node [above] at (7,1) {$h_7$};
  \end{tikzpicture}
\end{center}
then the rooted trees can be written as
\bea \{k_1:\{h_1: \{h_2, h_4\}; h_3\}; \{h_5, h_6\}\}~~~,~~~\{K_4:~h_8;~\{h_9, h_{10}: h_{11}; h_{12}\}\}~~~,~~~\{K_6: h_7\}~.~~~\label{D-exp-2} \eea
Applying it to $A^{\EYM}_{n,12}$ leads to
\bea & & ~~~A^{\YM}_{n+12}\left(1, ~~\{ h_1, \{h_2, h_4\}\shuffle h_3\}~~\shuffle  ~~\{h_5, h_6\}~~\right. \nn
& & ~~~~~~~~~~~\left. \shuffle ~~\{2,3,4,~h_8~\shuffle~\{h_9, h_{10},h_{11}\shuffle h_{12}\}~\shuffle ~\{5,6, ~h_7~\shuffle~\{7,...,n-1\}_R~~\}_R~~\}_R,~~n \right)~,~~~\label{D-exp-2-A} \eea
multiplied with $(k_1\cdot k_{h_1})(k_1\cdot k_{h_5})$. This example contains all crucial points we want to clarify, so let us give more explanations, especially about the similarity between shuffle structure in \eref{D-exp-2-A} and the rooted tree structure in \eref{D-exp-2}.
\begin{itemize}

\item Firstly, let us consider the tree with root $k_1$. It is connected to two branches $\{h_1: \{h_2, h_4\}; h_3\}$ and  $\{h_5, h_6\}$. Applying $\mathcal{T}_{1h_{1}2}$ and $\mathcal{T}_{1h_{5}2}$ will produce the structure
    \bea A(1, \{h_1\} \shuffle \{h_5\} \shuffle\{2,3,...,n-1\}_R, n)~,~~~\eea
    where the subscript $R$ is denoted for a "restricted shuffle", meaning that when making shuffle permutation for three sets, the first element of the third set should be placed after the first element of other two sets.  Applying $\mathcal{T}_{h_{5} h_6 n}$ from the first branch will  give us $\{h_5,h_6\}$ as
    \bea A(1, \{h_1\} \shuffle \{h_5,h_6\} \shuffle\{2,3,...,n-1\}_R, n)~,~~~\eea
    while applying insertion operators from the second branch will give
    $\{h_1, ~\{h_2, h_4\}\shuffle h_3\}$ as
\bea & & ~~~A^{\rm YM}_{n+12}\left(1, ~~\{ h_1, \{h_2, h_4\}\shuffle h_3\}~~\shuffle  ~~\{h_5, h_6\} \shuffle ~~\{2,3,...,n-1\}_R,~~n \right)~.~~~\label{D-exp-2-A-1} \eea

\item Second let us consider the rooted tree with root $K_4$, which also contains two branches. Applying $\mathcal{T}_{4h_{8}5}$ and $\mathcal{T}_{4h_{9}5}$ on the sub-structure $\{2,3,...,n-1\}_R$ in \eref{D-exp-2-A-1} results in
\bea & & A^{\rm YM}_{n+12}\left(1, ~~\{ h_1, \{h_2, h_4\}\shuffle h_3\}~~\shuffle  ~~\{h_5, h_6\}~~\right. \nn
& & ~~~~~~~~~~~~\left. \shuffle ~~\{2,3,4,h_8\shuffle\{h_9, h_{10},h_{11}\shuffle h_{12}\}\shuffle \{5,6,...,n-1\}_R~~\}_R,~~n \right)~.~~~\label{D-exp-2-A-2} \eea

\item Finally let us consider the remaining tree structure $\{K_6: h_7\}$ with root $K_6$. Applying $\mathcal{T}_{6h_{7}7}$ on the sub-structure $\{5,6,...,n-1\}_R$ in \eref{D-exp-2-A-2} will give us $\{5,6, h_7\shuffle\{7,...,n-1\}_R~~\}_R$ just as showed in \eref{D-exp-2-A}.
\\
\end{itemize}

\subsection{Applications of differential operators}
Having defined the corresponding differential operator $D_i$ for a vector in gauge invariant basis as in \eref{differential-operator}, we can apply it to the equation \eref{expansionGI-abbr} and get a linear equation for the expansion coefficient of a particular $B_i$ as well as other coefficients. However, for a vector with pseudo-loops, in general we will meet $D_i[B_j]\neq 0$ for some $j\neq i$. In this case, we get a set of linear equations. For an EYM amplitude with a large number of gravitons and gluons, the size of linear equations will become too large to be solved. Thus it is better to find a way to avoid solving a large number of linear equations.

To find a such method, we need to analyze the behaviors of different $B_j$ under the action of $D_i$, {\sl i.e.}, equations $D_i[B_j]\neq 0$ with different $B_j$'s under the same $D_i$.
By inspecting $D$-quivers and corresponding operators, we find that there are two types of problems which cause the difficulties of solving linear equations.

The first problem comes from a key observation that, while operators $\mathcal{T}_{a h_i(a+1)}$ or  $\mathcal{T}_{h_jh_in}$ is able to select a particular dashed line or solid line uniquely in the basis quiver, the operator $(k_1\cdot k_{h_i})\mathcal{T}_{1 h_i 2}$ fails to do so. As a consequence, the contributions of different basis quivers will mix together when they can produce the same $D$-quivers. The reason is that each pseudo-loop of the vectors' basis quiver has two possible ways of generating $D$-quivers, so it is possible that two basis quivers with pseudo-loops generate the same $D$-quiver. For example, let us consider the following four basis quivers $B_i$ which generate five $D$-quivers totally.
\begin{align}
  \begin{tikzpicture}
%    \node (B1) [] at (0,0) {$B_1$};
%    \node (D1) [] at (4,0) {$D_1$};
%    \draw [thick, ->] (B1)--(D1);
     \node (B4) [] at (5,0) {$=B_4$};
     \node (B3) [] at (5,1) {$=B_3$};
     \node (B2) [] at (5,2) {$=B_2$};
     \node (B1) [] at (5,3) {$=B_1$};
     \node (D1) [] at (10,3.5) {$D_1=$};
     \node (D2) [] at (10,2.5) {$D_2=$};
     \node (D3) [] at (10,1.5) {$D_3=$};
     \node (D4) [] at (10,0.5) {$D_4=$};
     \node (D5) [] at (10,-0.5) {$D_5=$};
     \draw [<->, thick] (B1)--(D1);
     \draw [<->, thick] (B1)--(D2);
     \draw [<->, thick] (B2)--(D2);
     \draw [<->, thick] (B2)--(D3);
     \draw [<->, thick] (B3)--(D3);
     \draw [<->, thick] (B3)--(D4);
     \draw [<->, thick] (B4)--(D4);
     \draw [<->, thick] (B4)--(D5);
     \node [below] at (0,0) {$h_1$};
     \node [below] at (1,0) {$h_2$};
     \node [below] at (2,0) {$h_3$};
     \node [below] at (3,0) {$h_4$};
     \node [below] at (4,0) {$h_5$};
     \node [below] at (0,1) {$h_1$};
     \node [below] at (1,1) {$h_2$};
     \node [below] at (2,1) {$h_3$};
     \node [below] at (3,1) {$h_4$};
     \node [below] at (4,1) {$h_5$};
     \node [below] at (0,2) {$h_1$};
     \node [below] at (1,2) {$h_2$};
     \node [below] at (2,2) {$h_3$};
     \node [below] at (3,2) {$h_4$};
     \node [below] at (4,2) {$h_5$};
     \node [below] at (0,3) {$h_1$};
     \node [below] at (1,3) {$h_2$};
     \node [below] at (2,3) {$h_3$};
     \node [below] at (3,3) {$h_4$};
     \node [below] at (4,3) {$h_5$};
     \node [below] at (11,-0.5) {$h_1$};
     \node [below] at (12,-0.5) {$h_2$};
     \node [below] at (13,-0.5) {$h_3$};
     \node [below] at (14,-0.5) {$h_4$};
     \node [below] at (15,-0.5) {$h_5$};
     \node [below] at (11,0.5) {$h_1$};
     \node [below] at (12,0.5) {$h_2$};
     \node [below] at (13,0.5) {$h_3$};
     \node [below] at (14,0.5) {$h_4$};
     \node [below] at (15,0.5) {$h_5$};
     \node [below] at (11,1.5) {$h_1$};
     \node [below] at (12,1.5) {$h_2$};
     \node [below] at (13,1.5) {$h_3$};
     \node [below] at (14,1.5) {$h_4$};
     \node [below] at (15,1.5) {$h_5$};
     \node [below] at (11,2.5) {$h_1$};
     \node [below] at (12,2.5) {$h_2$};
     \node [below] at (13,2.5) {$h_3$};
     \node [below] at (14,2.5) {$h_4$};
     \node [below] at (15,2.5) {$h_5$};
     \node [below] at (11,3.5) {$h_1$};
     \node [below] at (12,3.5) {$h_2$};
     \node [below] at (13,3.5) {$h_3$};
     \node [below] at (14,3.5) {$h_4$};
     \node [below] at (15,3.5) {$h_5$};
     \draw [thick, cyan] (0,3) to [out=45, in=180] (0.5,3.15) to [out=0, in=135] (1,3);
     \draw [thick, brown] (0,3) to [out=315, in=180] (0.5,2.85) to [out=0, in=225] (1,3);
     \draw [thick, cyan] (1,2) to [out=45, in=180] (1.5,2.15) to [out=0, in=135] (2,2);
     \draw [thick, brown] (1,2) to [out=315, in=180] (1.5,1.85) to [out=0, in=225] (2,2);
     \draw [thick, cyan] (2,1) to [out=45, in=180] (2.5,1.15) to [out=0, in=135] (3,1);
     \draw [thick, brown] (2,1) to [out=315, in=180] (2.5,0.85) to [out=0, in=225] (3,1);
     \draw [thick, cyan] (3,0) to [out=45, in=180] (3.5,0.15) to [out=0, in=135] (4,0);
     \draw [thick, brown] (3,0) to [out=315, in=180] (3.5,-0.15) to [out=0, in=225] (4,0);
     \draw [thick] (0,0)--(3,0) (0,1)--(2,1) (3,1)--(4,1) (0,2)--(1,2) (2,2)--(4,2) (1,3)--(4,3);
     \draw [thick] (11,-0.5)--(15,-0.5) (11,0.5)--(15,0.5) (11,1.5)--(15,1.5) (11,2.5)--(15,2.5) (11,3.5)--(15,3.5);
     \draw [thick, dashed, ->] (15,-0.5)--(15,-0.2);
     \draw [thick, dashed, ->] (14,0.5)--(14,0.8);
     \draw [thick, dashed, ->] (13,1.5)--(13,1.8);
     \draw [thick, dashed, ->] (12,2.5)--(12,2.8);
     \draw [thick, dashed, ->] (11,3.5)--(11,3.8);
     \node [] at (15,-0.1) {{\tiny $k_1$}};
     \node [] at (14,0.9) {{\tiny $k_1$}};
     \node [] at (13,1.9) {{\tiny $k_1$}};
     \node [] at (12,2.9) {{\tiny $k_1$}};
     \node [] at (11,3.9) {{\tiny $k_1$}};
     \draw [fill] (0,0) circle [radius=0.05] (1,0) circle [radius=0.05] (2,0) circle [radius=0.05] (3,0) circle [radius=0.05] (4,0) circle [radius=0.05] (0,1) circle [radius=0.05] (1,1) circle [radius=0.05] (2,1) circle [radius=0.05] (3,1) circle [radius=0.05] (4,1) circle [radius=0.05] (0,2) circle [radius=0.05] (1,2) circle [radius=0.05] (2,2) circle [radius=0.05] (3,2) circle [radius=0.05] (4,2) circle [radius=0.05] (0,3) circle [radius=0.05] (1,3) circle [radius=0.05] (2,3) circle [radius=0.05] (3,3) circle [radius=0.05] (4,3) circle [radius=0.05];
     \draw [fill] (11,-0.5) circle [radius=0.05] (12,-0.5) circle [radius=0.05] (13,-0.5) circle [radius=0.05] (14,-0.5) circle [radius=0.05] (15,-0.5) circle [radius=0.05] (11,0.5) circle [radius=0.05] (12,0.5) circle [radius=0.05] (13,0.5) circle [radius=0.05] (14,0.5) circle [radius=0.05] (15,0.5) circle [radius=0.05] (11,1.5) circle [radius=0.05] (12,1.5) circle [radius=0.05] (13,1.5) circle [radius=0.05] (14,1.5) circle [radius=0.05] (15,1.5) circle [radius=0.05] (11,2.5) circle [radius=0.05] (12,2.5) circle [radius=0.05] (13,2.5) circle [radius=0.05] (14,2.5) circle [radius=0.05] (15,2.5) circle [radius=0.05] (11,3.5) circle [radius=0.05] (12,3.5) circle [radius=0.05] (13,3.5) circle [radius=0.05] (14,3.5) circle [radius=0.05] (15,3.5) circle [radius=0.05];
  \end{tikzpicture}\label{B14}
\end{align}
Hence if we choose $D_2$ as the corresponding differential operator of the basis quiver $B_1$, then after applying $D_2$ to these five vectors, $B_2$ is also non-zero besides $B_1$, which means that the coefficients of $B_1, B_2$ are mixed together in the linear equation given by $D_2$.

The above phenomenon is general.
Assuming the basis quiver of a vector in the gauge invariant basis has a pseudo-loop ${\sf{F}}_{h_{\alpha_{2i-1}}h_{\alpha_{2i}}}$ connected with a solid line ${\sf{F}}_{h_{\beta}}^{h_{\a_{2i}}}$, and the corresponding differential operator of the pseudo-loop is $(k_1\cdot k_{h_{\alpha_{2i}}})\mathcal{T}_{h_{\alpha_{2i}}h_{\alpha_{2i-1}}n}\mathcal{T}_{1h_{\alpha_{2i}}2}$, then we can almost always find a new vector in the basis having a factor ${\sf{F}}_{h_{\b}h_{\alpha_{2i}}}{\sf{F}}_{h_{\a_{2i-1}}}^{h_{\a_{2i}}}$ \footnote{The new basis is gotten by the operation of exchanging two subscripts $h_{\a_{2i-1}}$ and $h_{\b}$.}, which is non-zero under the same differential operator. We can do this operation independently for each pseudo-loop in a vector. If there are $\kappa_i$ solid lines connecting to the node $h_{\a_{2i}}$,
the total number of vectors which is non-zero under the corresponding differential operator of the pseudo-loop will be $\left(\prod_{i=1}^{p} (\kappa_i+1)-1\right)$. The results of these vectors under the action of differential operator are $D_i[B_j]=1$ for $B_j$ being a vector of the set, the fact will be important in the later construction of linear combination of $D_i$'s.
%It is worth to notice that the number of pseudo-loops will be same in this type.

Now let us consider the second problem originating from identity \eref{FF-relation}. Although the basis quivers of some vectors will not produce the same $D$-quiver\footnote{Please recall that the collection of $D$-quivers is a subset of all $(\eps k)$-quivers.}, they could give the same $(\eps k)$-quiver by replacing a dashed line $(\eps\cdot K_a)$ or a solid line $(\eps\cdot k_{h_j})$ by $(\eps\cdot k_1)$. For example, applying $D_1$ on the following two basis quivers all yields non-zero results,
\begin{align}
\begin{tikzpicture}
%%% first quiver
\draw [fill] (0,0) circle [radius=0.05] (1,0) circle [radius=0.05] (2,0) circle [radius=0.05] (3,0) circle [radius=0.05] (4,0) circle [radius=0.05];
\draw [thick,->] (0,0)--(0.55,0);
\draw [thick] (0.5,0)--(1,0);
\draw [thick, dashed, ->] (1,0)--(1,0.5);
\draw [thick] (1,0)--(1.55,0);
\draw [thick,<-] (1.5,0)--(2,0);
\draw [thick] (2,0)--(2.55,0);
\draw [thick,<-] (2.5,0)--(3,0);
\draw [thick] (3,0)--(3.55,0);
\draw [thick,<-] (3.5,0)--(4,0);
\node [above] at (1,0.5) {$K_a$};
\node [] at (-0.6,0) {$\W B_1=$};
\node [below] at (0,0) {$h_1$};
\node [below] at (1,0) {$h_2$};
\node [below] at (2,0) {$h_3$};
\node [below] at (3,0) {$h_4$};
\node [below] at (4,0) {$h_5$};
\node [right] at (4.2,0) {,};
%%%% second quiver
\draw [fill] (7,0) circle [radius=0.05] (8,0) circle [radius=0.05] (9,0) circle [radius=0.05] (10,0) circle [radius=0.05] (11,0) circle [radius=0.05];
\draw [thick,->] (7,0)--(7.55,0);
\draw [thick] (7.5,0)--(8,0);
\draw [thick,  ->] (8,0)--(8,0.5);
\draw [thick] (8,0)--(8.55,0);
\draw [thick,<-] (8.5,0)--(9,0);
\draw [thick] (9,0)--(9.55,0);
\draw [thick,<-] (9.5,0)--(10,0);
\draw [thick] (10,0)--(10.55,0);
\draw [thick,<-] (10.5,0)--(11,0);
\node [above] at (8,0.5) {$h_i$};
\node [] at (6.4,0) {$\W B_2=$};
\node [below] at (7,0) {$h_1$};
\node [below] at (8,0) {$h_2$};
\node [below] at (9,0) {$h_3$};
\node [below] at (10,0) {$h_4$};
\node [below] at (11,0) {$h_5$};
\node [right] at (11.2,0) {.};
\end{tikzpicture}~~~\label{DK}
\end{align}
Note that $\W B_2$ can be a rooted tree by itself or a rooted tree obtained by split a pseudo-loop, while $\W B_1$ can only be a rooted tree obtained by split a pseudo-loop. Thus in this case a branch of disconnected component with a pseudo-loop is mixed with a disconnected component without pseudo-loop.
Explicitly, for a vector with a pseudo-loop ${\sf{F}}_{h_{\alpha_{2i-1}}h_{\alpha_{2i}}}$ and the corresponding operator for the pseudo-loop $(k_1\cdot k_{h_{\alpha_{2i}}}) \mathcal{T}_{h_{\alpha_{2i}}h_{\alpha_{2i-1}}n}\mathcal{T}_{1h_{\alpha_{2i}}2}$, we can always find some new vectors by replacing ${\sf{F}}_{h_{\alpha_{2i-1}}h_{\alpha_{2i}}}$
with ${\sf{F}}_{h_{\alpha_{2i-1}}}^{h_{\alpha_{2i}}} {\sf{F}}_{h_{\alpha_{2i}}}^{K_a}$,
$\forall a=2,...,n-1$ or ${\sf{F}}_{h_{\alpha_{2i-1}}}^{h_{\alpha_{2i}}} {\sf{F}}_{h_{\alpha_{2i}}}^{h_p}$ with  arbitrary $p\neq \a_{2i}$\footnote{When such replacement produces a real loop it should be excluded.}. Since the replacement for each pseudo-loop is independently, there are totally $(2^p-1)(n-2+m-1)$ new vectors, and applying $D_i$ to these new vectors would produce $(-k_{h_{\a_{2i}}}\cdot (k_1+ K_a))$  or $(-k_{h_{\a_{2i}}}\cdot k_{h_p})$ respectively according to \eref{FF-relation}. This is consistent with the counting of mass dimension. However, these new vectors have their corresponding differential operators \eref{DK} under which the original vector with a pseudo-loop vanishes. Thus the second problem is easy to deal with if we solve the linear equations of unknown coefficients in a proper order.
\\

We have discussed two types of problems in details and the second type is easily solved, then let us continue to discuss how to deal with the first one. The first type of problems originates from the fact that under the action of a differential operator several vectors with pseudo-loops in the gauge invariant basis do not vanish at the same time, then their coefficients are mixed together in the linear equations. Our solution is to construct a linear combination of differential operators such that under its action only one vector is non-vanishing. Let us start from the simple example \eref{B14}, and it is easy to get
\bea
D_2(b_1B_1+b_2B_2+b_3B_3+b_4B_4)& =&b_1+b_2~~~, ~~~
D_3(b_1B_1+b_2B_2+b_3B_3+b_4B_4) =b_2+b_3~,~~~ \nn
D_4(b_1B_1+b_2B_2+b_3B_3+b_4B_4)& =& b_3+b_4~~~, ~~~
D_5(b_1B_1+b_2B_2+b_3B_3+b_4B_4) =b_4~.~~~\label{DB-exa}
\eea
If we define some new differential operators as ${\cal D}_i := \sum_{a=i}^{4}(-1)^{a-1}D_{a+1}$, then
\bea {\cal D}_i[B_j]=(-)^{i-1}\delta_{ij}~~~,~~~i,j=1,2,3,4~.~~~\eea
It means ${\cal D}_i$ selects a unique vector from the entangled vectors, and the linear equations of the coefficients of these vectors are easily solved. Generalizing this example, we can construct the linear combination of differential operators as follows.
\begin{itemize}

\item For a given vector $B_i$, we can get many $D$-quivers in general, but we choose only one $D$-quiver freely. For example,
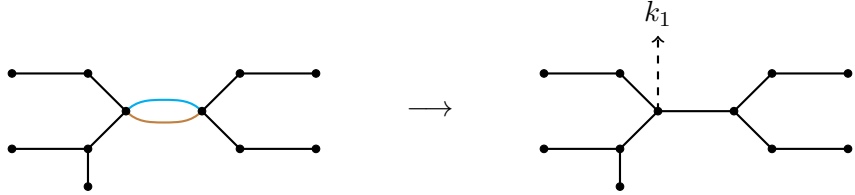
\begin{center}
  \begin{tikzpicture}
    \draw [thick, cyan] (1.5,0.5) to [out=45, in=180] (2,0.65) to [out=0, in=135] (2.5,0.5);
    \draw [thick, brown] (1.5,0.5) to [out=315, in=180] (2,0.35) to [out=0, in=225] (2.5,0.5);
    \draw [fill] (0,0) circle [radius=0.05] (0,1) circle [radius=0.05] (1,-0.5) circle [radius=0.05] (1,0) circle [radius=0.05] (1,1) circle [radius=0.05] (1.5,0.5) circle [radius=0.05] (2.5,0.5) circle [radius=0.05] (3,0) circle [radius=0.05] (3,1) circle [radius=0.05] (4,0) circle [radius=0.05] (4,1) circle [radius=0.05];
    \draw [fill] (7,0) circle [radius=0.05] (7,1) circle [radius=0.05] (8,-0.5) circle [radius=0.05] (8,0) circle [radius=0.05] (8,1) circle [radius=0.05] (8.5,0.5) circle [radius=0.05] (9.5,0.5) circle [radius=0.05] (10,0) circle [radius=0.05] (10,1) circle [radius=0.05] (11,0) circle [radius=0.05] (11,1) circle [radius=0.05];
    \draw [thick] (0,0)--(1,0)--(1.5,0.5)--(1,1)--(0,1) (1,0)--(1,-0.5) (4,0)--(3,0)--(2.5,0.5)--(3,1)--(4,1);
    \draw [thick] (7,0)--(8,0)--(8.5,0.5)--(8,1)--(7,1) (8,0)--(8,-0.5) (8.5,0.5)--(9.5,0.5) (11,0)--(10,0)--(9.5,0.5)--(10,1)--(11,1);
    \draw [thick, dashed, ->] (8.5,0.5)--(8.5,1.5);
    \node [above] at (8.5,1.5) {$k_1$};
    \node [] at (5.5,0.5) {$\longrightarrow$};
  \end{tikzpicture}
\end{center}

\item For the $D$-quiver whose root is $k_1$, there are two nodes coming from the original pseudo-loop. If the node connected with $k_1$ by a dashed line is denoted $h_{a}$, then another node denoted by $h_{b}$. We can separate this $D$-quiver into two parts in the node $h_a$ while assigning the line connecting $h_a$ and $h_b$ to $h_b$, and denote these two parts by ${\cal H}_a, {\cal H}_b$. For example,
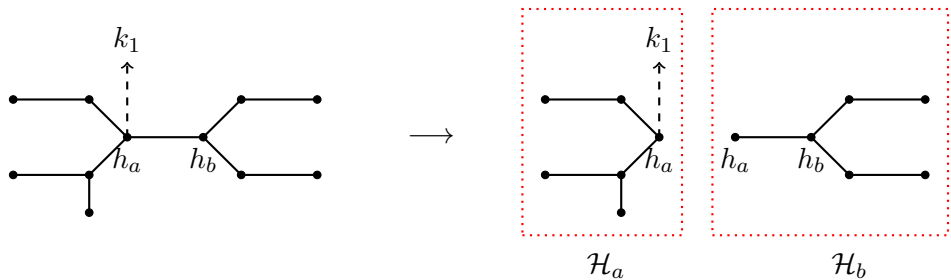
\begin{center}
  \begin{tikzpicture}
  \draw [fill] (0,0) circle [radius=0.05] (0,1) circle [radius=0.05] (1,-0.5) circle [radius=0.05] (1,0) circle [radius=0.05] (1,1) circle [radius=0.05] (1.5,0.5) circle [radius=0.05] (2.5,0.5) circle [radius=0.05] (3,0) circle [radius=0.05] (3,1) circle [radius=0.05] (4,0) circle [radius=0.05] (4,1) circle [radius=0.05];
  \draw [fill] (7,0) circle [radius=0.05]  (8,0) circle [radius=0.05] (8.5,0.5) circle [radius=0.05] (8,1) circle [radius=0.05] (7,1) circle [radius=0.05] (9.5,0.5) circle [radius=0.05] (10.5,0.5) circle [radius=0.05] (11,0) circle [radius=0.05] (11,1) circle [radius=0.05] (12,0) circle [radius=0.05] (12,1) circle [radius=0.05] (8,-0.5) circle [radius=0.05];
  \draw [thick] (0,0)--(1,0)--(1.5,0.5)--(1,1)--(0,1) (1,0)--(1,-0.5) (4,0)--(3,0)--(2.5,0.5)--(3,1)--(4,1) (1.5,0.5)--(2.5,0.5);
  \draw [thick, dashed, ->] (1.5,0.5)--(1.5,1.5);
  \draw [thick] (7,0)--(8,0)--(8.5,0.5)--(8,1)--(7,1) (9.5,0.5)--(10.5,0.5) (12,0)--(11,0)--(10.5,0.5)--(11,1)--(12,1) (8,0)--(8,-0.5);
  \draw [thick, dashed, ->] (8.5,0.5)--(8.5,1.5);
  \node [above] at (1.5,1.5) {$k_1$};
  \node [below] at (1.5,0.5) {$h_a$};
  \node [below] at (2.5,0.5) {$h_b$};
  \node [] at (5.5,0.5) {$\longrightarrow$};
  \node [above] at (8.5,1.5) {$k_1$};
  \node [below] at (8.5,0.5) {$h_a$};
  \node [below] at (9.5,0.5) {$h_a$};
  \node [below] at (10.5,0.5) {$h_b$};
  \draw [thick, dotted, red] (6.7,-0.8)--(8.8,-0.8)--(8.8,2.2)--(6.7,2.2)--cycle;
  \draw [thick, dotted, red] (9.2,-0.8)--(12.3,-0.8)--(12.3,2.2)--(9.2,2.2)--cycle;
  \node [] at (7.8,-1.2) {$\mathcal{H}_a$};
  \node [] at (11,-1.2) {$\mathcal{H}_b$};
  \end{tikzpicture}
\end{center}

\item In graph ${\cal H}_a$, $k_1$ is connected to $h_a$ and $k_1$ is the root with all lines' directions toward $k_1$. We can construct some new rooted trees by moving $k_1$ to other nodes and keeping $h_a$ being the root, then it is necessary to change the directions of lines. Each new rooted tree defines a differential operator denoted by $D_{{\cal H}_a,j}$ with $j=1,...,k$ where $k$ is the total number of nodes excluding $h_a$ in the rooted tree ${\cal H}_a$. Then we define a new differential operator of them by
    \bea {\cal D}_{{\cal H}_a}=\sum_{j=1}^k (-)^{s(j)}D_{{\cal H}_a,j}~,~~~\label{D-H-a} \eea
where $s(j)$ is the number of steps of moving $k_1$ from the node $h_a$ to the node $h_j$. For example,
\begin{center}
\begin{tikzpicture}
  \draw [fill] (0,0) circle [radius=0.05] (1,0) circle [radius=0.05] (1.5,0.5) circle [radius=0.05] (1,1) circle [radius=0.05] (0,1) circle [radius=0.05] (1,-0.5) circle [radius=0.05] (3,0) circle [radius=0.05] (4,0) circle [radius=0.05] (4.5,0.5) circle [radius=0.05] (4,1) circle [radius=0.05] (3,1) circle [radius=0.05] (4,-0.5) circle [radius=0.05] (6,0) circle [radius=0.05] (7,0) circle [radius=0.05] (7.5,0.5) circle [radius=0.05] (7,1) circle [radius=0.05] (6,1) circle [radius=0.05] (7,-0.5) circle [radius=0.05] (9,0) circle [radius=0.05] (10,0) circle [radius=0.05] (10.5,0.5) circle [radius=0.05] (10,1) circle [radius=0.05] (9,1) circle [radius=0.05] (10,-0.5) circle [radius=0.05] (12,0) circle [radius=0.05] (13,0) circle [radius=0.05] (13.5,0.5) circle [radius=0.05] (13,1) circle [radius=0.05] (12,1) circle [radius=0.05] (13,-0.5) circle [radius=0.05];
  \draw [thick] (0,0)--(1,0)--(1.5,0.5)--(1,1)--(0,1) (1,0)--(1,-0.5);
  \draw [thick] (3,0)--(4,0)--(4.5,0.5)--(4,1)--(3,1) (4,0)--(4,-0.5);
  \draw [thick] (6,0)--(7,0)--(7.5,0.5)--(7,1)--(6,1) (7,0)--(7,-0.5);
  \draw [thick] (9,0)--(10,0)--(10.5,0.5)--(10,1)--(9,1) (10,0)--(10,-0.5);
  \draw [thick] (12,0)--(13,0)--(13.5,0.5)--(13,1)--(12,1) (13,0)--(13,-0.5);
  \draw [thick, dashed, ->] (1,1)--(1,1.5);
  \draw [thick, dashed, ->] (3,1)--(3,1.5);
  \draw [thick, dashed, ->] (7,0)--(7.5,0);
  \draw [thick, dashed, ->] (9,0)--(9,-0.5);
  \draw [thick, dashed, ->] (13,-0.5)--(13.5,-0.5);
  \node [above] at (1,1.5) {$k_1$};
  \node [above] at (3,1.5) {$k_1$};
  \node [right] at (7.5,0) {$k_1$};
  \node [below] at (9,-0.5) {$k_1$};
  \node [right] at (13.5,-0.5) {$k_1$};
  \node [above] at (1.5,0.5) {$h_a$};
  \node [above] at (4.5,0.5) {$h_a$};
  \node [above] at (7.5,0.5) {$h_a$};
  \node [above] at (10.5,0.5) {$h_a$};
  \node [above] at (13.5,0.5) {$h_a$};
  \node [] at (-1.2,0.5) {$\mathcal{D}_{\mathcal{H}_a}=(-)$};
  \node [] at (2.3,0.5) {$+(-)^2$};
  \node [] at (5.3,0.5) {$+(-)$};
  \node [] at (8.3,0.5) {$+(-)^2$};
  \node [] at (11.3,0.5) {$+(-)^2$};
\end{tikzpicture}
\end{center}

\item Multiplying ${\cal D}_{{\cal H}_a}$ with the differential operator corresponding to ${\cal H}_b$ gives us the expected operator that will select only one particular vector from the set of the vectors entangled with the original vector. For example we get the linear combination
\begin{center}
  \begin{tikzpicture}
    \draw [fill] (0,0) circle [radius=0.05]  (0.5,0) circle [radius=0.05] (1,0.5) circle [radius=0.05] (0.5,1) circle [radius=0.05] (0,1) circle [radius=0.05] (1.5,0.5) circle [radius=0.05] (2,0) circle [radius=0.05] (2,1) circle [radius=0.05] (2.5,0) circle [radius=0.05] (2.5,1) circle [radius=0.05] (0.5,-0.5) circle [radius=0.05];
    \draw [fill] (3.5,0) circle [radius=0.05]  (4,0) circle [radius=0.05] (4.5,0.5) circle [radius=0.05] (4,1) circle [radius=0.05] (3.5,1) circle [radius=0.05] (5,0.5) circle [radius=0.05] (5.5,0) circle [radius=0.05] (5.5,1) circle [radius=0.05] (6,0) circle [radius=0.05] (6,1) circle [radius=0.05] (4,-0.5) circle [radius=0.05];
    \draw [fill] (7,0) circle [radius=0.05]  (7.5,0) circle [radius=0.05] (8,0.5) circle [radius=0.05] (7.5,1) circle [radius=0.05] (7,1) circle [radius=0.05] (8.5,0.5) circle [radius=0.05] (9,0) circle [radius=0.05] (9,1) circle [radius=0.05] (9.5,0) circle [radius=0.05] (9.5,1) circle [radius=0.05] (7.5,-0.5) circle [radius=0.05];
    \draw [fill] (10.5,0) circle [radius=0.05]  (11,0) circle [radius=0.05] (11.5,0.5) circle [radius=0.05] (11,1) circle [radius=0.05] (10.5,1) circle [radius=0.05] (12,0.5) circle [radius=0.05] (12.5,0) circle [radius=0.05] (12.5,1) circle [radius=0.05] (13,0) circle [radius=0.05] (13,1) circle [radius=0.05] (11,-0.5) circle [radius=0.05];
    \draw [fill] (14,0) circle [radius=0.05]  (14.5,0) circle [radius=0.05] (15,0.5) circle [radius=0.05] (14.5,1) circle [radius=0.05] (14,1) circle [radius=0.05] (15.5,0.5) circle [radius=0.05] (16,0) circle [radius=0.05] (16,1) circle [radius=0.05] (16.5,0) circle [radius=0.05] (16.5,1) circle [radius=0.05] (14.5,-0.5) circle [radius=0.05];
    \draw [thick] (0,0)--(0.5,0)--(1,0.5)--(0.5,1)--(0,1) (2.5,0)--(2,0)--(1.5,0.5)--(2,1)--(2.5,1) (0.5,0)--(0.5,-0.5) (1,0.5)--(1.5,0.5);
    \draw [thick] (3.5,0)--(4,0)--(4.5,0.5)--(4,1)--(3.5,1) (6,0)--(5.5,0)--(5,0.5)--(5.5,1)--(6,1) (4,0)--(4,-0.5) (4.5,0.5)--(5,0.5);
    \draw [thick] (7,0)--(7.5,0)--(8,0.5)--(7.5,1)--(7,1) (9.5,0)--(9,0)--(8.5,0.5)--(9,1)--(9.5,1) (7.5,0)--(7.5,-0.5) (8,0.5)--(8.5,0.5);
    \draw [thick] (10.5,0)--(11,0)--(11.5,0.5)--(11,1)--(10.5,1) (13,0)--(12.5,0)--(12,0.5)--(12.5,1)--(13,1) (11,0)--(11,-0.5) (11.5,0.5)--(12,0.5);
    \draw [thick] (14,0)--(14.5,0)--(15,0.5)--(14.5,1)--(14,1) (16.5,0)--(16,0)--(15.5,0.5)--(16,1)--(16.5,1) (14.5,0)--(14.5,-0.5) (15,0.5)--(15.5,0.5);
    \draw [thick, dashed, ->] (0.5,1)--(0.5,1.5);
    \draw [thick, dashed, ->] (3.5,1)--(3.5,1.5);
    \draw [thick, dashed, ->] (7.5,0)--(8,0);
    \draw [thick, dashed, ->] (10.5,0)--(10.5,-0.5);
    \draw [thick, dashed, ->] (14.5,-0.5)--(15,-0.5);
    \node [above] at (0.5,1.5) {$k_1$};
    \node [above] at (3.5,1.5) {$k_1$};
    \node [right] at (8,0) {$k_1$};
    \node [below] at (10.5,-0.5) {$k_1$};
    \node [right] at (15,-0.5) {$k_1$};
    \node [] at (-0.5,0.5) {$-$};
    \node [] at (3,0.5) {$+$};
    \node [] at (6.5,0.5) {$-$};
    \node [] at (10,0.5) {$+$};
    \node [] at (13.5,0.5) {$+$};
  \end{tikzpicture}
\end{center}

\item  A basis quiver of a vector would have many disconnected components, and for each disconnected component with a pseudo-loop we can apply the same procedure to it and similarly construct a corresponding operator $\mathcal{D}$ as a linear combination of some operators $D$. Multiplying all these operators $\mathcal{D}$ with those operators obtained from disconnected components without pseudo-loops, we get the final differential operator which will select a particular vector $B_i$ in gauge invariant basis without the first type of problems.

\end{itemize}
We should emphasize that, after obtaining these differential operators by the above method, if we apply them to the expansion there are still some troubles resulting from the second type of problems. It suggests that we should solve coefficients of vectors with fewer pseudo-loops first. We also remark that, although we have provided the method to solve the problem of mixing of some vectors in solving the linear equations of coefficients, when the size of linear equations is small it is quite favorable to solve them directly using the original differential operators defined in \eref{differential-operator}. The reason is that, while it is much simpler for computing coefficients of the mixed vectors by using differential operators constructed by the above method, it may be complicated for the cases we meet in the second type of problems since some vectors with less pseudo-loops are non-vanishing under the actio of these operators for the second type of problems.

%%%%%%%%%%%%%%%%%%%%%%%%%
\subsection{Algorithm for the evaluation of expansion coefficients}
%%%%%%%%%%%%%%%%%%%%%%%%%%%%

After clarifying the structure of differential operators, the next step is to apply them to the computing of expansion coefficients for the generic expansion formula (\ref{expansionGI}). For vectors of gauge invariant basis defined in \eref{basis-def-2}, the algorithm is implemented order by order, starting from $p=0$ to the largest value $p$. For a given $p$, we start from the largest $r$ to $r=0$. The value of $p$ denotes the number of pseudo-loops in a vector, hence when $p=0$ the basis quiver contains only solid and dashed lines without any pseudo-loop. Such vector can be mapped to an unique $D$-quiver representing the following differential operator,
\begin{equation}
\left(\prod_{i=1}^{q} \mathcal{T}_{h_{\beta'_{i}}h_{\beta_{i}}n}\right) \left( \prod_{i=1}^{r} \mathcal{T}_{a_{\gamma_i}h_{\gamma_{i}}(a_{\gamma_i+1})}\right)~~,~~q,r \in \mathbb{N}~~,~~q+r=m~.~~~\label{differential-operator-p0}
\end{equation}
Recalling identities (\ref{Tkhk-relation}), (\ref{Tkhk-relation2}) and (\ref{Thhk-relation}), a vector $B_j$ is non-vanishing only when its $D_j$-quiver is the same as that given by \eref{differential-operator-p0}. Thus the differential operator (\ref{differential-operator-p0}) uniquely selects one vector in gauge invariant basis while all others vanish, and the expansion coefficient can be solved by an univariate linear equation. Furthermore, the differential operator is normalized to one,
\begin{equation}
\left(\prod_{i=1}^{q} \mathcal{T}_{h_{\beta'_{i}}h_{\beta_{i}}n}\right) \left( \prod_{i=1}^{r} \mathcal{T}_{a_{\gamma_i}h_{\gamma_{i}}(a_{\gamma_i+1})}\right)~\left[~\left(\prod_{i=1}^{q}{\sf{F}}_{h_{\beta_i}}^{h_{\beta'_i}}\right) \left( \prod_{i=1}^{r} {\sf{F}}_{h_{\gamma_i}}^{a_{\gamma_i}}\right)~\right]=1~,~~~
\end{equation}
hence the expansion coefficient can be directly computed by applying differential operator (\ref{differential-operator-p0}) on the EYM amplitude, leading to
\begin{equation}
\mathcal{C}[\mathsf{F}_{h_{\beta_1}}^{h_{\beta'_1}}\cdots \mathsf{F}_{h_{\beta_q}}^{h_{\beta'_q}}\mathsf{F}_{h_{\gamma_1}}^{a_{\gamma_1}}\cdots \mathsf{F}_{h_{\gamma_r}}^{a_{\gamma_r}}]=\left(\prod_{i=1}^{q} \mathcal{T}_{h_{\beta'_{i}}h_{\beta_{i}}n}\right) \left( \prod_{i=1}^{r} \mathcal{T}_{a_{\gamma_i}h_{\gamma_{i}}(a_{\gamma_i+1})}\right)~A^{\EYM}_{n,m}(1,2,\ldots,n;h_1,\ldots, h_m)~.~~~\label{sol-coef-0}
\end{equation}
Note that $\mathcal{T}_{a_{\gamma_i}h_{\gamma_{i}}(a_{\gamma_i+1})}$ inserts $h_{\gamma_i}$ between $a_{\gamma_i}$ and $a_{\gamma_i}+1$ relative to the color-ordering, while $\mathcal{T}_{h_{\beta'_{i}}h_{\beta_{i}}n}$ inserts $h_{\beta_i}$ between $n$ and another graviton $h_{\beta'_i}$. Hence in the resulting Yang-Mills amplitudes, the legs $h_i$'s can never appear in the positions before $2$ or after $n$, and all Yang-Mills amplitudes are in the BCJ basis with legs $1, 2, n$ fixed. An example of evaluating \eref{sol-coef-0} has been discussed in \eref{D-exp-2-A}.

Heading to $p=1$ case, the differential operator for vector with one pseudo-loop is defined as\footnote{As mentioned, using the simple rule (\ref{differential-operator}) we might need to solve algebraic systems of linear equations. While using a more complicated combination of differential operators as (\ref{D-H-a}), the algebraic system is decoupled to univariate linear equations.},
\begin{equation}
\left((k_1\cdot k_{\alpha_2})\mathcal{T}_{h_{\alpha_{2}}h_{\alpha_{1}}n}\mathcal{T}_{1h_{\alpha_{2}}2}\right) \left(\prod_{i=1}^{q} \mathcal{T}_{h_{\beta'_{i}}h_{\beta_{i}}n}\right) \left( \prod_{i=1}^{r} \mathcal{T}_{a_{\gamma_i}h_{\gamma_{i}}(a_{\gamma_i+1})}\right)~~~,~~~q,r \in \mathbb{N}~~,~~q+r=m-2~,~~~\label{differential-operator-p1}
\end{equation}
with indices following convention (\ref{basis-def-indices}), and the total number of differential operators is
\begin{equation*}
  \frac{m!}{2(m-2)!}(n+m-3)^{m-2}~.~~~
\end{equation*}
In differential operators (\ref{differential-operator-p1}), the insertion operator $\mathcal{T}_{1h_{\alpha_2}2}$ will contribute a derivative $\partial_{\epsilon_{h_{\alpha_2}}\cdot k_1}$ relating to momentum $k_1$. In its quiver, there is only one branch with root $k_1$, and as we have analyzed, applying these differential operators on vectors will produce non-zero results only if the $D$-quiver of vector contains only one or no branch with root $k_1$. So all vectors with two or more pseudo-loops will vanish under (\ref{differential-operator-p1}). Furthermore, when applying (\ref{differential-operator-p1}) on vectors without pseudo-loop, there could be non-zero contribution. However it is not an issue since all coefficients of such vectors have been solved {\sl a priori} by differential operators (\ref{differential-operator-p0}) and they enter into the linear equations as known parameters.

For vectors with one pseudo-loop, there are in general more than one vectors being non-vanishing under a specific differential operator (\ref{differential-operator-p1}), as shown in \eref{DB-exa}. So we need to apply a complete set of differential operators to generate an algebraic system of linear equations, and solving expansion coefficients from this algebraic system. Alternatively, we can also apply the differential operator  constructed by rule \eref{D-H-a}, {\sl i.e.}, a special linear combination of differential operators in (\ref{differential-operator-p1}). Then an expansion coefficient can be determined by an univariate linear equation again. Nevertheless, we can compute the coefficient of vector with one pseudo-loop as,
\begin{eqnarray}
&& \mathcal{C}[\mathsf{F}_{h_{\alpha_1}h_{\alpha_2}}\mathsf{F}_{h_{\beta_1}}^{h_{\beta'_1}}\cdots \mathsf{F}_{h_{\beta_q}}^{h_{\beta'_q}}\mathsf{F}_{h_{\gamma_1}}^{a_{\gamma_1}}\cdots \mathsf{F}_{h_{\gamma_r}}^{a_{\gamma_r}}]\label{sol-coef-1}\\
&&~~~~= \left\{ \sum \left((k_1\cdot k_{\alpha_2})\mathcal{T}_{h_{\alpha_{2}}h_{\alpha_{1}}n}\mathcal{T}_{1h_{\alpha_{2}}2}\right) \left(\prod_{i=1}^{q} \mathcal{T}_{h_{\beta'_{i}}h_{\beta_{i}}n}\right) \left( \prod_{i=1}^{r} \mathcal{T}_{a_{\gamma_i}h_{\gamma_{i}}(a_{\gamma_i+1})}\right) \right\}~A^{\EYM}_{n,m}(1,2,\ldots,n;h_1,\ldots, h_m)\nonumber\\
&&~~~~~~~~~~~~~~~~~~~~~~~~~~~~~~~~~~~~~~~~~~~~~~~~~~~~~+\Big( \mbox{Contributions~from~basis~with~no~pseudo-loops}\Big)~,~~~\nonumber
\end{eqnarray}
where the summation in curly bracket represents a linear combination of differential operators constructed following the rule (\ref{D-H-a}). Note that the insertion operator $\mathcal{T}_{1h_{\alpha_2}2}$ inserts $h_{\alpha_2}$ in between $1$ and $2$, so the resulting Yang-Mills amplitudes are no longer in the BCJ basis with legs $1, 2, n$ fixed. BCJ relations are required in this step to write all Yang-Mills amplitudes into BCJ basis. While Yang-Mills amplitudes from contributions of vectors with no pseudo-loops are still in BCJ basis.

Now let us proceed to the vectors with $p$ pseudo-loops. According to the same argument with one pseudo-loop, by applying corresponding differential operator, all vectors with $(p+1)$ or more pseudo-loops will vanish. While for different vectors with $p$ pseudo-loops, a linear combination of differential operators constructed by rule \eref{D-H-a} is able to uniquely select a vector from all other vectors with $p$ pseudo-loops. However, these differential operators still produce non-zero results when applying on vectors with $(p-1)$ or fewer pseudo-loops. In order to solve these linear equations, all coefficients of vectors with $(p-1)$ or fewer pseudo-loops should be solved {\sl a priori} and enter these linear equations as known parameters. This inspires us to solve linear equations order by order from vectors with $p=0$ to $p=\lfloor \frac{m}{2}\rfloor$ pseudo-loops.

The differential operators relating to vectors with $p$ pseudo-loops in gauge invariant basis are given as,
\begin{equation}
\left( \prod_{i=1}^{p} (k_1\cdot k_{\alpha_{2i}})\mathcal{T}_{h_{\alpha_{2i}}h_{\alpha_{2i-1}}n}\mathcal{T}_{1h_{\alpha_{2i}}2}\right) \left(\prod_{i=1}^{q} \mathcal{T}_{h_{\beta'_{i}}h_{\beta_{i}}n}\right) \left( \prod_{i=1}^{r} \mathcal{T}_{a_{\gamma_i}h_{\gamma_{i}}(a_{\gamma_i+1})}\right)~,~q,r~\in~\mathbb{N}~,~q+r=m-2p~,~~~\label{differential-operator-pp}
\end{equation}
with indices following convention (\ref{basis-def-indices}), and the total number of differential operators is
\begin{equation*}
  \frac{m!}{p!~ 2^p~(m-2p)!} (n+m-3)^{m-2p}~.~~~
\end{equation*}
The expansion coefficients of vectors with $p$ pseudo-loops reads,
\begin{small}
\begin{eqnarray}
&& \mathcal{C}[\mathsf{F}_{h_{\alpha_{1}}h_{\alpha_{2}}}\cdots \mathsf{F}_{h_{\alpha_{2p-1}}h_{\alpha_{2p}}}\mathsf{F}_{h_{\beta_1}}^{h_{\beta'_1}}\cdots \mathsf{F}_{h_{\beta_q}}^{h_{\beta'_q}}\mathsf{F}_{h_{\gamma_1}}^{a_{\gamma_1}}\cdots \mathsf{F}_{h_{\gamma_r}}^{a_{\gamma_r}}]\\
&&{\tiny =\left\{ \sum \left( \prod_{i=1}^{p} (k_1\cdot k_{h_{\alpha_{2i}}}) \mathcal{T}_{h_{\alpha_{2i}}h_{\alpha_{2i-1}}n}\mathcal{T}_{1h_{\alpha_{2i}}2}\right) \left(\prod_{i=1}^{q} \mathcal{T}_{h_{\beta'_{i}}h_{\beta_{i}}n}\right) \left( \prod_{i=1}^{r} \mathcal{T}_{a_{\gamma_i}h_{\gamma_{i}}(a_{\gamma_i+1})}\right) \right\}}~A^{\EYM}_{n,m}(1,2,\ldots,n;h_1,\ldots, h_m)\nonumber\\
&&~~~~~~~~~~~~~~~~~~~~~~~~~~~~~~~~~~~~~~~+\Big( \mbox{Contributions~from~basis~with~}~(p-1)~\mbox{or~fewer~pseudo-loops}\Big)~.~~~\nonumber
\end{eqnarray}\end{small}
Again, the insertion operator $\prod_{i=1}^{p}\mathcal{T}_{1h_{\alpha_{2i}}2}$ inserts $h_{\alpha_2i}$'s in between legs $1$ and $2$, and we need to rewrite the resulting Yang-Mills amplitudes into BCJ basis by BCJ relations.

The algorithm for evaluation of expansion coefficients can be summarized as follows,
\begin{quote}
~\hfill {\sl\footnotesize -~-~Start of Algorithm~-~-} \hfill~

{\bf S{\footnotesize TEP 0}}: Apply differential operators (\ref{differential-operator-p0}) on EYM expansion formula (\ref{expansionGI}) to generate $(m+n-3)^{m}$ linear equations, and solve expansion coefficients from these equations\footnote{In fact, solving equations is not necessary in this step. The expansion coefficients have been uniquely determined by (\ref{sol-coef-0}), and the remaining thing to do is to explicitly work out the differential operators on $A^{\EYM}_{n,m}$ according to (\ref{sol-coef-0}).}. The result is directly given by (\ref{sol-coef-0}).

{\bf S{\footnotesize TEP 1}}: Substitute solutions of Step-0 back to formula (\ref{expansionGI}), then apply differential operators (\ref{differential-operator-p1}) on the resulting formula to generate linear equations. Solve expansion coefficients from these equations, and rewrite Yang-Mills amplitude into BCJ basis by BCJ relation.

~\hfill \vdots \hfill~

{\bf S{\footnotesize TEP $p$}}:  Substitute solutions of all previous steps back to formula (\ref{expansionGI}), then apply differential operators (\ref{differential-operator-pp}) on the resulting formula to generate linear equations. Solve expansion coefficients from these equations, and rewrite Yang-Mills amplitude into BCJ basis by BCJ relation.

~\hfill \vdots \hfill~

{\bf S{\footnotesize TEP $\lfloor \frac{m}{2} \rfloor$}}: Repeat the previous step but with $p=\lfloor \frac{m}{2} \rfloor$ differential operators.

~\hfill {\sl\footnotesize -~-~End of Algorithm~-~-} \hfill~
\end{quote}
The total number of repeated steps in the algorithm depends on the number of gravitons but not the gluons, while the total number of equations is much more sensitive to $m$ than to $n$. Table \ref{Table-count-eqns} shows the number of linear equations to be solved in the algorithm for some EYM amplitudes. Comparing the total number of equations for a fixed $m$, for example $A^{\EYM}_{5,4}$ and $A^{\EYM}_{15,4}$, we see the latter is about 44 times larger than the former when number of gluon increases ten. While comparing the total number of equations for a fixed $n$, for example $A^{\EYM}_{5,4}$ and $A^{\EYM}_{5,8}$, we see the latter is about 85902 times larger than the former when number of graviton increases four. Hence the size of algebraic system is significantly controlled by $m$. One also notice that the number of equations decreases rapidly as moving to the next step in the algorithm. A large amount of equations are solved in Step-0, where expansion coefficients are explicitly defined by acting differential operators on EYM amplitudes. So in some sense it is trivial. For step $p\neq 0$ in the algorithm, the number of equations decreases significantly compared to the previous step, however non-trivial contributions from previous steps and BCJ relations would make results involving. Nevertheless, in each step the linear equation system is decoupled, and an expansion coefficient is trivially solved via an univariate linear equation.
\begin{table}[h]
  \centering
  \begin{tabular}{|c|c|c|c|c|c|c|c|c|c|}
  \hline
  ~~~ & $A^{\EYM}_{5,4}$ &  $A^{\EYM}_{10,4}$  &  $A^{\EYM}_{15,4}$  &  $A^{\EYM}_{5,6}$  &  $A^{\EYM}_{10,6}$  &  $A^{\EYM}_{15,6}$  &  $A^{\EYM}_{5,8}$  &  $A^{\EYM}_{10,8}$  &  $A^{\EYM}_{15,8}$  \\
  \hline
  Step-0 & 1296 & 14641 & 65536 & 262144 & 4826809 & 34012224 & 100000000 & 2562890625 & 25600000000  \\ \hline
  Step-1 & 216 & 726 & 1536 & 61440 & 428415 & 1574640 & 28000000 & 318937500 & 1792000000 \\ \hline
  Step-2 & 3 & 3 & 3 & 2880 & 7605 & 14580 & 2100000 & 10631250 & 33600000 \\ \hline
  Step-3 & 0 & 0 & 0 & 15 & 15 & 15 & 42000 & 94500 & 168000 \\ \hline
  Step-4 & 0 & 0 & 0 & 0 & 0 & 0 & 105 & 105 & 105  \\ \hline
  Total & 1515 & 15370 & 67075 & 326479 & 5262844 & 35601459 & 130142105 & 2892553980 & 27425768105 \\
  \hline
\end{tabular}
  \caption{The number of linear equations in each step and in total for some EYM amplitudes. }\label{Table-count-eqns}
\end{table}
%

%%%%%%%%%%%%%%%%%%
\section{Demonstration of EYM amplitude expansion in gauge invariant vector space} \label{section-examples}
%%%%%%%%%%%%%%%%%

In order to demonstrate the EYM amplitude expansion in gauge invariant basis and the algorithm for determining expansion coefficients, in this section we present the expansion of EYM amplitudes with up to four gravitons. Expansion of EYM amplitudes with one, two and three gravitons to Yang-Mills amplitudes in BCJ basis has been discussed in paper \cite{Feng:2019tvb}, however here it receives a more systematic analysis in the language of gauge invariant vector space. While expansion of EYM amplitude with four gravitons to Yang-Mills amplitudes in BCJ basis is a new result.

%%%%%%%%%%%%%%%%%%
\subsection{The expansion of EYM amplitude with one and two gravitons}
%%%%%%%%%%%%%%%%%

Let us start with $A^{\EYM}_{n,1}(1,\ldots, n;h_1)$. This amplitude lives in the gauge invariant vector space $\mathcal{W}_{n+1,1}$, and the dimension of this space is $(n-2)$ according to (\ref{dimension-W}). Hence $A^{\EYM}_{n,1}$ can be expanded in a complete set of gauge invariant basis with $(n-2)$ gauge invariant vectors, as
\begin{equation}
\mathsf{F}_{h_1}^{a_1}=\frac{k_1\cdot f_{h_1}\cdot K_{a_1}}{k_1\cdot k_{h_1}}~~~,~~~a_1=2,\ldots,n-1~.~~~
\end{equation}
The expansion coefficient according to (\ref{sol-coef-0}) is calculated as,
\begin{equation}
\mathcal{C}[\mathsf{F}_{h_1}^{a_1}]=\mathcal{T}_{a_1h_1(a_1+1)}~A^{\EYM}_{n,1}(1,\ldots,n;h_1)=A^{\YM}_{n+1}(1,\ldots,a_1,h_1,a_1+1,\ldots,n)~,~~~
\end{equation}
where the graviton $h_1$ is transformed to a gluon and inserted between $a_1, a_1+1$ by $\mathcal{T}_{a_1h_1(a_1+1)}$. Hence
\begin{equation}
A^{\EYM}_{n,1}(1,\ldots,n;h_1)=\sum_{a_1=2}^{n-1}\mathcal{C}[\mathsf{F}_{h_1}^{a_1}] \mathsf{F}_{h_1}^{a_1}=\sum_{a_1=2}^{n-1}\frac{k_1\cdot f_{h_1}\cdot K_{a_1}}{k_1\cdot k_{h_1}}A^{\YM}_{n+1}(1,\ldots,a_1,h_1,a_1+1,\ldots,n)~.~~~
\end{equation}
In comparison with the result in \cite{Feng:2019tvb}, we can reformulate above result as,
\begin{equation}
A^{\EYM}_{n,1}(1,\ldots,n;h_1)=\sum_{\shuffle} \frac{k_1\cdot f_{h_1}\cdot Y_{h_1}}{k_1\cdot k_{h_1}} A^{\YM}_{n+1}(1,2,\{3,\ldots,n-1\}\shuffle \{h_1\},n)~,~~~
\end{equation}
where the shuffle permutation $\shuffle$ is defined in \eref{shuffle} and $Y_p$ as well as $X_p$ are defined in \eref{XpYp-def}.

Let us continue to $A^{\EYM}_{n,2}(1,\ldots,n;h_1,h_2)$. The dimension of gauge invariant vector space $\mathcal{W}_{n+2,2}$ is $\dim~\mathcal{W}_{n+2,2}=(n-1)^2+1$. The vectors in gauge invariant basis and their quiver representations are shown below as,
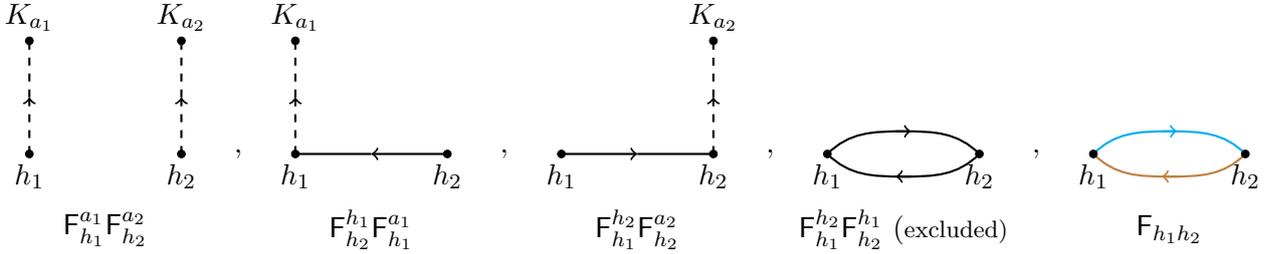
\begin{center}
\begin{tikzpicture}
  \draw [thick, ->, dashed] (0,0)--(0,0.75);
  \draw [thick, dashed] (0,0.75)--(0,1.5);
  \draw [thick, ->, dashed] (2,0)--(2,0.75);
  \draw [thick, dashed] (2,0.75)--(2,1.5);
  \draw [thick, ->, dashed] (3.5,0)--(3.5,0.75);
  \draw [thick, dashed] (3.5,0.75)--(3.5,1.5);
  \draw [thick, ->] (5.5,0)--(4.5,0);
  \draw [thick] (4.5,0)--(3.5,0);
  \draw [thick, ->, dashed] (9,0)--(9,0.75);
  \draw [thick, dashed] (9,0.75)--(9,1.5);
  \draw [thick, ->] (7,0)--(8,0);
  \draw [thick] (8,0)--(9,0);
  \draw [thick] (10.5,0) to [out=45, in=180] (11.5,0.3) to [out=0, in=135] (12.5,0);
  \draw [thick] (10.5,0) to [out=315, in=180] (11.5,-0.3) to [out=0, in=225] (12.5,0);
  \draw [thick, ->] (11.6,-0.3)--(11.4,-0.3);
  \draw [thick, ->] (11.4,0.3)--(11.6,0.3);
  \draw [thick, cyan] (14,0) to [out=45, in=180] (15,0.3) to [out=0, in=135] (16,0);
  \draw [thick, brown] (14,0) to [out=315, in=180] (15,-0.3) to [out=0, in=225] (16,0);
  \draw [thick, ->, brown] (15.1,-0.3)--(14.9,-0.3);
  \draw [thick, ->, cyan] (14.9,0.3)--(15.1,0.3);
  \draw [fill] (0,0) circle [radius=0.05] (2,0) circle [radius=0.05] (3.5,0) circle [radius=0.05] (5.5,0) circle [radius=0.05] (7,0) circle [radius=0.05] (9,0) circle [radius=0.05] (10.5,0) circle [radius=0.05] (12.5,0) circle [radius=0.05] (14,0) circle [radius=0.05] (16,0) circle [radius=0.05] (0,1.5) circle [radius=0.05] (2,1.5) circle [radius=0.05] (3.5,1.5) circle [radius=0.05] (9,1.5) circle [radius=0.05];
  \node [below] at (0,0) {$h_1$};
  \node [below] at (2,0) {$h_2$};
  \node [below] at (3.5,0) {$h_1$};
  \node [below] at (5.5,0) {$h_2$};
  \node [below] at (7,0) {$h_1$};
  \node [below] at (9,0) {$h_2$};
  \node [below] at (10.5,0) {$h_1$};
  \node [below] at (12.5,0) {$h_2$};
  \node [below] at (14,0) {$h_1$};
  \node [below] at (16,0) {$h_2$};
  \node [above] at (0,1.5) {$K_{a_1}$};
  \node [above] at (2,1.5) {$K_{a_2}$};
  \node [above] at (3.5,1.5) {$K_{a_1}$};
  \node [above] at (9,1.5) {$K_{a_2}$};
  \node [] at (2.75,0) {$,$};
  \node [] at (6.25,0) {$,$};
  \node [] at (9.75,0) {$,$};
  \node [] at (13.25,0) {$,$};
  \node [] at (1,-1) {$\mathsf{F}_{h_1}^{a_1}\mathsf{F}_{h_2}^{a_2}$};
  \node [] at (4.5,-1) {$\mathsf{F}_{h_2}^{h_1}\mathsf{F}_{h_1}^{a_1}$};
  \node [] at (8,-1) {$\mathsf{F}_{h_1}^{h_2}\mathsf{F}_{h_2}^{a_2}$};
  \node [] at (11.5,-1) {$\mathsf{F}_{h_1}^{h_2}\mathsf{F}_{h_2}^{h_1}$ (\footnotesize excluded)};
  \node [] at (15,-1) {$\mathsf{F}_{h_1h_2}$};
\end{tikzpicture}
\end{center}
where $2\leq a_1,a_2\leq n-1$. $\mathsf{F}_{h_1}^{h_2}\mathsf{F}_{h_2}^{h_1}$ is a real loop and should be excluded from the basis, while there is only one vector with pseudo-loop. Following the algorithm, Step-0 is to compute the coefficients of expansion basis with no pseudo-loops, {\sl i.e.}, $\mathsf{F}_{h_1}^{a_1}\mathsf{F}_{h_2}^{a_2}$, $\mathsf{F}_{h_2}^{h_1}\mathsf{F}_{h_1}^{a_1}$ and $\mathsf{F}_{h_1}^{h_2}\mathsf{F}_{h_2}^{a_2}$,  by formula (\ref{sol-coef-0}). Applying differential operators $\mathcal{T}_{a_1h_1(a_1+1)}\mathcal{T}_{a_2h_2(a_2+1)}$ and $\mathcal{T}_{h_{\sigma_2}h_{\sigma_1}n}\mathcal{T}_{a_{\sigma_2}h_{\sigma_2}(a_{\sigma_2}+1)}$ on $A^{\EYM}_{n,2}$ respectively leads to
\begin{eqnarray}
&& \mathcal{C}[\mathsf{F}_{h_{1}}^{a_1}\mathsf{F}_{h_{2}}^{a_2}]
=\left\{\begin{array}{l}
A^{\YM}_{n+2}(1,\ldots,a_{\sigma_1},h_{\sigma_1},a_{\sigma_1}+1,\ldots,a_{\sigma_2},h_{\sigma_2},a_{\sigma_2}+1,\ldots,n)~~~,~~~a_{\sigma_1}<a_{\sigma_2}\\
\sum_{\sigma \in S_2}A^{\YM}_{n+2}(1,\ldots,a_1,h_{\sigma_1},h_{\sigma_2},a_1+1,\ldots,n)~~~,~~~~~~~~~~~~~~~~~~~~a_1=a_2
\end{array}\right.~,~~~\label{sol-m2-coef-FkFk}\\
&& \mathcal{C}[\mathsf{F}_{h_{\sigma_1}}^{h_{\sigma_2}}\mathsf{F}_{h_{\sigma_2}}^{a_{\sigma_2}}]=\sum_{\shuffle} A^{\YM}_{n+2}(1,2,\ldots,a_{\sigma_2},h_{\sigma_2},\{a_{\sigma_2}+1,\ldots,n-1\}\shuffle \{h_{\sigma_1}\},n)~,~~~\label{sol-m2-coef-FhFk}
\end{eqnarray}
where $\sigma=\{\sigma_1, \sigma_2\}$ is a permutation of $\{1,2\}$, and the summation is over all elements of $S_2$. In Step-1, we substitute above solutions back to the expansion formula and get,
\begin{equation}
A^{\EYM}_{n,2}=\sum_{a_1,a_2=2}^{n-1}\mathcal{C}[\mathsf{F}_{h_1}^{a_1}\mathsf{F}_{h_2}^{a_2}] \mathsf{F}_{h_1}^{a_1}\mathsf{F}_{h_2}^{a_2}
+\sum_{a=2}^{n-1}\left(\mathcal{C}[\mathsf{F}_{h_2}^{h_1}\mathsf{F}_{h_1}^{a}] \mathsf{F}_{h_2}^{h_1}\mathsf{F}_{h_1}^{a}
+\mathcal{C}[\mathsf{F}_{h_1}^{h_2}\mathsf{F}_{h_2}^{a}] \mathsf{F}_{h_1}^{h_2}\mathsf{F}_{h_2}^{a} \right)+\mathcal{C}[\mathsf{F}_{h_1h_2}] \mathsf{F}_{h_1h_2}~,~~~\nonumber
\end{equation}
and there is only one unknown variable $\mathcal{C}[\mathsf{F}_{h_1h_2}]$. If applying differential operator $(k_1\cdot k_{h_2})\mathcal{T}_{h_2h_1n}\mathcal{T}_{1h_22}$ on both sides of above formula, in the RHS the non-vanishing contribution comes from vectors $\mathsf{F}_{h_1}^{h_2}\mathsf{F}_{h_2}^{a}$ and $\mathsf{F}_{h_1h_2}$, and according to \eref{FF-relation}, \eref{ThhkT1h2-relation} we get
\begin{equation}
(k_1\cdot k_{h_2})\mathcal{T}_{h_2h_1n}\mathcal{T}_{1h_22}~\mathsf{F}_{h_1}^{h_2}\mathsf{F}_{h_2}^{a_2}= -k_{h_2}\cdot (k_1+K_{a_2})~~~,~~~
(k_1\cdot k_{h_2})\mathcal{T}_{h_2h_1n}\mathcal{T}_{1h_22}~ \mathsf{F}_{h_1h_2}=1~.~~~
\end{equation}
In the LHS we get,
\begin{equation}
\mathcal{T}_{h_2h_1n}\mathcal{T}_{1h_22}~A^{\EYM}_{n,2}=A^{\YM}_{n+2}(1,\{h_2,h_1\}
\shuffle \{2,\ldots,n-1\}_R, n)=A^{\YM}_{n+2}(1,h_2,\{h_1\}
\shuffle \{2,\ldots,n-1\}, n)~.~~~
\end{equation}
Then we arrive at
\bea &&\mathcal{C}[\mathsf{F}_{h_1h_2}] =  (k_{h_2}\cdot k_1)A^{\YM}_{n+2}(1,h_2,\{h_1\}
\shuffle \{2,\ldots,n-1\}, n)-\sum_{a=2}^{n-1}\mathcal{C}[\mathsf{F}_{h_1}^{h_2}\mathsf{F}_{h_2}^{a}] {(-k_{h_2}\cdot (k_1+K_{a_2}))}\label{2p-C-1}\\
& & =\begin{array}{l}
        (k_{h_2}\cdot k_1)A^{\YM}_{n+2}(1,\{h_2,h_1\}
\shuffle \{2,\ldots,n-1\}_R, n)+{(k_{h_2}\cdot Y_{h_2})} A^{\YM}_{n+2}(1,2,\{3,...,n-1\}\shuffle \{h_2, h_1\},n)  \\

      \end{array}~.~~~\nonumber\eea
Yang-Mills amplitudes in the second term is already in the BCJ basis with legs $1,2,n$ fixed while those in the first term is not. So we need to rewrite the first term in BCJ basis as,
\begin{eqnarray}
&&A^{\YM}_{n+2}(1,h_2,\{h_1\}
\shuffle \{2,\ldots,n-1\}, n)=A^{\YM}_{n+2}(1,h_2,h_1,2,\ldots,n)+A^{\YM}_{n+2}(1,h_2,2,\{3,\ldots.n-1\}\shuffle \{h_1\},n)~\nonumber\\
&&=\sum_{\shuffle} \frac{(k_{h_2}\cdot X_{h_2}-k_1\cdot k_{h_2})(k_{h_1}\cdot X_{h_1})-(k_{h_2}\cdot X_{h_2}){\cal K}_{1h_1h_2}}{(k_1\cdot k_{h_2}) {\cal K}_{1h_1h_2}}A^{\YM}_{n+2}(1,2,\{3,\ldots,n-1\}\shuffle \{h_2,h_1\},n)\nonumber\\
&&~~~~~~~~+\sum_{\shuffle} \frac{(k_{h_2}\cdot X_{h_2})(k_{h_1}\cdot X_{h_1}-k_1\cdot k_{h_1})}{(k_1\cdot k_{h_2}) {\cal K}_{1h_1h_2}}A^{\YM}_{n+2}(1,2,\{3,\ldots,n-1\}\shuffle \{h_1,h_2\},n) ~,~~~\label{sol-m2-coef-Fhh-pre}
\end{eqnarray}
with ${\cal K}_{a_1\cdots a_m}=\sum_{1\le i<j\le m} k_{a_i}\cdot k_{a_j}$. Combining above results together, we finally obtain
\begin{eqnarray}
\mathcal{C}[\mathsf{F}_{h_1h_2}]&=&~(k_1\cdot k_{h_2})(\mathcal{T}_{h_2h_1n}\mathcal{T}_{1h_22}~A^{\EYM}_{n,2})~+~\sum_{a_2=2}^{n-1}\Big(k_{h_2}\cdot (k_1+K_{a_2})\Big)\mathcal{C}[\mathsf{F}_{h_1}^{h_2}\mathsf{F}_{h_2}^{a_2}]~ \label{sol-m2-coef-Fhh}\\
&=&\sum_{\{\sigma_1,\sigma_2\}\in S_2}\sum_{\shuffle} \frac{(k_{h_{\sigma_1}}\cdot X_{h_{\sigma_1}}-k_1\cdot k_{h_{\sigma_1}})(k_{h_{\sigma_2}}\cdot X_{h_{\sigma_2}})}{{\cal K}_{1h_1h_2}}A^{\YM}_{n+2}(1,2,\{3,\ldots,n-1\}\shuffle \{h_{\sigma_1},h_{\sigma_2}\},n)~.~~~\nonumber
\end{eqnarray}
Summing over all expansion basis with corresponding coefficients (\ref{sol-m2-coef-FkFk}), (\ref{sol-m2-coef-FhFk}) and (\ref{sol-m2-coef-Fhh}), we get the expected EYM amplitude expansion. In fact, all contributions of vectors with no pseudo-loops computed in Step-0 can be rearranged in a compact expression as,
\begin{eqnarray}
&&\sum_{a_1=2}^{n-1}\sum_{a_2=2}^{n-1}\mathcal{C}[\mathsf{F}_{h_1}^{a_1}\mathsf{F}_{h_2}^{a_2}]\mathcal{B}[\mathsf{F}_{h_1}^{a_1}\mathsf{F}_{h_2}^{a_2}]+\sum_{a_1=2}^{n-1}\mathcal{C}[\mathsf{F}_{h_2}^{h_1}\mathsf{F}_{h_1}^{a_1}]\mathcal{B}[\mathsf{F}_{h_2}^{h_1}\mathsf{F}_{h_1}^{a_1}]+\sum_{a_2=2}^{n-1}\mathcal{C}[\mathsf{F}_{h_1}^{h_2}\mathsf{F}_{h_2}^{a_2}]\mathcal{B}[\mathsf{F}_{h_1}^{h_2}\mathsf{F}_{h_2}^{a_2}]\nonumber\\
&&~~~~~~~~~~~~~~~=\sum_{\shuffle} \frac{k_1\cdot f_{h_1}\cdot X_{h_1}}{k_1\cdot k_{h_1}}\frac{k_1\cdot f_{h_2}\cdot X_{h_2}}{k_1\cdot k_{h_2}}A^{\YM}_{n+2}(1,2,\{3,\ldots,n-1\}\shuffle \{h_1\}\shuffle \{h_2\},n)~,~~~\label{step-0-2h}\end{eqnarray}
and as we shall see, this is a general property for EYM amplitudes with arbitrary gravitons. After rearrangement of terms, we can rewrite the expansion of EYM amplitude with two gravitons in a rather compact form as,
\begin{eqnarray}
&& A^{\EYM}_{n,2}(1,2,\ldots,n;h_1,h_2)=\sum_{\shuffle} \frac{k_1\cdot f_{h_1}\cdot X_{h_1}}{k_1\cdot k_{h_1}}\frac{k_1\cdot f_{h_2}\cdot X_{h_2}}{k_1\cdot k_{h_2}}A^{\YM}_{n+2}(1,2,\{3,\ldots,n-1\}\shuffle \{h_1\}\shuffle \{h_2\},n)\nonumber\\
&&~+\sum_{\{\sigma_1,\sigma_2\}\in S_2}\sum_{\shuffle} \mathsf{F}_{h_1h_2}\frac{(k_{h_{\sigma_1}}\cdot X_{h_{\sigma_1}}-k_1\cdot k_{h_{\sigma_1}})(k_{h_{\sigma_2}}\cdot X_{h_{\sigma_2}})}{{\cal K}_{1h_1h_2}}A^{\YM}_{n+2}(1,2,\{3,\ldots,n-1\}\shuffle \{h_{\sigma_1},h_{\sigma_2}\},n)~.~~~\nonumber
\end{eqnarray}
%

%%%%%%%%%%%%%%%%%%
\subsection{The expansion of EYM amplitude with three gravitons}
%%%%%%%%%%%%%%%%%
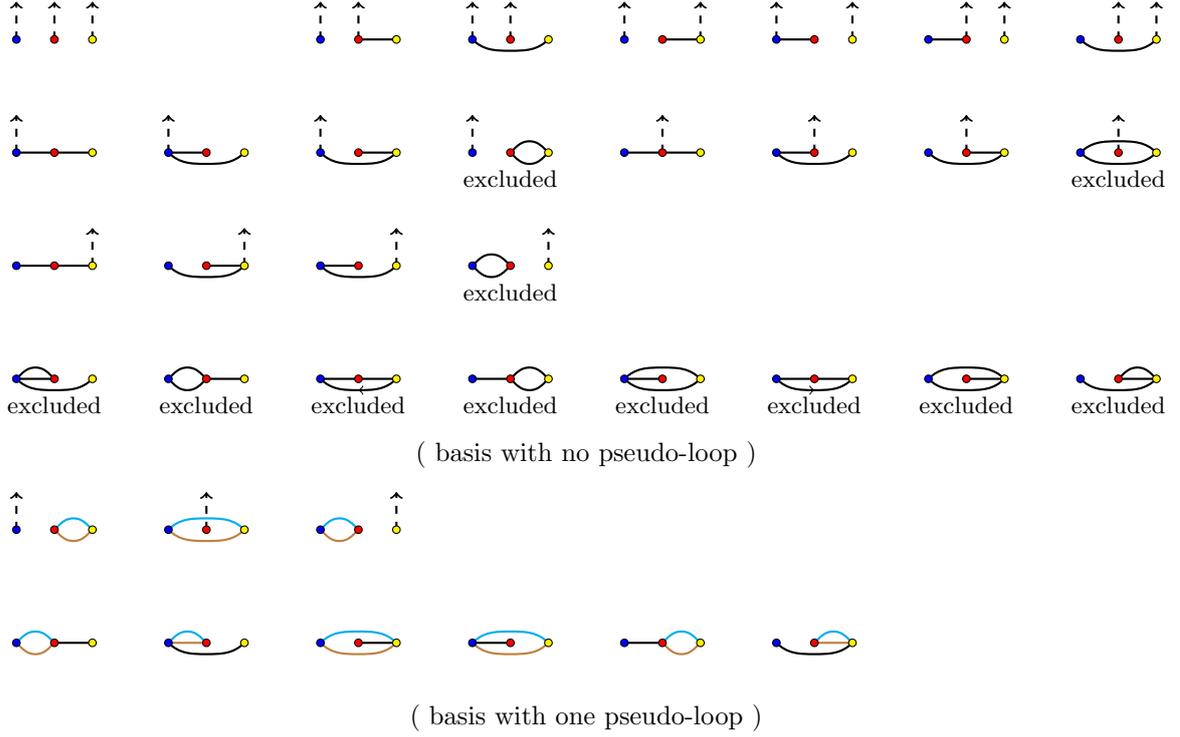
\begin{figure}
  \centering
\begin{tikzpicture}
  \draw [thick, dashed, ->] (0,0)--(0,0.5);
  \draw [thick, dashed, ->] (0.5,0)--(0.5,0.5);
  \draw [thick, dashed, ->] (1,0)--(1,0.5);
  \draw [thick, dashed, ->] (4,0)--(4,0.5);
  \draw [thick, dashed, ->] (4.5,0)--(4.5,0.5);
  \draw [thick, dashed, ->] (6,0)--(6,0.5);
  \draw [thick, dashed, ->] (6.5,0)--(6.5,0.5);
  \draw [thick, dashed, ->] (8,0)--(8,0.5);
  \draw [thick, dashed, ->] (9,0)--(9,0.5);
  \draw [thick, dashed, ->] (10,0)--(10,0.5);
  \draw [thick, dashed, ->] (11,0)--(11,0.5);
  \draw [thick, dashed, ->] (12.5,0)--(12.5,0.5);
  \draw [thick, dashed, ->] (13,0)--(13,0.5);
  \draw [thick, dashed, ->] (14.5,0)--(14.5,0.5);
  \draw [thick, dashed, ->] (15,0)--(15,0.5);
  \draw [thick] (4.5,0)--(5,0) (8.5,0)--(9,0) (10,0)--(10.5,0) (12,0)--(12.5,0);
  \draw [thick] (6,0) to [out=315, in=180] (6.5,-0.15) to [out=0, in=225] (7,0);
  \draw [thick] (14,0) to [out=315, in=180] (14.5,-0.15) to [out=0, in=225] (15,0);
  \draw [thick, dashed, ->] (0,-1.5)--(0,-1);
  \draw [thick, dashed, ->] (2,-1.5)--(2,-1);
  \draw [thick, dashed, ->] (4,-1.5)--(4,-1);
  \draw [thick, dashed, ->] (6,-1.5)--(6,-1);
  \draw [thick, dashed, ->] (8.5,-1.5)--(8.5,-1);
  \draw [thick, dashed, ->] (10.5,-1.5)--(10.5,-1);
  \draw [thick, dashed, ->] (12.5,-1.5)--(12.5,-1);
  \draw [thick, dashed, ->] (14.5,-1.5)--(14.5,-1);
  \draw [thick] (0,-1.5)--(1,-1.5) (2,-1.5)--(2.5,-1.5) (4.5,-1.5)--(5,-1.5) (8,-1.5)--(9,-1.5) (10,-1.5)--(10.5,-1.5) (12.5,-1.5)--(13,-1.5);
  \draw [thick] (2,-1.5) to [out=315, in=180] (2.5,-1.65) to [out=0, in=225] (3,-1.5);
  \draw [thick] (4,-1.5) to [out=315, in=180] (4.5,-1.65) to [out=0, in=225] (5,-1.5);
  \draw [thick] (10,-1.5) to [out=315, in=180] (10.5,-1.65) to [out=0, in=225] (11,-1.5);
  \draw [thick] (12,-1.5) to [out=315, in=180] (12.5,-1.65) to [out=0, in=225] (13,-1.5);
  \draw [thick] (14,-1.5) to [out=315, in=180] (14.5,-1.65) to [out=0, in=225] (15,-1.5);
  \draw [thick] (14,-1.5) to [out=45, in=180] (14.5,-1.35) to [out=0, in=135] (15,-1.5);
  \draw [thick] (6.5,-1.5) to [out=315, in=180] (6.75,-1.65) to [out=0, in=225] (7,-1.5);
  \draw [thick] (6.5,-1.5) to [out=45, in=180] (6.75,-1.35) to [out=0, in=135] (7,-1.5);
  \draw [thick, dashed, ->] (1,-3)--(1,-2.5);
  \draw [thick, dashed, ->] (3,-3)--(3,-2.5);
  \draw [thick, dashed, ->] (5,-3)--(5,-2.5);
  \draw [thick, dashed, ->] (7,-3)--(7,-2.5);
  \draw [thick] (0,-3)--(1,-3) (2.5,-3)--(3,-3) (4,-3)--(4.5,-3);
  \draw [thick] (2,-3) to [out=315, in=180] (2.5,-3.15) to [out=0, in=225] (3,-3);
  \draw [thick] (4,-3) to [out=315, in=180] (4.5,-3.15) to [out=0, in=225] (5,-3);
  \draw [thick] (6,-3) to [out=315, in=180] (6.25,-3.15) to [out=0, in=225] (6.5,-3);
  \draw [thick] (6,-3) to [out=45, in=180] (6.25,-2.85) to [out=0, in=135] (6.5,-3);
  \draw [thick] (2.5,-4.5)--(3,-4.5) (4,-4.5)--(5,-4.5) (6,-4.5)--(6.5,-4.5) (8,-4.5)--(8.5,-4.5) (10,-4.5)--(11,-4.5) (12.5,-4.5)--(13,-4.5) (0,-4.5)--(0.5,-4.5) (14.5,-4.5)--(15,-4.5);
  \draw [thick] (0,-4.5) to [out=315, in=180] (0.5,-4.65) to [out=0, in=225] (1,-4.5);
  \draw [thick] (4,-4.5) to [out=315, in=180] (4.5,-4.65) to [out=0, in=225] (5,-4.5);
  \draw [thick] (8,-4.5) to [out=315, in=180] (8.5,-4.65) to [out=0, in=225] (9,-4.5);
  \draw [thick] (10,-4.5) to [out=315, in=180] (10.5,-4.65) to [out=0, in=225] (11,-4.5);
  \draw [thick] (12,-4.5) to [out=315, in=180] (12.5,-4.65) to [out=0, in=225] (13,-4.5);
  \draw [thick] (14,-4.5) to [out=315, in=180] (14.5,-4.65) to [out=0, in=225] (15,-4.5);
  \draw [thick] (8,-4.5) to [out=45, in=180] (8.5,-4.35) to [out=0, in=135] (9,-4.5);
  \draw [thick] (12,-4.5) to [out=45, in=180] (12.5,-4.35) to [out=0, in=135] (13,-4.5);
  \draw [fill=blue] (0,0) circle [radius=0.05] (4,0) circle [radius=0.05] (6,0) circle [radius=0.05] (8,0) circle [radius=0.05] (10,0) circle [radius=0.05] (12,0) circle [radius=0.05] (14,0) circle [radius=0.05];
  \draw [fill=red] (0.5,0) circle [radius=0.05] (4.5,0) circle [radius=0.05] (6.5,0) circle [radius=0.05] (8.5,0) circle [radius=0.05] (10.5,0) circle [radius=0.05] (12.5,0) circle [radius=0.05] (14.5,0) circle [radius=0.05];
  \draw [fill=yellow] (1,0) circle [radius=0.05] (5,0) circle [radius=0.05] (7,0) circle [radius=0.05] (9,0) circle [radius=0.05] (11,0) circle [radius=0.05] (13,0) circle [radius=0.05] (15,0) circle [radius=0.05];
  \draw [thick] (2,-4.5) to [out=315, in=180] (2.25,-4.65) to [out=0, in=225] (2.5,-4.5);
  \draw [thick] (2,-4.5) to [out=45, in=180] (2.25,-4.35) to [out=0, in=135] (2.5,-4.5);
  \draw [thick] (6.5,-4.5) to [out=315, in=180] (6.75,-4.65) to [out=0, in=225] (7,-4.5);
  \draw [thick] (6.5,-4.5) to [out=45, in=180] (6.75,-4.35) to [out=0, in=135] (7,-4.5);
  \draw [thick] (0,-4.5) to [out=45, in=180] (0.25,-4.35) to [out=0, in=135] (0.5,-4.5);
  \draw [thick] (14.5,-4.5) to [out=45, in=180] (14.75,-4.35) to [out=0, in=135] (15,-4.5);
  \draw [->] (4.6,-4.65)--(4.5,-4.65);
  \draw [->] (10.4,-4.65)--(10.5,-4.65);
  \draw [thick, dashed, ->] (0,-6.5)--(0,-6);
  \draw [thick, dashed, ->] (2.5,-6.5)--(2.5,-6);
  \draw [thick, dashed, ->] (5,-6.5)--(5,-6);
  \draw [thick, brown] (0.5,-6.5) to [out=315, in=180] (0.75,-6.65) to [out=0, in=225] (1,-6.5);
  \draw [thick, cyan] (0.5,-6.5) to [out=45, in=180] (0.75,-6.35) to [out=0, in=135] (1,-6.5);
  \draw [thick, brown] (4,-6.5) to [out=315, in=180] (4.25,-6.65) to [out=0, in=225] (4.5,-6.5);
  \draw [thick, cyan] (4,-6.5) to [out=45, in=180] (4.25,-6.35) to [out=0, in=135] (4.5,-6.5);
  \draw [thick, brown] (2,-6.5) to [out=315, in=180] (2.5,-6.65) to [out=0, in=225] (3,-6.5);
  \draw [thick, cyan] (2,-6.5) to [out=45, in=180] (2.5,-6.35) to [out=0, in=135] (3,-6.5);
  \draw [thick] (0.5,-8)--(1,-8) (4.5,-8)--(5,-8) (6,-8)--(6.5,-8) (8,-8)--(8.5,-8);
  \draw [thick, brown] (2,-8)--(2.5,-8) (10.5,-8)--(11,-8);
  \draw [thick, brown] (0,-8) to [out=315, in=180] (0.25,-8.15) to [out=0, in=225] (0.5,-8);
  \draw [thick, cyan] (0,-8) to [out=45, in=180] (0.25,-7.85) to [out=0, in=135] (0.5,-8);
  \draw [thick, brown] (8.5,-8) to [out=315, in=180] (8.75,-8.15) to [out=0, in=225] (9,-8);
  \draw [thick, cyan] (8.5,-8) to [out=45, in=180] (8.75,-7.85) to [out=0, in=135] (9,-8);
  \draw [thick, cyan] (2,-8) to [out=45, in=180] (2.25,-7.85) to [out=0, in=135] (2.5,-8);
  \draw [thick, cyan] (10.5,-8) to [out=45, in=180] (10.75,-7.85) to [out=0, in=135] (11,-8);
  \draw [thick, brown] (4,-8) to [out=315, in=180] (4.5,-8.15) to [out=0, in=225] (5,-8);
  \draw [thick, cyan] (4,-8) to [out=45, in=180] (4.5,-7.85) to [out=0, in=135] (5,-8);
  \draw [thick, brown] (6,-8) to [out=315, in=180] (6.5,-8.15) to [out=0, in=225] (7,-8);
  \draw [thick, cyan] (6,-8) to [out=45, in=180] (6.5,-7.85) to [out=0, in=135] (7,-8);
  \draw [thick] (2,-8) to [out=315, in=180] (2.5,-8.15) to [out=0, in=225] (3,-8);
  \draw [thick] (10,-8) to [out=315, in=180] (10.5,-8.15) to [out=0, in=225] (11,-8);
  \draw [fill=blue] (0,-1.5) circle [radius=0.05] (2,-1.5) circle [radius=0.05]  (4,-1.5) circle [radius=0.05] (6,-1.5) circle [radius=0.05] (8,-1.5) circle [radius=0.05] (10,-1.5) circle [radius=0.05] (12,-1.5) circle [radius=0.05] (14,-1.5) circle [radius=0.05];
  \draw [fill=red] (0.5,-1.5) circle [radius=0.05] (2.5,-1.5) circle [radius=0.05]  (4.5,-1.5) circle [radius=0.05] (6.5,-1.5) circle [radius=0.05] (8.5,-1.5) circle [radius=0.05] (10.5,-1.5) circle [radius=0.05] (12.5,-1.5) circle [radius=0.05] (14.5,-1.5) circle [radius=0.05];
  \draw [fill=yellow] (1,-1.5) circle [radius=0.05] (3,-1.5) circle [radius=0.05]  (5,-1.5) circle [radius=0.05] (7,-1.5) circle [radius=0.05] (9,-1.5) circle [radius=0.05] (11,-1.5) circle [radius=0.05] (13,-1.5) circle [radius=0.05] (15,-1.5) circle [radius=0.05];
  \draw [fill=blue] (0,-3) circle [radius=0.05] (2,-3) circle [radius=0.05] (4,-3) circle [radius=0.05] (6,-3) circle [radius=0.05];
  \draw [fill=red] (0.5,-3) circle [radius=0.05] (2.5,-3) circle [radius=0.05] (4.5,-3) circle [radius=0.05] (6.5,-3) circle [radius=0.05];
  \draw [fill=yellow] (1,-3) circle [radius=0.05] (3,-3) circle [radius=0.05] (5,-3) circle [radius=0.05] (7,-3) circle [radius=0.05];
  \draw [fill=blue] (0,-4.5) circle [radius=0.05] (2,-4.5) circle [radius=0.05] (4,-4.5) circle [radius=0.05] (6,-4.5) circle [radius=0.05] (8,-4.5) circle [radius=0.05] (10,-4.5) circle [radius=0.05] (12,-4.5) circle [radius=0.05] (14,-4.5) circle [radius=0.05];
  \draw [fill=red] (0.5,-4.5) circle [radius=0.05] (2.5,-4.5) circle [radius=0.05] (4.5,-4.5) circle [radius=0.05] (6.5,-4.5) circle [radius=0.05] (8.5,-4.5) circle [radius=0.05] (10.5,-4.5) circle [radius=0.05] (12.5,-4.5) circle [radius=0.05] (14.5,-4.5) circle [radius=0.05];
  \draw [fill=yellow] (1,-4.5) circle [radius=0.05] (3,-4.5) circle [radius=0.05] (5,-4.5) circle [radius=0.05] (7,-4.5) circle [radius=0.05] (9,-4.5) circle [radius=0.05] (11,-4.5) circle [radius=0.05] (13,-4.5) circle [radius=0.05] (15,-4.5) circle [radius=0.05];
  \draw [fill=blue] (0,-6.5) circle [radius=0.05] (2,-6.5) circle [radius=0.05] (4,-6.5) circle [radius=0.05];
  \draw [fill=red] (0.5,-6.5) circle [radius=0.05] (2.5,-6.5) circle [radius=0.05] (4.5,-6.5) circle [radius=0.05];
  \draw [fill=yellow] (1,-6.5) circle [radius=0.05] (3,-6.5) circle [radius=0.05] (5,-6.5) circle [radius=0.05];
  \draw [fill=blue] (0,-8) circle [radius=0.05] (2,-8) circle [radius=0.05]  (4,-8) circle [radius=0.05] (6,-8) circle [radius=0.05] (8,-8) circle [radius=0.05] (10,-8) circle [radius=0.05];
  \draw [fill=red] (0.5,-8) circle [radius=0.05] (2.5,-8) circle [radius=0.05]  (4.5,-8) circle [radius=0.05] (6.5,-8) circle [radius=0.05] (8.5,-8) circle [radius=0.05] (10.5,-8) circle [radius=0.05];
  \draw [fill=yellow] (1,-8) circle [radius=0.05] (3,-8) circle [radius=0.05]  (5,-8) circle [radius=0.05] (7,-8) circle [radius=0.05] (9,-8) circle [radius=0.05] (11,-8) circle [radius=0.05];
  \node [] at (7.5, -5.5) {{\small (~basis with no pseudo-loop~)}};
  \node [] at (7.5, -9) {{\small (~basis with one pseudo-loop~)}};
  \node [below] at (6.5,-1.6) {{\footnotesize excluded}};
  \node [below] at (14.5,-1.6) {{\footnotesize excluded}};
  \node [below] at (6.5,-3.1) {{\footnotesize excluded}};
  \node [below] at (0.5,-4.6) {{\footnotesize excluded}};
  \node [below] at (2.5,-4.6) {{\footnotesize excluded}};
  \node [below] at (4.5,-4.6) {{\footnotesize excluded}};
  \node [below] at (6.5,-4.6) {{\footnotesize excluded}};
  \node [below] at (8.5,-4.6) {{\footnotesize excluded}};
  \node [below] at (10.5,-4.6) {{\footnotesize excluded}};
  \node [below] at (12.5,-4.6) {{\footnotesize excluded}};
  \node [below] at (14.5,-4.6) {{\footnotesize excluded}};
\end{tikzpicture}
  \caption{Quiver representation of gauge invariant basis for $A^{\EYM}_{n,3}$. For simplicity, $h_1,h_2$ and $h_3$ are denoted as blue, red and yellow dots respectively. Arrows always flow from starting points of solid line toward pseudo-loops or the ending points of dashed line, so they are omitted unless causing confusion. The ending point of dashed line is $K_{a_i}$ depending on the $h_i$ it connects, and $2\leq a_1,a_2,a_3 \leq n-1$. Quivers with real loops are excluded.  }\label{Fig-m3-quiver}
\end{figure}
The EYM amplitude $A^{\EYM}_{n,3}(1,\ldots,n-1;h_1,h_2,h_3)$ lives in the gauge invariant vector space $\mathcal{W}_{n+3,3}$. Since $\dim\mathcal{W}_{n+3,3}=n^3+3n$, it is supposed to be expanded into $(n^3+3n)$ terms. Among these gauge invariant vectors, there are $3(n-2)+8$ terms containing real loops and should be excluded. So we need to compute $(n^3-2)$ expansion coefficients. The expansion basis and their quiver representations are shown in Fig.\ref{Fig-m3-quiver}. Following the algorithm, in Step-0 we consider the gauge invariant vectors with no pseudo-loops by formula (\ref{sol-coef-0}). Applying differential operators $\mathcal{T}_{a_1h_1(a_1+1)}\mathcal{T}_{a_2h_2(a_2+1)}\mathcal{T}_{a_3h_3(a_3+1)}$ on the expansion formula of $A^{\EYM}_{n,3}$, we immediately get
\begin{equation}
\mathcal{C}[\mathsf{F}_{h_1}^{a_1}\mathsf{F}_{h_2}^{a_2}\mathsf{F}_{h_3}^{a_3}]=
\left\{\begin{array}{l}
A^{\YM}_{n+3}(1,2,\ldots,a_{\sigma_1},h_{\sigma_1},\ldots,a_{\sigma_2},h_{\sigma_2},\ldots,a_{\sigma_3},h_{\sigma_3},\ldots,n)~~,~~~~~~~~~~~~~~~~~~~a_{\sigma_1}<a_{\sigma_2}<a_{\sigma_3} \\
\sum_{\rho\{\sigma_2,\sigma_3\}\in S_2}A^{\YM}_{n+3}(1,2,\ldots,a_{\sigma_1},h_{\sigma_1},\ldots,a_{\sigma_2},h_{\rho_2},h_{\rho_3},a_{\sigma_2}+1,\ldots,n)~~,~~a_{\sigma_1}<a_{\sigma_2}=a_{\sigma_3} \\
\sum_{\rho\{\sigma_1,\sigma_2\}\in S_2}A^{\YM}_{n+3}(1,2,\ldots,a_{\sigma_1},h_{\rho_1},h_{\rho_2},a_{\sigma_1}+1,\ldots,a_{\sigma_3},h_{\sigma_3},\ldots,n)~~,~~a_{\sigma_1}=a_{\sigma_2}<a_{\sigma_3} \\
\sum_{\rho\{\sigma_1,\sigma_2,\sigma_3\}\in S_3}A^{\YM}_{n+3}(1,2,\ldots,a_{\sigma_1},h_{\rho_1},h_{\rho_2},h_{\rho_3},a_{\sigma_1}+1,\ldots,n)~~,~~~~~~~~~~a_{\sigma_1}=a_{\sigma_2}=a_{\sigma_3}
\end{array}
 \right.~,~~~\nonumber
\end{equation}
where $\rho\{\sigma_1,\cdots,\sigma_m\}$ is a permutation of $\{\sigma_1,\cdots,\sigma_m\}$, and the summation is over all elements of $S_m$.
Applying $\mathcal{T}_{h_{\beta'_1}h_{\beta_1}n}\mathcal{T}_{a_{\gamma_1}h_{\gamma_1}(a_{\gamma_1}+1)}\mathcal{T}_{a_{\gamma_2}h_{\gamma_2}(a_{\gamma_2}+1)}$ on $A^{\EYM}_{n,3}$, we get
\begin{eqnarray}
&& \mathcal{C}[\mathsf{F}_{h_{\beta_1}}^{h_{\beta'_1}}\mathsf{F}_{h_{\gamma_1}}^{a_{\gamma_1}}\mathsf{F}_{h_{\gamma_2}}^{a_{\gamma_2}}]\nonumber\\
&&
=\left\{\begin{array}{l}
\sum_{\shuffle}A^{\YM}_{n+3}(1,2,\ldots,a_{\gamma_1},h_{\gamma_1},\{ a_{\gamma_1}+1,\ldots,a_{\gamma_2},h_{\gamma_2},\ldots,n-1\}\shuffle \{h_{\beta_1}\},n)~~,~~a_{\gamma_1}<a_{\gamma_2}~,~\beta'_1=\gamma_1 \\
\sum_{\shuffle}A^{\YM}_{n+3}(1,2,\ldots,a_{\gamma_1},h_{\gamma_1},\ldots,a_{\gamma_2},h_{\gamma_2},\{a_{\gamma_2}+1,\ldots,n-1\}\shuffle \{h_{\beta_1}\},n)~~,~~a_{\gamma_1}<a_{\gamma_2}~,~\beta'_1=\gamma_2 \\
\begin{array}{l}
\sum_{\shuffle}A^{\YM}_{n+3}(1,2,\ldots,a_{\gamma_1},h_{\gamma_1},\{h_{\gamma_2},a_{\gamma_1}+1,\ldots,n-1\}\shuffle \{h_{\beta_1}\},n)\\
~~~~~~~+\sum_{\shuffle}A^{\YM}_{n+3}(1,2,\ldots,a_{\gamma_1},h_{\gamma_2},h_{\gamma_1},\{a_{\gamma_1}+1,\ldots,n-1\}\shuffle \{h_{\beta_1}\},n)\end{array}~~,~~a_{\gamma_1}=a_{\gamma_2}~,~\beta'_1=\gamma_1
\end{array}
 \right.~,~~~\nonumber
\end{eqnarray}
where $\{\beta_1\}\cup \{\gamma_1,\gamma_2\}$ is a splitting of $\{1,2,3\}$.
Applying $\mathcal{T}_{h_{\beta'_1}h_{\beta_1}n}\mathcal{T}_{h_{\beta'_2}h_{\beta_2}n}\mathcal{T}_{a_{\gamma_1}h_{\gamma_1}(a_{\gamma_1}+1)}$ on $A^{\EYM}_{n,3}$, we get
\begin{eqnarray}
&&\mathcal{C}[\mathsf{F}_{h_{\beta_1}}^{h_{\beta'_1}}\mathsf{F}_{h_{\beta_2}}^{h_{\beta'_2}}\mathsf{F}_{h_{\gamma_1}}^{a_{\gamma_1}}]\nonumber\\
&&~~~~=
\left\{\begin{array}{l}
\sum_{\shuffle} A^{\YM}_{n+3}(1,2\ldots,a_{\gamma_1},h_{\gamma_1},\{a_{\gamma_1}+1,\ldots,n-1\}\shuffle\{h_{\beta_2},h_{\beta_1}\},n)~~,~~~~~~~\beta'_1=\beta_2~,~\beta'_2=\gamma_1\\
\sum_{\shuffle}A^{\YM}_{n+3}(1,2\ldots,a_{\gamma_1},h_{\gamma_1},\{a_{\gamma_1}+1,\ldots,n-1\}\shuffle\{h_{\beta_2}\}\shuffle\{h_{\beta_1}\},n)~~,~~\beta'_1=\gamma_1~,~\beta'_2=\gamma_1
\end{array}\right.~,~~~\nonumber
\end{eqnarray}
with $\{\beta_1,\beta_2\}\cup \{\gamma_1\}=\{1,2,3\}$. As mentioned, after summing over all above results produced in Step-0, we get a compact expression,
\begin{eqnarray}
&& \mbox{[Step~0]}~=\label{step-0-3h}\\
&&\sum_{\shuffle} \frac{k_1\cdot f_{h_1}\cdot X_{h_1}}{k_1\cdot k_{h_1}}\frac{k_1\cdot f_{h_2}\cdot X_{h_2}}{k_1\cdot k_{h_2}} \frac{k_1\cdot f_{h_3}\cdot X_{h_3}}{k_1\cdot k_{h_3}}A^{\YM}_{n+3}(1,2,\{3,\ldots,n-1\}\shuffle \{h_1\}\shuffle \{h_2\}\shuffle\{h_3\},n)~.~~~\nonumber
\end{eqnarray}
Recalling the compact expression \eref{step-0-2h} for EYM amplitude with two gravitons, we confirm that the total contribution of Step-0 is always possible to be written in a compact form.

Then we proceed to Step-1, and compute the expansion coefficients for vectors with one pseudo-loop. After substituting solutions in Step-0 back to the expansion formula, we get
\begin{equation}
A^{\EYM}_{n,3}-[\mbox{Step~0}]
=\sum_{a_{\gamma_1}=2\atop \{\alpha_1,\alpha_2,\gamma_1\}}^{n-1}   \mathcal{C}[\mathsf{F}_{h_{\alpha_1}h_{\alpha_2}}\mathsf{F}_{h_{\gamma_1}}^{a_{\gamma_1}}]    \mathsf{F}_{h_{\alpha_1}h_{\alpha_2}}\mathsf{F}_{h_{\gamma_1}}^{a_{\gamma_1}}
+\sum_{ \{\alpha_1,\alpha_2,\beta_1\}\atop \beta'_1\neq \beta_1}    \mathcal{C}[\mathsf{F}_{h_{\alpha_1}h_{\alpha_2}}\mathsf{F}_{h_{\beta_1}}^{h_{\beta'_1}}]   \mathsf{F}_{h_{\alpha_1}h_{\alpha_2}}\mathsf{F}_{h_{\beta_1}}^{h_{\beta'_1}}~,~~~\label{sol-m3-coef-FhhFk-pre}
\end{equation}
where the first summation runs over all possible splitting $\{\alpha_1,\alpha_2,\gamma_1\}$ of $\{1,2,3\}$, while the second summation not only runs over all splitting $\{\alpha_1,\alpha_2,\beta_1\}$ but also all possible values of $\beta'_1$. Terms in the first summation correspond to the first three quivers with one pseudo-loop in Fig.\ref{Fig-m3-quiver}, while terms in the second summation correspond to the remaining six quivers. As mentioned, for a fixed value of $p$, we should start from terms with larger $r$, {\sl i.e.}, terms in the first summation. As argued in the previous section, when applying a defined differential operator, only the corresponding vector survives and all others vanish. This means there is no mixing contributions between different pseudo-loop of the first type. For example, applying  differential operator $\mathcal{T}_{h_3h_2n}\mathcal{T}_{1h_32}\mathcal{T}_{a_1h_1(a_1+1)}$ on formula (\ref{sol-m3-coef-FhhFk-pre}), the only surviving vector with one pseudo-loop is $\mathsf{F}_{h_2h_3}\mathsf{F}_{h_1}^{a_1}$. However vectors with no pseudo-loops would contribute, and from our previous general argument we can determine the non-vanishing vectors to be $\mathsf{F}_{h_2}^{h_3}\mathsf{F}^{K_a}_{h_3}\mathsf{F}_{h_1}^{a_1}$ and $\mathsf{F}_{h_2}^{h_3}\mathsf{F}^{h_1}_{h_3}\mathsf{F}_{h_1}^{a_1}$. Hence we get
\begin{eqnarray}
\mathcal{C}[\mathsf{F}_{h_2h_3}\mathsf{F}_{h_1}^{a_1}]&=&\Big(~(k_1\cdot k_{h_3})(\mathcal{T}_{h_3h_2n}\mathcal{T}_{1h_32}\mathcal{T}_{a_1h_1(a_1+1)}~A^{\EYM}_{n,3})~\Big)\nonumber\\
&&~~~~~~+\Big(~\sum_{a_3=2}^{n-1}(k_{h_3}\cdot X_{h_3})\mathcal{C}[\mathsf{F}_{h_2}^{h_3}\mathsf{F}_{h_1}^{a_1}\mathsf{F}_{h_3}^{a_3}]~\Big)
+\Big(~(k_{h_3}\cdot k_{h_1})\mathcal{C}[\mathsf{F}_{h_2}^{h_3}\mathsf{F}_{h_3}^{h_1}\mathsf{F}_{h_1}^{a_1}]~\Big)~,~~~
\end{eqnarray}
where the relation \eref{FF-relation} has been used. Working it out explicitly, we get
\bea  \mathcal{C}[\mathsf{F}_{h_2h_3}\mathsf{F}_{h_1}^{a_1}]&= &(k_1\cdot k_{h_3})
A(1, h_3, \{h_2\} \shuffle\{2,...,a_1, h_1, a_1+1,...,n-1\}, n) \nn
& &~~~~~~ +(k_{h_3}\cdot X_{h_3}) A(1, 2, \{3,...,a_1, h_1, a_1+1,...,n-1\}\shuffle\{h_3, h_2\},n)
\nn & & ~~~~~~~~~~~~~~~+ (k_{h_3}\cdot k_{h_1}) A(1, 2,...,a_1,h_1, \{h_3, h_2\}\shuffle\{a_1+1,...,n-1\}, n)~.~~~\eea
Terms in the first and second lines are similar to the one given in \eref{2p-C-1}, hence we can borrow the result \eref{sol-m2-coef-Fhh} to here and immediately work out the summation as,
\bea & & \sum_{a_1=2}^{n-1} \mathcal{C}[\mathsf{F}_{h_2h_3}\mathsf{F}_{h_1}^{a_1}]\mathsf{F}_{h_2h_3}\mathsf{F}_{h_1}^{a_1}
= \mathsf{F}_{h_2h_3}{ (k_1\cdot f_{h_1}\cdot Y_{h_1})(k_{h_3}\cdot k_{h_1})\over(k_1\cdot k_{h_1})} A^{\YM}_{n+2}(1, 2,\{h_1, h_3, h_2\}\shuffle\{3,...,n-1\}, n)\nn
& & ~~~~~~+ \sum_{\{\sigma_1,\sigma_2\}\in S_2(h_2, h_3)} \mathsf{F}_{h_2h_3}{ (k_1\cdot f_{h_1}\cdot Y_{h_1})\over(k_1\cdot k_{h_1})}\frac{(k_{h_{\sigma_1}}\cdot X_{h_{\sigma_1}}-k_1\cdot k_{h_{\sigma_1}})(k_{h_{\sigma_2}}\cdot X_{h_{\sigma_2}})}{{\cal K}_{1h_3h_2}}\nn
& &~~~~~~~~~~~~~~~~~~~~~~~~~~~~~~~~~~~~~~~~~~~~~~~~~A^{\YM}_{n+2}(1,2,\{3,\ldots,n-1\}\shuffle\{h_1\}\shuffle \{h_{\sigma_1},h_{\sigma_2}\},n)~.~~~\label{3p-temp-1}
 \eea
The other two terms with $r=1$ can be obtained by permutation of above result.

Now we move to the vectors with $p=1,r=0$. As discussed, a defined differential operator (\ref{differential-operator-p1}) would possibly mix contributions of many vectors with one pseudo-loop, and in general we should solve an algebraic system of linear equations to compute all of them. However, in the current simple example we can intentionally choose a differential operator to avoid the mixing of vectors.  For instance, in order to compute the coefficient of vector $\mathsf{F}_{h_2h_3}\mathsf{F}_{h_1}^{h_2}$ we should choose the differential operator $\mathcal{T}_{h_3h_2n}\mathcal{T}_{1h_32}\mathcal{T}_{h_2h_1n}$. If instead we choose the other differential operator $\mathcal{T}_{h_2h_3n}\mathcal{T}_{1h_2 2}\mathcal{T}_{h_2h_1n}$, both vectors $\mathsf{F}_{h_1h_2}\mathsf{F}_{h_3}^{h_2}$ and $\mathsf{F}_{h_2h_3}\mathsf{F}_{h_1}^{h_2}$ would be non-vanishing and their contributions will mix together. Hence we apply $\mathcal{T}_{h_3h_2n}\mathcal{T}_{1h_32}\mathcal{T}_{h_2h_1n}$ on formula (\ref{sol-m3-coef-FhhFk-pre}), and compute the coefficient as,
\begin{eqnarray}
&&\mathcal{C}[\mathsf{F}_{h_2h_3}\mathsf{F}_{h_1}^{h_2}]=\Big(~(k_1\cdot k_{h_3})(\mathcal{T}_{h_2h_1n}\mathcal{T}_{h_3h_2n}\mathcal{T}_{1h_32}~A^{\EYM}_{n,3})~\Big)
+\Big(~\sum_{a_3=2}^{n-1}(k_{h_3}\cdot X_{h_3})\mathcal{C}[\mathsf{F}_{h_1}^{h_2}\mathsf{F}_{h_2}^{h_3}\mathsf{F}_{h_3}^{a_3}]~\Big)\label{3p-temp-2}\\
& &=(k_1\cdot k_{h_3}) A^{\YM}_{n+3}(1, \{h_3, h_2, h_1\}\shuffle\{2,...,n-1\}_R, n)+ (k_{h_3}\cdot X_{h_3}) A^{\YM}_{n+3}(1, 2,\{h_3, h_2, h_1\}\shuffle\{3,...,n-1\}, n)~.~~~\nonumber
\end{eqnarray}
Yang-Mills amplitudes in the second term are already in BCJ basis with legs $1,2,n$ fixed, while those in the first term should be rewritten to BCJ basis by applying BCJ relations. Similar computations can be inferred from \eref{2p-C-1} and
\eref{3p-temp-2}, and consequently all coefficients of vectors with one pseudo-loop can be computed. Summing up all above results we get the complete expansion of $A^{\EYM}_{n,3}$, which is consistent with results given in \cite{Feng:2019tvb}.

%%%%%%%%%%%%%%%%%%
\subsection{The expansion of EYM amplitude with four gravitons}
%%%%%%%%%%%%%%%%%
%
\begin{figure}
\centering
\begin{tikzpicture}
  \draw [thick, dashed, ->] (0,0)--(0,0.5);
  \draw [thick, dashed, ->] (0.5,0)--(0.5,0.5);
  \draw [thick, dashed, ->] (1,0)--(1,0.5);
  \draw [thick, dashed, ->] (1.5,0)--(1.5,0.5);
  \draw [thick, dashed, ->] (2.5,0)--(2.5,0.5);
  \draw [thick, dashed, ->] (3,0)--(3,0.5);
  \draw [thick, dashed, ->] (3.5,0)--(3.5,0.5);
  \draw [thick, dashed, ->] (5,0)--(5,0.5);
  \draw [thick, dashed, ->] (6,0)--(6,0.5);
  \draw [thick, dashed, ->] (7.5,0)--(7.5,0.5);
  \draw [thick, dashed, ->] (8.5,0)--(8.5,0.5);
  \draw [thick, dashed, ->] (10,0)--(10,0.5);
  \draw [thick, dashed, ->] (10.5,0)--(10.5,0.5);
  \draw [thick, dashed, ->] (12.5,0)--(12.5,0.5);
  \draw [thick, dashed, ->] (13,0)--(13,0.5);
  \draw [thick] (3.5,0)--(4,0) (5.5,0)--(6.5,0) (7.5,0)--(8,0) (8.5,0)--(9,0) (10.5,0)--(11.5,0);
  \draw [thick] (13.5,0) to [out=45, in=180] (13.75,0.15) to [out=0, in=135] (14,0);
  \draw [thick] (13.5,0) to [out=315, in=180] (13.75,-0.15) to [out=0, in=225] (14,0);
  \draw [thick, dashed, ->] (0.5,-1.5)--(0.5,-1);
  \draw [thick, dashed, ->] (3,-1.5)--(3,-1);
  \draw [thick, dashed, ->] (5,-1.5)--(5,-1);
  \draw [thick, dashed, ->] (7.5,-1.5)--(7.5,-1);
  \draw [thick, dashed, ->] (10,-1.5)--(10,-1);
  \draw [thick, dashed, ->] (12.5,-1.5)--(12.5,-1);
  \draw [thick, dashed, ->] (15,-1.5)--(15,-1);
  \draw [thick] (0,-1.5)--(1,-1.5) (2.5,-1.5)--(4,-1.5) (5,-1.5)--(6,-1.5) (7.5,-1.5)--(9,-1.5) (10,-1.5)--(10.5,-1.5) (13.5,-1.5)--(14,-1.5) (15.5,-1.5)--(16.5,-1.5);
  \draw [thick] (11,-1.5) to [out=45, in=180] (11.25,-1.35) to [out=0, in=135] (11.5,-1.5);
  \draw [thick] (11,-1.5) to [out=315, in=180] (11.25,-1.65) to [out=0, in=225] (11.5,-1.5);
  \draw [thick] (13,-1.5) to [out=45, in=180] (13.25,-1.35) to [out=0, in=135] (13.5,-1.5);
  \draw [thick] (13,-1.5) to [out=315, in=180] (13.25,-1.65) to [out=0, in=225] (13.5,-1.5);
  \draw [thick] (0.5,-1.5) to [out=45, in=180] (1,-1.35) to [out=0, in=135] (1.5,-1.5);
  \draw [thick] (5.5,-1.5) to [out=45, in=180] (6,-1.35) to [out=0, in=135] (6.5,-1.5);
  \draw [thick] (15.5,-1.5) to [out=45, in=180] (16,-1.35) to [out=0, in=135] (16.5,-1.5);
  \draw [thick] (0.5,-3)--(1,-3) (2.5,-3)--(3,-3) (3.5,-3)--(4,-3) (5.5,-3)--(6.5,-3) (10,-3)--(11.5,-3) (12.5,-3)--(14,-3);
  \draw [thick] (0,-3) to [out=45, in=180] (0.25,-2.85) to [out=0, in=135] (0.5,-3);
  \draw [thick] (0,-3) to [out=315, in=180] (0.25,-3.15) to [out=0, in=225] (0.5,-3);
  \draw [thick] (3,-3) to [out=45, in=180] (3.25,-2.85) to [out=0, in=135] (3.5,-3);
  \draw [thick] (3,-3) to [out=315, in=180] (3.25,-3.15) to [out=0, in=225] (3.5,-3);
  \draw [thick] (5,-3) to [out=45, in=180] (5.25,-2.85) to [out=0, in=135] (5.5,-3);
  \draw [thick] (5,-3) to [out=315, in=180] (5.25,-3.15) to [out=0, in=225] (5.5,-3);
  \draw [thick] (7.5,-3) to [out=45, in=180] (7.75,-2.85) to [out=0, in=135] (8,-3);
  \draw [thick] (7.5,-3) to [out=315, in=180] (7.75,-3.15) to [out=0, in=225] (8,-3);
  \draw [thick] (8.5,-3) to [out=45, in=180] (8.75,-2.85) to [out=0, in=135] (9,-3);
  \draw [thick] (8.5,-3) to [out=315, in=180] (8.75,-3.15) to [out=0, in=225] (9,-3);
  \draw [thick] (0.5,-3) to [out=45, in=180] (1,-2.85) to [out=0, in=135] (1.5,-3);
  \draw [thick] (10,-3) to [out=45, in=180] (10.5,-2.85) to [out=0, in=135] (11,-3);
  \draw [thick] (12.5,-3) to [out=45, in=180] (13.25,-2.8) to [out=0, in=135] (14,-3);
  \draw [thick, dashed, ->] (0,-5)--(0,-4.5);
  \draw [thick, dashed, ->] (0.5,-5)--(0.5,-4.5);
  \draw [thick, dashed, ->] (2.5,-5)--(2.5,-4.5);
  \draw [thick, dashed, ->] (5,-5)--(5,-4.5);
  \draw [thick] (2.5,-5)--(3,-5) (6,-5)--(6.5,-5) (8,-5)--(8.5,-5) (10,-5)--(10.5,-5) (11,-5)--(11.5,-5) (13,-5)--(14,-5);
  \draw [thick, cyan] (1,-5) to [out=45, in=180] (1.25,-4.85) to [out=0, in=135] (1.5,-5);
  \draw [thick, brown] (1,-5) to [out=315, in=180] (1.25,-5.15) to [out=0, in=225] (1.5,-5);
  \draw [thick, cyan] (3.5,-5) to [out=45, in=180] (3.75,-4.85) to [out=0, in=135] (4,-5);
  \draw [thick, brown] (3.5,-5) to [out=315, in=180] (3.75,-5.15) to [out=0, in=225] (4,-5);
  \draw [thick, cyan] (5.5,-5) to [out=45, in=180] (5.75,-4.85) to [out=0, in=135] (6,-5);
  \draw [thick, brown] (5.5,-5) to [out=315, in=180] (5.75,-5.15) to [out=0, in=225] (6,-5);
  \draw [thick, cyan] (7.5,-5) to [out=45, in=180] (7.75,-4.85) to [out=0, in=135] (8,-5);
  \draw [thick, brown] (7.5,-5) to [out=315, in=180] (7.75,-5.15) to [out=0, in=225] (8,-5);
  \draw [thick, cyan] (10.5,-5) to [out=45, in=180] (10.75,-4.85) to [out=0, in=135] (11,-5);
  \draw [thick, brown] (10.5,-5) to [out=315, in=180] (10.75,-5.15) to [out=0, in=225] (11,-5);
  \draw [thick, cyan] (12.5,-5) to [out=45, in=180] (12.75,-4.85) to [out=0, in=135] (13,-5);
  \draw [thick, brown] (12.5,-5) to [out=315, in=180] (12.75,-5.15) to [out=0, in=225] (13,-5);
  \draw [thick, cyan] (15,-5) to [out=45, in=180] (15.25,-4.85) to [out=0, in=135] (15.5,-5);
  \draw [thick, brown] (15,-5) to [out=315, in=180] (15.25,-5.15) to [out=0, in=225] (15.5,-5);
  \draw [thick] (16,-5) to [out=45, in=180] (16.25,-4.85) to [out=0, in=135] (16.5,-5);
  \draw [thick] (16,-5) to [out=315, in=180] (16.25,-5.15) to [out=0, in=225] (16.5,-5);
  \draw [thick] (8,-5) to [out=45, in=180] (8.5,-4.85) to [out=0, in=135] (9,-5);
  \draw [thick, cyan] (0,-7) to [out=45, in=180] (0.25,-6.85) to [out=0, in=135] (0.5,-7);
  \draw [thick, brown] (0,-7) to [out=315, in=180] (0.25,-7.15) to [out=0, in=225] (0.5,-7);
  \draw [thick, cyan] (1,-7) to [out=45, in=180] (1.25,-6.85) to [out=0, in=135] (1.5,-7);
  \draw [thick, brown] (1,-7) to [out=315, in=180] (1.25,-7.15) to [out=0, in=225] (1.5,-7);
  \draw [fill] (0,0) circle [radius=0.05]  (0.5,0) circle [radius=0.05]  (1,0) circle [radius=0.05] (1.5,0) circle [radius=0.05] (2.5,0) circle [radius=0.05] (3,0) circle [radius=0.05] (3.5,0) circle [radius=0.05] (4,0) circle [radius=0.05] (5,0) circle [radius=0.05] (5.5,0) circle [radius=0.05] (6,0) circle [radius=0.05] (6.5,0) circle [radius=0.05] (7.5,0) circle [radius=0.05] (8,0) circle [radius=0.05] (8.5,0) circle [radius=0.05] (9,0) circle [radius=0.05] (10,0) circle [radius=0.05] (10.5,0) circle [radius=0.05] (11,0) circle [radius=0.05] (11.5,0) circle [radius=0.05] (12.5,0) circle [radius=0.05] (13,0) circle [radius=0.05] (13.5,0) circle [radius=0.05] (14,0) circle [radius=0.05];
  \draw [fill] (0,-1.5) circle [radius=0.05]  (0.5,-1.5) circle [radius=0.05]  (1,-1.5) circle [radius=0.05] (1.5,-1.5) circle [radius=0.05] (2.5,-1.5) circle [radius=0.05] (3,-1.5) circle [radius=0.05] (3.5,-1.5) circle [radius=0.05] (4,-1.5) circle [radius=0.05] (5,-1.5) circle [radius=0.05] (5.5,-1.5) circle [radius=0.05] (6,-1.5) circle [radius=0.05] (6.5,-1.5) circle [radius=0.05] (7.5,-1.5) circle [radius=0.05] (8,-1.5) circle [radius=0.05] (8.5,-1.5) circle [radius=0.05] (9,-1.5) circle [radius=0.05] (10,-1.5) circle [radius=0.05] (10.5,-1.5) circle [radius=0.05] (11,-1.5) circle [radius=0.05] (11.5,-1.5) circle [radius=0.05] (12.5,-1.5) circle [radius=0.05] (13,-1.5) circle [radius=0.05] (13.5,-1.5) circle [radius=0.05] (14,-1.5) circle [radius=0.05] (15,-1.5) circle [radius=0.05] (15.5,-1.5) circle [radius=0.05] (16,-1.5) circle [radius=0.05] (16.5,-1.5) circle [radius=0.05];
  \draw [fill] (0,-3) circle [radius=0.05]  (0.5,-3) circle [radius=0.05]  (1,-3) circle [radius=0.05] (1.5,-3) circle [radius=0.05] (2.5,-3) circle [radius=0.05] (3,-3) circle [radius=0.05] (3.5,-3) circle [radius=0.05] (4,-3) circle [radius=0.05] (5,-3) circle [radius=0.05] (5.5,-3) circle [radius=0.05] (6,-3) circle [radius=0.05] (6.5,-3) circle [radius=0.05] (7.5,-3) circle [radius=0.05] (8,-3) circle [radius=0.05] (8.5,-3) circle [radius=0.05] (9,-3) circle [radius=0.05] (10,-3) circle [radius=0.05] (10.5,-3) circle [radius=0.05] (11,-3) circle [radius=0.05] (11.5,-3) circle [radius=0.05] (12.5,-3) circle [radius=0.05] (13,-3) circle [radius=0.05] (13.5,-3) circle [radius=0.05] (14,-3) circle [radius=0.05];
  \draw [fill] (0,-5) circle [radius=0.05]  (0.5,-5) circle [radius=0.05]  (1,-5) circle [radius=0.05] (1.5,-5) circle [radius=0.05] (2.5,-5) circle [radius=0.05] (3,-5) circle [radius=0.05] (3.5,-5) circle [radius=0.05] (4,-5) circle [radius=0.05] (5,-5) circle [radius=0.05] (5.5,-5) circle [radius=0.05] (6,-5) circle [radius=0.05] (6.5,-5) circle [radius=0.05] (7.5,-5) circle [radius=0.05] (8,-5) circle [radius=0.05] (8.5,-5) circle [radius=0.05] (9,-5) circle [radius=0.05] (10,-5) circle [radius=0.05] (10.5,-5) circle [radius=0.05] (11,-5) circle [radius=0.05] (11.5,-5) circle [radius=0.05] (12.5,-5) circle [radius=0.05] (13,-5) circle [radius=0.05] (13.5,-5) circle [radius=0.05] (14,-5) circle [radius=0.05] (15,-5) circle [radius=0.05] (15.5,-5) circle [radius=0.05] (16,-5) circle [radius=0.05] (16.5,-5) circle [radius=0.05];
  \draw [fill] (0,-7) circle [radius=0.05]  (0.5,-7) circle [radius=0.05]  (1,-7) circle [radius=0.05] (1.5,-7) circle [radius=0.05];
  \node [] at (8, -4) {{\small (~basis with no pseudo-loop~)}};
  \node [] at (8, -6) {{\small (~basis with one pseudo-loop~)}};
  \node [] at (8, -8) {{\small (~basis with two pseudo-loops~)}};
  \node [below] at (13.25,-0.1) {{\footnotesize excluded}};
  \node [below] at (10.75,-1.6) {{\footnotesize excluded}};
  \node [below] at (13.25,-1.6) {{\footnotesize excluded}};
  \node [below] at (15.75,-1.6) {{\footnotesize excluded}};
  \node [below] at (0.75,-3.1) {{\footnotesize excluded}};
  \node [below] at (3.25,-3.1) {{\footnotesize excluded}};
  \node [below] at (5.75,-3.1) {{\footnotesize excluded}};
  \node [below] at (8.25,-3.1) {{\footnotesize excluded}};
  \node [below] at (10.75,-3.1) {{\footnotesize excluded}};
  \node [below] at (13.25,-3.1) {{\footnotesize excluded}};
  \node [below] at (15.75,-5.1) {{\footnotesize excluded}};
\end{tikzpicture}
  \caption{Quiver representation of gauge invariant basis for $A^{\EYM}_{n,4}$. For presentation purpose, we only show quivers of distinct topologies, and a graph here denotes several graphs with black dots specifying any possible independent labels $(h_1,h_2,h_3,h_4)$. Arrows always flow from starting point of solid line towards pseudo-loops or the ending point of dashed line, and they are omitted unless causing confusion. The ending point of dashed line is $K_{a_i}$ depending on the $h_i$ it connects, and $2\leq a_1,a_2,a_3, a_4 \leq n-1$. Quiver graphs with real solid loops are excluded.  }\label{Fig-m4-quiver}
\end{figure}
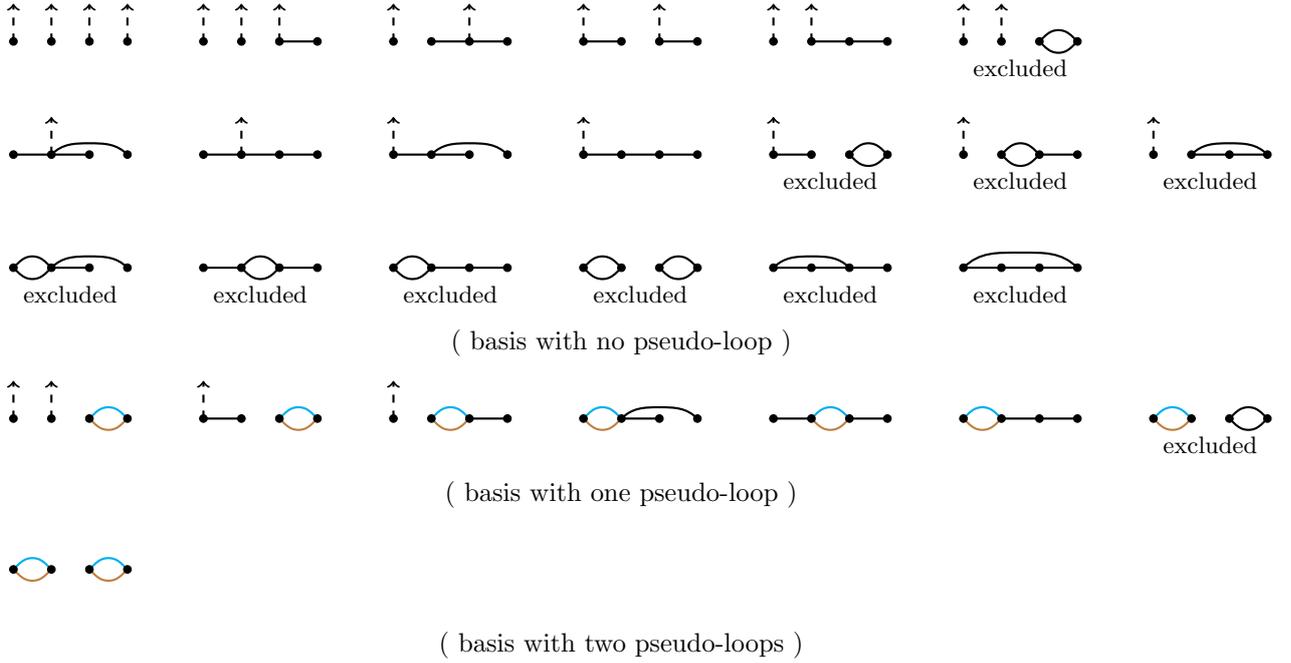
EYM amplitude $A^{\EYM}_{n,4}(1,\ldots,n;h_1,h_2,h_3,h_4)$ lives in gauge invariant vector space $\mathcal{W}_{n+4,4}$, and it can be expanded as linear combination of $\dim\mathcal{W}_{n+4,4}=(n+1)^4+6(n+1)^2+3$ vectors. All vectors in gauge invariant basis and their quivers are shown in Fig.\ref{Fig-m4-quiver}. Among them, there are in total $6\times (n-2)^2+44(n-2)+87 $ vectors with real loops which should be excluded. For the remaining vectors, we can compute their expansion coefficients following the algorithm. Again in Step-0, we compute the coefficients of vectors with no pseudo-loops by formula
(\ref{sol-coef-0}). We shall not write down the explicit coefficient for each basis but present the summation of them in a compact expression as\footnote{Note that the result of Step-0 can be similarly generalized to arbitrary points.},
\begin{equation}[\mbox{Step~0}]=\sum_{\shuffle} \left(\prod_{i=1}^{4}\frac{k_1\cdot f_{h_i}\cdot X_{h_i}}{k_1\cdot k_{h_i}}\right)A^{\YM}_{n+4}(1,2,\{3,\ldots,n-1\}\shuffle \{h_1\}\shuffle \{h_2\}\shuffle\{h_3\}\shuffle\{h_4\},n)~.~~~\end{equation}

Then let us continue with Step-1, to compute expansion coefficients of vectors with one pseudo-loop. As shown in Fig.\ref{Fig-m4-quiver}, there are in total seven distinct topologies, and the last one should be excluded. For the other six topologies, according to rules (\ref{differential-operator}) we assign each of them with a differential operator respectively, and represent differential operators in quiver representation as
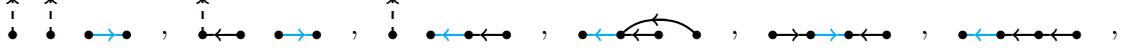
\begin{center}
\begin{tikzpicture}
 \draw [thick, dashed, ->] (0,0)--(0,0.5);
 \draw [thick, dashed, ->] (0.5,0)--(0.5,0.5);
 \draw [thick, dashed, ->] (2.5,0)--(2.5,0.5);
 \draw [thick, dashed, ->] (5,0)--(5,0.5);
 \draw [thick, ->, cyan] (1,0)--(1.35,0);
 \draw [thick, cyan] (1.35,0)--(1.5,0);
 \draw [thick, ->] (3,0)--(2.65,0);
 \draw [thick] (2.65,0)--(2.5,0);
 \draw [thick, ->, cyan] (3.5,0)--(3.85,0);
 \draw [thick, cyan] (3.85,0)--(4,0);
 \draw [thick, ->, cyan] (6,0)--(5.65,0);
 \draw [thick, cyan] (5.65,0)--(5.5,0);
 \draw [thick, ->] (6.5,0)--(6.15,0);
 \draw [thick] (6.15,0)--(6,0);
 \draw [thick, ->, cyan] (8,0)--(7.65,0);
 \draw [thick, cyan] (7.65,0)--(7.5,0);
 \draw [thick, ->] (8.5,0)--(8.15,0);
 \draw [thick] (8.15,0)--(8,0);
 \draw [thick, ->] (10,0)--(10.35,0);
 \draw [thick] (10.35,0)--(10.5,0);
 \draw [thick, ->, cyan] (10.5,0)--(10.85,0);
 \draw [thick, cyan] (10.85,0)--(11,0);
 \draw [thick, ->] (11.5,0)--(11.15,0);
 \draw [thick] (11.15,0)--(11,0);
 \draw [thick, ->, cyan] (13,0)--(12.65,0);
 \draw [thick, cyan] (12.65,0)--(12.5,0);
 \draw [thick, ->] (13.5,0)--(13.15,0);
 \draw [thick] (13.15,0)--(13,0);
 \draw [thick, ->] (14,0)--(13.65,0);
 \draw [thick] (13.65,0)--(13.5,0);
 \draw [thick] (8,0) to [out=45, in=180] (8.5,0.2) to [out=0, in=135] (9,0);
 \draw [thick, ->] (8.6,0.2)--(8.4,0.2);
 \draw [fill] (0,0) circle [radius=0.05]  (0.5,0) circle [radius=0.05]  (1,0) circle [radius=0.05] (1.5,0) circle [radius=0.05] (2.5,0) circle [radius=0.05] (3,0) circle [radius=0.05] (3.5,0) circle [radius=0.05] (4,0) circle [radius=0.05] (5,0) circle [radius=0.05] (5.5,0) circle [radius=0.05] (6,0) circle [radius=0.05] (6.5,0) circle [radius=0.05] (7.5,0) circle [radius=0.05] (8,0) circle [radius=0.05] (8.5,0) circle [radius=0.05] (9,0) circle [radius=0.05] (10,0) circle [radius=0.05] (10.5,0) circle [radius=0.05] (11,0) circle [radius=0.05] (11.5,0) circle [radius=0.05] (12.5,0) circle [radius=0.05] (13,0) circle [radius=0.05] (13.5,0) circle [radius=0.05] (14,0) circle [radius=0.05];
 \node [] at (2,0) {$,$};
 \node [] at (4.5,0) {$,$};
 \node [] at (7,0) {$,$};
 \node [] at (9.5,0) {$,$};
 \node [] at (12,0) {$,$};
 \node [] at (14.5,0) {$,$};
\end{tikzpicture}
\end{center}
where without ambiguity we have ignored the dashed line $(h_a1)$ corresponding to $(k_1k_{h_a})\mathcal{T}_{1h_a2}$, which is always linked to the ending point of the cyan line. The first two quivers of differential operators are consistent with the rules  (\ref{map-D-rule}), and they are sufficient to distinguish the corresponding vectors uniquely. For the third and fourth quivers of differential operators, noticing the choice of direction of cyan line we know that they are also able to determine the expansion coefficients without mixing contributions from other vectors with one pseudo-loop. However, the last two types of vectors do mix together under the defined differential operators. It can be seen that, with the sixth quiver of differential operators it is able to distinguish the sixth type of vectors. However with the fifth quiver of differential operators, contributions from the fifth type of vectors would be mixed up with those from the sixth type of vectors. Although we can disentangle all vectors by constructing linear combination of differential operators as in formula \eref{D-H-a}, in the current simple example we have alternative way of solving equations. By firstly solving the coefficients of vectors of the sixth topology and then solving the vectors of the fifth topology but with the former solutions as known inputs, we are able to compute all coefficients order by order. Furthermore, we want to emphasize that, the differential operators also pick up contributions from vectors with no pseudo-loops, and we should compute all coefficients of vectors with no pseudo-loops before computing of vectors with one pseudo-loop.

Let us analyze these six topologies one by one. For the first topology, the corresponding differential operator also picks up following contributions in Step-0,
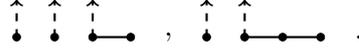
\begin{center}
\begin{tikzpicture}
  \draw [thick, dashed, ->] (0,0)--(0,0.5);
  \draw [thick, dashed, ->] (0.5,0)--(0.5,0.5);
  \draw [thick, dashed, ->] (1,0)--(1,0.5);
  \draw [thick, dashed, ->] (2.5,0)--(2.5,0.5);
  \draw [thick, dashed, ->] (3,0)--(3,0.5);
  \draw [thick] (1,0)--(1.5,0) (3,0)--(4,0);
 \draw [fill] (0,0) circle [radius=0.05]  (0.5,0) circle [radius=0.05]  (1,0) circle [radius=0.05] (1.5,0) circle [radius=0.05] (2.5,0) circle [radius=0.05] (3,0) circle [radius=0.05] (3.5,0) circle [radius=0.05] (4,0) circle [radius=0.05];
 \node [] at (2,0) {$,$};
 \node [] at (4.5,0) {$.$};
\end{tikzpicture}
\end{center}
For instance, using differential operator $(k_1\cdot k_{h_4})\mathcal{T}_{h_4h_3n}\mathcal{T}_{1h_42}\mathcal{T}_{a_1h_1(a_1+1)}\mathcal{T}_{a_2h_2(a_2+1)}$ we can compute the coefficient of $\mathcal{B}[\mathsf{F}_{h_3h_4}\mathsf{F}_{h_1}^{a_1}\mathsf{F}_{h_2}^{a_2}]$ as
\begin{eqnarray}
&&\mathcal{C}[\mathsf{F}_{h_3h_4}\mathsf{F}_{h_1}^{a_1}\mathsf{F}_{h_2}^{a_2}]=\Big(~(k_1\cdot k_{h_4})(\mathcal{T}_{h_4h_3n}\mathcal{T}_{1h_42}\mathcal{T}_{a_1h_1(a_1+1)}\mathcal{T}_{a_2h_2(a_2+1)}~A^{\EYM}_{n,4})~\Big)\nonumber\\
&&+\Big(~\sum_{a_4=2}^{n-1}(k_{h_4}\cdot X_{h_4})\mathcal{C}[\mathsf{F}_{h_3}^{h_4}\mathsf{F}_{h_1}^{a_1}\mathsf{F}_{h_2}^{a_2}\mathsf{F}_{h_4}^{a_4}]~\Big)+\Big(~(k_{h_1}\cdot k_{h_4})\mathcal{C}[\mathsf{F}_{h_3}^{h_4}\mathsf{F}_{h_4}^{h_1}\mathsf{F}_{h_1}^{a_1}\mathsf{F}_{h_2}^{a_2}]+(k_{h_2}\cdot k_{h_4})\mathcal{C}[\mathsf{F}_{h_3}^{h_4}\mathsf{F}_{h_4}^{h_2}\mathsf{F}_{h_1}^{a_1}\mathsf{F}_{h_2}^{a_2}]~\Big)~.~~~\nonumber
\end{eqnarray}
Applying differential operator on $A^{\EYM}_{n,4}$ produces Yang-Mills amplitudes
\begin{equation*}
  A^{\YM}_{n+4}(1,h_4,\{2,\ldots,a_1,h_1,\ldots,a_2,h_2,\ldots,n-1\}\shuffle \{h_3\},n)~,~~~
\end{equation*}
and using BCJ relations they can be rewritten into BCJ basis.

For the second topology, the corresponding differential operator picks up following contributions in Step-0,
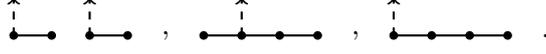
\begin{center}
\begin{tikzpicture}
   \draw [thick, dashed, ->] (0,0)--(0,0.5);
   \draw [thick, dashed, ->] (1,0)--(1,0.5);
   \draw [thick, dashed, ->] (3,0)--(3,0.5);
   \draw [thick, dashed, ->] (5,0)--(5,0.5);
   \draw [thick] (0,0)--(0.5,0) (1,0)--(1.5,0) (2.5,0)--(4,0) (5,0)--(6.5,0);

   \draw [fill] (0,0) circle [radius=0.05]  (0.5,0) circle [radius=0.05]  (1,0) circle [radius=0.05] (1.5,0) circle [radius=0.05] (2.5,0) circle [radius=0.05] (3,0) circle [radius=0.05] (3.5,0) circle [radius=0.05] (4,0) circle [radius=0.05] (5,0) circle [radius=0.05] (5.5,0) circle [radius=0.05] (6,0) circle [radius=0.05] (6.5,0) circle [radius=0.05];

   \node [] at (2,0) {$,$};
   \node [] at (4.5,0) {$,$};
   \node [] at (7,0) {$.$};
\end{tikzpicture}
\end{center}
For instance, using differential operator $(k_1\cdot k_{h_4})\mathcal{T}_{h_4h_3n}\mathcal{T}_{1h_42}\mathcal{T}_{h_1h_2n}\mathcal{T}_{a_1h_1(a_1+1)}$ we can compute the coefficient of $\mathcal{B}[\mathsf{F}_{h_3h_4}\mathsf{F}_{h_2}^{h_1}\mathsf{F}_{h_1}^{a_1}]$ as
\begin{eqnarray}
&&\mathcal{C}[\mathsf{F}_{h_3h_4}\mathsf{F}_{h_2}^{h_1}\mathsf{F}_{h_1}^{a_1}]=\Big(~(k_1\cdot k_{h_4})(\mathcal{T}_{h_4h_3n}\mathcal{T}_{1h_42}\mathcal{T}_{h_1h_2n}\mathcal{T}_{a_1h_1(a_1+1)}~A^{\EYM}_{n,4})~\Big)\nonumber\\
&&+\Big(~\sum_{a_4=2}^{n-1}(k_{h_4}\cdot X_{h_4})\mathcal{C}[\mathsf{F}_{h_2}^{h_1}\mathsf{F}_{h_3}^{h_4}\mathsf{F}_{h_1}^{a_1}\mathsf{F}_{h_4}^{a_4}]~\Big)+\Big(~(k_{h_1}\cdot k_{h_4})\mathcal{C}[\mathsf{F}_{h_2}^{h_1}\mathsf{F}_{h_3}^{h_4}\mathsf{F}_{h_4}^{h_1}\mathsf{F}_{h_1}^{a_1}]~\Big)+\Big(~(k_{h_2}\cdot k_{h_4})\mathcal{C}[\mathsf{F}_{h_2}^{h_1}\mathsf{F}_{h_3}^{h_4}\mathsf{F}_{h_4}^{h_2}\mathsf{F}_{h_1}^{a_1}]~\Big)~.~~~\nonumber
\end{eqnarray}
Applying differential operator on $A^{\EYM}_{n,4}$ produces Yang-Mills amplitudes
\begin{equation*}
  A^{\YM}_{n+4}(1,h_4,\{~2,\ldots,a_1,h_1,\{a_1+1,\ldots,n-1\}\shuffle \{h_2\}~\}\shuffle \{h_3\},n)~.~~~
\end{equation*}
%
%and using BCJ relations they can be rewritten into BCJ basis.

For the third topology, the corresponding differential operator picks up following contributions in Step-0,
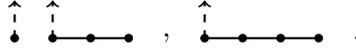
\begin{center}
\begin{tikzpicture}
   \draw [thick, dashed, ->] (0,0)--(0,0.5);
   \draw [thick, dashed, ->] (0.5,0)--(0.5,0.5);
   \draw [thick, dashed, ->] (2.5,0)--(2.5,0.5);
   \draw [thick] (0.5,0)--(1.5,0) (2.5,0)--(4,0);

   \draw [fill] (0,0) circle [radius=0.05]  (0.5,0) circle [radius=0.05]  (1,0) circle [radius=0.05] (1.5,0) circle [radius=0.05] (2.5,0) circle [radius=0.05] (3,0) circle [radius=0.05] (3.5,0) circle [radius=0.05] (4,0) circle [radius=0.05];

   \node [] at (2,0) {$,$};
   \node [] at (4.5,0) {$.$};
\end{tikzpicture}
\end{center}
For instance, using differential operator $(k_1\cdot k_{h_4})\mathcal{T}_{h_4h_3n}\mathcal{T}_{1h_42}\mathcal{T}_{h_3h_2n}\mathcal{T}_{a_1h_1(a_1+1)}$ we can compute the coefficient of $\mathcal{B}[\mathsf{F}_{h_3h_4}\mathsf{F}_{h_2}^{h_3}\mathsf{F}_{h_1}^{a_1}]$ as
\begin{eqnarray}
\mathcal{C}[\mathsf{F}_{h_3h_4}\mathsf{F}_{h_2}^{h_3}\mathsf{F}_{h_1}^{a_1}]&=&\Big(~(k_1\cdot k_{h_4})(\mathcal{T}_{h_4h_3n}\mathcal{T}_{1h_42}\mathcal{T}_{h_3h_2n}\mathcal{T}_{a_1h_1(a_1+1)}~A^{\EYM}_{n,4})~\Big)\nonumber\\
&&~~~~~~+\Big(~\sum_{a_4=2}^{n-1}(k_{h_4}\cdot X_{h_4})\mathcal{C}[\mathsf{F}_{h_2}^{h_3}\mathsf{F}_{h_3}^{h_4}\mathsf{F}_{h_1}^{a_1}\mathsf{F}_{h_4}^{a_4}]~\Big)+\Big(~(k_{h_1}\cdot k_{h_4})\mathcal{C}[\mathsf{F}_{h_2}^{h_3}\mathsf{F}_{h_3}^{h_4}\mathsf{F}_{h_4}^{h_1}\mathsf{F}_{h_1}^{a_1}]~\Big)~.~~~
\end{eqnarray}
Applying differential operator on $A^{\EYM}_{n,4}$ produces Yang-Mills amplitudes
\begin{equation*}
  A^{\YM}_{n+4}(1,h_4,\{2,\ldots,a_1,h_1,\ldots,n-1\}\shuffle \{h_3,h_2\},n)~.~~~
\end{equation*}
%
%and using BCJ relations they can be rewritten into BCJ basis.

For the fourth topology, the corresponding differential operator picks up following contributions in Step-0,
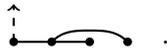
\begin{center}
\begin{tikzpicture}
   \draw [thick, dashed, ->] (0,0)--(0,0.5);
   \draw [thick] (0,0)--(1,0);
   \draw [thick] (0.5,0) to [out=45, in=180] (1,0.15) to [out=0, in=135] (1.5,0);
   \draw [fill] (0,0) circle [radius=0.05]  (0.5,0) circle [radius=0.05]  (1,0) circle [radius=0.05] (1.5,0) circle [radius=0.05];
   \node [] at (2,0) {$.$};
\end{tikzpicture}
\end{center}
For instance, using differential operator $(k_1\cdot k_{h_4})\mathcal{T}_{h_4h_3n}\mathcal{T}_{1h_42}\mathcal{T}_{h_3h_1n}\mathcal{T}_{h_3h_2n}$ we can compute the coefficient of $\mathcal{B}[\mathsf{F}_{h_3h_4}\mathsf{F}_{h_1}^{h_3}\mathsf{F}_{h_2}^{h_3}]$ as
\begin{eqnarray}
\mathcal{C}[\mathsf{F}_{h_3h_4}\mathsf{F}_{h_1}^{h_3}\mathsf{F}_{h_2}^{h_3}]&=&\Big(~(k_1\cdot k_{h_4})(\mathcal{T}_{h_4h_3n}\mathcal{T}_{1h_42}\mathcal{T}_{h_3h_1n}\mathcal{T}_{h_3h_2n}~A^{\EYM}_{n,4})~\Big)+\Big(~\sum_{a_4=2}^{n-1}(k_{h_4}\cdot X_{h_4})\mathcal{C}[\mathsf{F}_{h_1}^{h_3}\mathsf{F}_{h_2}^{h_3}\mathsf{F}_{h_3}^{h_4}\mathsf{F}_{h_4}^{a_4}]~\Big)~.~~~\nonumber
\end{eqnarray}
Applying differential operator on $A^{\EYM}_{n,4}$ produces Yang-Mills amplitudes
\begin{equation*}
  \sum_{\{\sigma_1,\sigma_2\}\in S_2}A^{\YM}_{n+4}(1,h_4,\{2,\ldots,n-1\}\shuffle \{h_3,h_{\sigma_1},h_{\sigma_2}\},n)~.~~~
\end{equation*}
%
%and using BCJ relations they can be rewritten into BCJ basis.

According to our discussion, we will consider the sixth topology before the fifth. The corresponding differential operator picks up following contributions in Step-0,
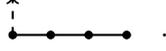
\begin{center}
\begin{tikzpicture}
   \draw [thick, dashed, ->] (0,0)--(0,0.5);
   \draw [thick] (0,0)--(1.5,0);

   \draw [fill] (0,0) circle [radius=0.05]  (0.5,0) circle [radius=0.05]  (1,0) circle [radius=0.05] (1.5,0) circle [radius=0.05];
   \node [] at (2,0) {$.$};
\end{tikzpicture}
\end{center}
For instance, using differential operator $(k_1\cdot k_{h_4})\mathcal{T}_{h_4h_3n}\mathcal{T}_{1h_42}\mathcal{T}_{h_2h_1n}\mathcal{T}_{h_3h_2n}$ we can compute the coefficient of $\mathcal{B}[\mathsf{F}_{h_3h_4}\mathsf{F}_{h_1}^{h_2}\mathsf{F}_{h_2}^{h_3}]$ as
\begin{eqnarray}
\mathcal{C}[\mathsf{F}_{h_3h_4}\mathsf{F}_{h_1}^{h_2}\mathsf{F}_{h_2}^{h_3}]&=&\Big(~(k_1\cdot k_{h_4})(\mathcal{T}_{h_4h_3n}\mathcal{T}_{1h_42}\mathcal{T}_{h_2h_1n}\mathcal{T}_{h_3h_2n}~A^{\EYM}_{n,4})~\Big)+\Big(~\sum_{a_4=2}^{n-1}(k_{h_4}\cdot X_{h_4})\mathcal{C}[\mathsf{F}_{h_1}^{h_2}\mathsf{F}_{h_2}^{h_3}\mathsf{F}_{h_3}^{h_4}\mathsf{F}_{h_4}^{a_4}]~\Big)~.~~~\nonumber
\end{eqnarray}
Applying differential operator on $A^{\EYM}_{n,4}$ produces Yang-Mills amplitudes
\begin{equation*}
  A^{\YM}_{n+4}(1,h_4,\{2,\ldots,n-1\}\shuffle \{h_3,h_2,h_1\},n)~.~~~
\end{equation*}

Then come to the last piece. Besides the contribution from the sixth topology, the differential operator corresponding to the fifth topology also picks up following contributions in Step-0,
\begin{center}
\begin{tikzpicture}
   \draw [thick, dashed, ->] (0.5,0)--(0.5,0.5);
   \draw [thick] (0,0)--(1.5,0);

   \draw [fill] (0,0) circle [radius=0.05]  (0.5,0) circle [radius=0.05]  (1,0) circle [radius=0.05] (1.5,0) circle [radius=0.05];
   \node [] at (2,0) {$.$};
\end{tikzpicture}
\end{center}
Let's consider an example, the differential operator $(k_1\cdot k_{h_4})\mathcal{T}_{h_4h_3n}\mathcal{T}_{1h_42}\mathcal{T}_{h_3h_1n}\mathcal{T}_{h_4h_2n}$. We can use it to compute the coefficient of $\mathcal{B}[\mathsf{F}_{h_3h_4}\mathsf{F}_{h_1}^{h_3}\mathsf{F}_{h_2}^{h_4}]$ as
\begin{eqnarray}
&&\mathcal{C}[\mathsf{F}_{h_3h_4}\mathsf{F}_{h_1}^{h_3}\mathsf{F}_{h_2}^{h_4}]
=\\
&&\Big(~(k_1\cdot k_{h_4})(\mathcal{T}_{h_4h_3n}\mathcal{T}_{1h_42}\mathcal{T}_{h_3h_1n}\mathcal{T}_{h_4h_2n}~A^{\EYM}_{n,4})~\Big)
 +\Big(~\sum_{a_4=2}^{n-1}(k_{h_4}\cdot X_{h_4})\mathcal{C}[\mathsf{F}_{h_1}^{h_3}\mathsf{F}_{h_2}^{h_4}\mathsf{F}_{h_3}^{h_4}\mathsf{F}_{h_4}^{a_4}]~\Big) -\mathcal{C}[\mathsf{F}_{h_2h_4}F_{h_3}^{h_4}F_{h_1}^{h_3}]~.~~~\nonumber
\end{eqnarray}
Applying differential operator on $A^{\EYM}_{n,4}$ produces Yang-Mills amplitudes
\begin{equation*}
  A^{\YM}_{n+4}(1,h_4,\{2,\ldots,n-1\}\shuffle \{h_3,h_1\}\shuffle \{h_2\},n)~.~~~
\end{equation*}
%
%and using BCJ relations they can be rewritten into BCJ basis.
Above computations provide all expansion coefficients for gauge invariant basis with one pseudo-loop based on the solutions in Step-0 and the BCJ relations.

Let us continue to Step-2, where there are only three different vectors $\mathcal{B}[\mathsf{F}_{h_1h_2}\mathsf{F}_{h_3h_4}]$, $\mathcal{B}[\mathsf{F}_{h_1h_3}\mathsf{F}_{h_2h_4}]$ and $\mathcal{B}[\mathsf{F}_{h_1h_4}\mathsf{F}_{h_2h_3}]$. According to the rule, we define differential operators for them respectively as
\begin{equation}
\mathcal{T}_{h_2h_1n}\mathcal{T}_{h_4h_3n}\mathcal{T}_{1h_22}\mathcal{T}_{1h_42}~~~,~~~\mathcal{T}_{h_3h_1n}\mathcal{T}_{h_4h_2n}\mathcal{T}_{1h_32}\mathcal{T}_{1h_44}~~~,~~~\mathcal{T}_{h_4h_1n}\mathcal{T}_{h_3h_2n}\mathcal{T}_{1h_32}\mathcal{T}_{1h_42}~.~~~
\end{equation}
It can be checked directly that each differential operator picks up only one vector with two pseudo-loops, while it also picks up following contributions in Step-0 and Step-1,
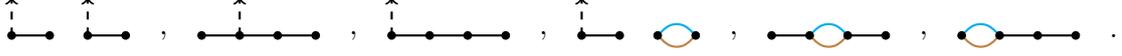
\begin{center}
\begin{tikzpicture}
   \draw [thick, dashed, ->] (0,0)--(0,0.5);
   \draw [thick, dashed, ->] (1,0)--(1,0.5);
   \draw [thick, dashed, ->] (3,0)--(3,0.5);
   \draw [thick, dashed, ->] (5,0)--(5,0.5);
   \draw [thick, dashed, ->] (7.5,0)--(7.5,0.5);
   \draw [thick] (0,0)--(0.5,0) (1,0)--(1.5,0) (2.5,0)--(4,0) (5,0)--(6.5,0) (7.5,0)--(8,0) (10,0)--(10.5,0) (11,0)--(11.5,0) (13,0)--(14,0);
  \draw [thick, cyan] (8.5,0) to [out=45, in=180] (8.75,0.15) to [out=0, in=135] (9,0);
  \draw [thick, brown] (8.5,0) to [out=315, in=180] (8.75,-0.15) to [out=0, in=225] (9,0);
  \draw [thick, cyan] (10.5,0) to [out=45, in=180] (10.75,0.15) to [out=0, in=135] (11,0);
  \draw [thick, brown] (10.5,0) to [out=315, in=180] (10.75,-0.15) to [out=0, in=225] (11,0);
  \draw [thick, cyan] (12.5,0) to [out=45, in=180] (12.75,0.15) to [out=0, in=135] (13,0);
  \draw [thick, brown] (12.5,0) to [out=315, in=180] (12.75,-0.15) to [out=0, in=225] (13,0);

   \draw [fill] (0,0) circle [radius=0.05]  (0.5,0) circle [radius=0.05]  (1,0) circle [radius=0.05] (1.5,0) circle [radius=0.05] (2.5,0) circle [radius=0.05] (3,0) circle [radius=0.05] (3.5,0) circle [radius=0.05] (4,0) circle [radius=0.05] (5,0) circle [radius=0.05] (5.5,0) circle [radius=0.05] (6,0) circle [radius=0.05] (6.5,0) circle [radius=0.05] (7.5,0) circle [radius=0.05] (8,0) circle [radius=0.05] (8.5,0) circle [radius=0.05] (9,0) circle [radius=0.05] (10,0) circle [radius=0.05] (10.5,0) circle [radius=0.05] (11,0) circle [radius=0.05] (11.5,0) circle [radius=0.05] (12.5,0) circle [radius=0.05] (13,0) circle [radius=0.05] (13.5,0) circle [radius=0.05] (14,0) circle [radius=0.05];

   \node [] at (2,0) {$,$};
   \node [] at (4.5,0) {$,$};
   \node [] at (7,0) {$,$};
   \node [] at (9.5,0) {$,$};
   \node [] at (12,0) {$,$};
   \node [] at (14.5,0) {$.$};
\end{tikzpicture}
\end{center}
For instance, Using differential operator $\mathcal{T}_{h_2h_1n}\mathcal{T}_{h_4h_3n}\mathcal{T}_{1h_22}\mathcal{T}_{1h_42}$ we can compute the coefficient of $\mathcal{B}[\mathsf{F}_{h_1h_2}\mathsf{F}_{h_3h_4}]$ as,
\begin{eqnarray}
\mathcal{C}[\mathsf{F}_{h_1h_2}\mathsf{F}_{h_3h_4}]&=&\Big((k_1\cdot k_{h_2})(k_1\cdot k_{h_4})(\mathcal{T}_{h_2h_1n}\mathcal{T}_{h_4h_3n}\mathcal{T}_{1h_22}\mathcal{T}_{1h_42}~A^{\EYM}_{n,4})\Big)+\Big([\mbox{Step-0}]+[\mbox{Step-1}]\Big)\Big|_{\mathsf{F}_{h_1h_2}\mathsf{F}_{h_3h_4}}~,~~~\nonumber
\end{eqnarray}
%
%
%\begin{eqnarray}
%\mathcal{C}[\mathsf{F}_{h_1h_2}\mathsf{F}_{h_3h_4}]&=&\Big(~(k_1\cdot k_{h_2})(k_1\cdot %k_{h_4})(\mathcal{T}_{h_2h_1n}\mathcal{T}_{h_4h_3n}\mathcal{T}_{1h_22}\mathcal{T}_{1h_42}~A^{\EYM}_{n,4})\Big)\nonumber\\
%&&~~~~~~~~~-\Big(~(k_1\cdot k_{h_2})(k_1\cdot %k_{h_4})(\mathcal{T}_{h_2h_1n}\mathcal{T}_{h_4h_3n}\mathcal{T}_{1h_22}\mathcal{T}_{1h_42}~\left([\mbox{Step-0}]+[\mbox{Step-1}]\right) )~\Big)~,~~~
%\end{eqnarray}
%
where
\begin{eqnarray}
&& [\mbox{Step-0}]\Big|_{\mathsf{F}_{h_1h_2}\mathsf{F}_{h_3h_4}}=\Big(\sum_{a_2=2}^{n-1}\sum_{a_4=2}^{n-1} (k_{h_2}\cdot X_{h_2})(k_{h_4}\cdot X_{h_4})\mathcal{C}[\mathsf{F}_{h_1}^{h_2}\mathsf{F}_{h_3}^{h_4}\mathsf{F}_{h_2}^{a_2}\mathsf{F}_{h_4}^{a_4}]\Big)\\
&& +\Big(\sum_{a_2=2}^{n-1} \sum_{i=1,2} (k_{h_2}\cdot X_{h_2})(k_{h_4}\cdot k_{h_i})\mathcal{C}[\mathsf{F}_{h_1}^{h_2}\mathsf{F}_{h_3}^{h_4}\mathsf{F}_{h_4}^{h_i}\mathsf{F}_{h_2}^{a_2}]\Big)+\Big(\sum_{a_4=2}^{n-1} \sum_{i=3,4} (k_{h_4}\cdot X_{h_4})(k_{h_2}\cdot k_{h_i})\mathcal{C}[\mathsf{F}_{h_1}^{h_2}\mathsf{F}_{h_2}^{h_i}\mathsf{F}_{h_3}^{h_4}\mathsf{F}_{h_4}^{a_4}]\Big)~,~~~\nonumber
\end{eqnarray}
is the contribution from expansion in Step-0, and
\begin{eqnarray}
&& [\mbox{Step-1}]\Big|_{\mathsf{F}_{h_1h_2}\mathsf{F}_{h_3h_4}}=\Big(\sum_{a_2=2}^{n-1}(k_{h_2}\cdot X_{h_2})\mathcal{C}[\mathsf{F}_{h_3h_4}\mathsf{F}_{h_1}^{h_2}\mathsf{F}_{h_2}^{a_2}]+\sum_{i=3,4}(k_{h_2}\cdot k_{h_i})\mathcal{C}[\mathsf{F}_{h_3h_4}\mathsf{F}_{h_1}^{h_2}\mathsf{F}_{h_2}^{h_i}]\Big)\\
&&~~~~+\Big(\sum_{a_4=2}^{n-1}(k_{h_4}\cdot X_{h_4})\mathcal{C}[\mathsf{F}_{h_1h_2}\mathsf{F}_{h_3}^{h_4}\mathsf{F}_{h_4}^{a_4}]+\sum_{i=1,2}(k_{h_4}\cdot k_{h_i})\mathcal{C}[\mathsf{F}_{h_1h_2}\mathsf{F}_{h_3}^{h_4}\mathsf{F}_{h_4}^{h_i}]\Big)+\Big((k_{h_2}\cdot k_{h_4})\mathcal{C}[\mathsf{F}_{h_2h_4}\mathsf{F}_{h_1}^{h_2}\mathsf{F}_{h_3}^{h_4}]\Big)~,~~~\nonumber
\end{eqnarray}
is the contribution from expansion in Step-1. While applying differential operator on $A^{\EYM}_{n,4}$ produces Yang-Mills amplitudes,
\begin{equation}
\begin{array}{l}
\sum_{\{\sigma_2,\sigma_4\}\in S_2}A^{\YM}_{n+4}(1,h_{\sigma_2},h_{\sigma_4},\{2,\ldots,n-1\}\shuffle\{h_1\}\shuffle \{h_3\},n)\\
~~+A^{\YM}_{n+4}(1,h_2,h_1,h_4,\{2,\ldots,n-1\}\shuffle \{h_3\},n)+A^{\YM}_{n+4}(1,h_4,h_3,h_2,\{2,\ldots,n-1\}\shuffle \{h_1\},n)
\end{array}~.~~~
\end{equation}
Then using BCJ relations for $A^{\YM}(1,\alpha_1,2,\ldots,n)$, $A^{\YM}(1,\alpha_1,\alpha_2,2,\ldots,n)$, $A^{\YM}(1,\alpha_1,\alpha_2,\alpha_3,2,\ldots,n)$ and $A^{\YM}(1,\alpha_1,\alpha_2,\alpha_3,\alpha_4,2,\ldots,n)$ we can rewrite all Yang-Mills amplitude into BCJ basis with legs $1,2,n$ fixed. Collecting all above results, we get the required EYM amplitude expansion. Because the final result is complicated we would not present the explicit expression for $A^{\EYM}_{n,4}(1,\cdots,n;h_1,h_2,h_3,h_4)$. However we have numerically checked the algorithm up to $A^{\EYM}_{6,4}$ and find agreement with  CHY formalism.

%%%%%%%%%%%%%%%%%%
\section{Conclusion}
\label{section-conclusion}
%%%%%%%%%%%%%%%%%

There are already quite a lot well-formulated results for expansion of EYM amplitudes to Yang-Mills amplitudes in KK basis, however a compact expression or even a recursive formula for expansion to Yang-Mills amplitudes in BCJ basis is still in pursuit. The latter expansion is generally much more complicated as conventionally expected. In the KK basis the expansion coefficients of Yang-Mills amplitudes are only polynomials of polarizations and momenta, and they are constrained to explicit compact expressions by gauge invariance. In the BCJ basis, the expansion coefficients of Yang-Mills amplitudes are instead rational functions, whose explicit form is much more difficult to determine. This is the reason that we consider using differential operators to determine expansion coefficients in paper \cite{Feng:2019tvb}.

This paper is motivated by the problem of expanding EYM amplitudes to Yang-Mills amplitudes in BCJ basis by differential operators. We have implemented an algorithm to systematically perform the expansion and compute the expansion coefficients. However the EYM amplitude is not directly expanded to BCJ basis but instead to a basis in gauge invariant vector space, as schematically shown in formula \eref{expansionGI-abbr}. After determining the expansion coefficients, we transform Yang-Mills amplitudes to BCJ basis by BCJ relations. Expanding EYM amplitude in a manifest gauge invariant form for both expansion basis and their coefficients is a very interesting point of view, and differential operators can be naturally introduced into the problem. It contributes to our major results.

The first major part of this paper is devoted to the construction of gauge invariant basis and their corresponding differential operators. A systematic algorithm is built upon the properties of applying differential operators on different basis. To construct a complete set of manifestly gauge invariant polynomials as the expansion basis, we start from the most general vector space $\mathcal{V}_{n,m}$ with $m\leq n$, where all possible polynomials of Lorentz contractions among polarizations and momenta live in this space, obeying some additional conditions. Then we define some linear mapping $\mathcal{G}_i$, which is a realization of gauge invariant condition for a polarization. By taking the interaction of kernels of all possible $\mathcal{G}_i$'s, we construct the gauge invariant sub-space $\mathcal{W}_{n,m}$ from $\mathcal{V}_{n,m}$, which is the vector space containing all gauge invariant polynomials. This is also the space where the expansion basis of EYM amplitude lives. We present the formula for computing the dimension of $\mathcal{W}_{n,m}$, which indicates the number of gauge invariant vectors a EYM amplitude would be expanded to. We also find that the gauge invariant vectors can be realized by linear combinations of multiplications of fundamental $f$-terms. Above results at the end help us to construct a linearly independent and complete basis combinatorially for EYM amplitude expansion.

After clarifying the structure of gauge invariant expansion basis, we further construct differential operators from multiplication of insertion operators. The differential operators are constructed such that when applying a differential operator on an expression only one particular vector in gauge invariant basis is non-vanishing while all others vanishing. In order to do so, we start with analyzing the structures of gauge invariant basis and find the quiver representation for them. With the help of quiver representation, we summarize all possible components appearing in gauge invariant vectors, and provide mapping rules for writing a differential operator directly from a gauge invariant vector, as multiplication of three basic types of insertion operators. Based on above results, an algorithm for expansion of EYM amplitudes is implemented, with the idea of solving algebraic systems of linear equations order by order. To demonstrate the algorithm, we present the expansions of EYM amplitudes with up to four gravitons in the language of gauge invariant basis, which are all consistent with CHY formalism numerically.

Although the algorithm for expanding tree-level single-trace EYM amplitude to Yang-Mills amplitudes in BCJ basis has been laid down thoroughly in this paper, it still inspires further works to do in future. Firstly, the expansion coefficients of BCJ basis demands an explicit and possibly compact formulation. It is a rather difficult problem, but we have found some clues in results \eref{2p-C-1} and
\eref{3p-temp-2} already, and hope it could help to figure out the general picture. Secondly, in this paper we only deal with single-trace EYM amplitudes, while discussions can be generalized to multi-trace EYM amplitudes by using trace operator ${\cal T}_{\eps_i\eps_j}$. We think this generalization should be straightforward.

Thirdly, in this paper we are focusing on EYM amplitudes, so the parameters of vector space $\mathcal{V}_{n,m}$ is constrained to $m<n$. However, the case $m=n$ is also very interesting in physics since Yang-Mills amplitudes live in this space. Another interesting example is the deformed Yang-Mills theory with $F^3$ term \cite{He:2016iqi, Garozzo:2018uzj}. Although the dimension of $\mathcal{W}_{n,m}$ still holds for $m=n$, the explicit form of vectors in gauge invariant basis should be reconsidered since we are not able to trivially exclude momentum $k_n$ in all expression by momentum conservation. Furthermore, for Yang-Mills amplitude an additional constraint should be applied to the vector space, {\sl i.e.}, there should be at least one $(\eps\cdot \eps)$ contraction, and let us denote the vector space by $\widetilde{W}_{n,m}$. The new vector space $\widetilde{W}_{n,m}$ can help us to understand the implication of gauge invariance in Yang-Mills amplitudes more deeply, along the line of former studies in papers \cite{Boels:2016xhc, Arkani-Hamed:2016rak, Rodina:2016jyz}. It is also a curious problem about how to write Yang-Mills amplitudes in a manifestly gauge invariant form. Maybe it can also help us to understand more about the Pfaffian in the integrand of CHY formula, and provide a new point of view for BCJ relations.

%%%%%%%%%%%%%%%%%%%%%%%
\section*{Acknowledgments}
%%%%%%%%%%%%%%%%%%%%%%%%

We are grateful to Kang Zhou, ZhongJie Huang and Yiwen Lin for discussions about this work. Xiao-Di Li would like to thank Yi-Jian Du for his enlightening discussions and kind hospitality in Wuhan University. B.F. is supported by Qiu-Shi Funding and the
National Natural Science Foundation of China (NSFC) with Grant No.11935013, No.11575156.
R.H. is supported by the National Natural Science Foundation of China (NSFC) with Grant No.11805102, Natural Science Foundation of Jiangsu Province with Grant No.BK20180724, and {\sl ShuangChuang Talent Program} of Jiangsu Province.

\appendix

%%%%%%%%%%%%%%%%%%%
\section{Proof of propositions about the gauge invariant vector space} \label{appendix-proof}
%%%%%%%%%%%%%%%%%%%%%

\noindent {\bf Proof of proposition 1}: We want to prove the following splitting formula of linear maps $\mathcal{G}_1,\mathcal{G}_2$,
\begin{equation}
\mathrm{Ker}~\mathcal{G}_1+\mathrm{Ker}~\mathcal{G}_2=\mathrm{Ker}~\mathcal{G}_1\mathcal{G}_2~.~~~
\label{App-pro-1}
\end{equation}
In order to do so, it is suffice to show
\begin{equation}
\mathrm{Ker}~\mathcal{G}_1+\mathrm{Ker}~\mathcal{G}_2 \subseteq \mathrm{Ker}~\mathcal{G}_1\mathcal{G}_2~~~\mbox{and}~~~\mathrm{Ker}~ \mathcal{G}_1+\mathrm{Ker}~ \mathcal{G}_2 \supseteq \mathrm{Ker}~ \mathcal{G}_1\mathcal{G}_2~.~~~
\end{equation}

The proof of $\mathrm{Ker}~\mathcal{G}_1+\mathrm{Ker}~\mathcal{G}_2 \subseteq \mathrm{Ker}~\mathcal{G}_1\mathcal{G}_2$ is trivial. For each $v\in \mathrm{Ker}~\mathcal{G}_1+\mathrm{Ker}~\mathcal{G}_2$, it can always be written as
\begin{equation*}
v=v_1+v_2~~~,~~~  v_i\in \mathrm{Ker}~\mathcal{G}_i~~\mbox{and}~~~\mathcal{G}_iv_i=0~.~~~
\end{equation*}
Thus the action of $\mathcal{G}_1\mathcal{G}_2$ on $v$ is
\begin{equation}
\mathcal{G}_1\mathcal{G}_2 v=\mathcal{G}_1\mathcal{G}_2 v_1+\mathcal{G}_1\mathcal{G}_2 v_2=\mathcal{G}_2(\mathcal{G}_1 v_1)+\mathcal{G}_1(\mathcal{G}_2 v_2)=0~,~~~
\end{equation}
where we have used the  commutative of ${\cal G}_i$, {\sl i.e.}, $\mathcal{G}_1\mathcal{G}_2=\mathcal{G}_2\mathcal{G}_1$. Hence $v\in \mathrm{Ker}~\mathcal{G}_1\mathcal{G}_2$, and consequently $\mathrm{Ker}~\mathcal{G}_1+\mathrm{Ker}~\mathcal{G}_2 \subseteq \mathrm{Ker}~\mathcal{G}_1\mathcal{G}_2$.

The proof of $\mathrm{Ker}~ \mathcal{G}_1+\mathrm{Ker}~ \mathcal{G}_2 \supseteq \mathrm{Ker}~ \mathcal{G}_1\mathcal{G}_2$ is not so easy and we will prove it by induction. Let us start from the vector space $\mathcal{V}_{n,2}$, {\sl i.e.}, containing only two polarizations $\epsilon_1,\epsilon_2$. A polynomial $\mathfrak{h}_{n,2}$ in $\mathcal{V}_{n,2}$ can be written as
\begin{equation}
\mathfrak{h}_{n,2}=\alpha_1(\epsilon_1\cdot \epsilon_2)+\sum_{i,j=1}^{n-1}\alpha_{2}^{ij}(\epsilon_1\cdot k_i)(\epsilon_2\cdot k_j)~,~~~
\end{equation}
where momentum conservation has been used to eliminate the appearance of $k_n$. For $\mathfrak{h}_{n,2}\in~\mathrm{Ker}~\mathcal{G}_1\mathcal{G}_2$, by imposing $\mathcal{G}_1\mathcal{G}_2\mathfrak{h}_{n,2}=0$ we get
\begin{equation}
\mathcal{G}_1\mathcal{G}_2\mathfrak{h}_{n,2}=\mathfrak{h}_{n,2}|_{{\epsilon_1\to k_1 \atop \epsilon_2\to k_2}}=\alpha_1(k_1\cdot k_2)+\sum_{i,j=1}^{n-1}\alpha_2^{ij}(k_1\cdot k_i)(k_2\cdot k_j)=0~.~~~
\end{equation}
From above equation we can solve $\alpha_1$ and substitute it back to $\mathfrak{h}_{n,2}$. After reorganization of terms, we get
\begin{equation}
\mathfrak{h}_{n,2}=\sum_{i,j=1}^{n-1} \alpha_2^{ij} \frac{\epsilon_2\cdot f_1\cdot k_i}{k_1\cdot k_2}(k_2\cdot k_j)+\sum_{i,j=1}^{n-1} \alpha_2^{ij} \frac{k_1\cdot f_2\cdot k_j}{k_1\cdot k_2}(\epsilon_1\cdot k_i):=v_1+v_2~.~~~
\end{equation}
Since the appearance of $f_i$, it is easy to see that ${\cal G}_i v_i=0$.
Hence $v_1\in \mathrm{Ker}~\mathcal{G}_1$ and $v_2\in \mathrm{Ker}~\mathcal{G}_2$. This shows that if $\mathfrak{h}_{n,2}\in~\mathrm{Ker}~\mathcal{G}_1\mathcal{G}_2$, there is also $\mathfrak{h}_{n,2}\in~\mathrm{Ker}~\mathcal{G}_1+\mathrm{Ker}~ \mathcal{G}_2$.

Now let us assume that for all vector spaces $\mathcal{V}_{n,s}, s<m$, if a polynomial $\mathfrak{h}_{n,s}\in \mathrm{Ker}~\mathcal{G}_1\mathcal{G}_2$, then it can always be separated into two parts, one part belonging to $\mathrm{Ker}~\mathcal{G}_1$ and the other belonging to $\mathrm{Ker}~\mathcal{G}_2$. For a polynomial $\mathfrak{h}_{n,m}$ in the vector space $\mathcal{V}_{n,m}$, it can be expanded to
\begin{equation}
\mathfrak{h}_{n,m}=\sum_{i=1}^{m-1}(\epsilon_m\cdot \epsilon_i) T_{mi}+\sum_{i=1}^{m-1}(\epsilon_m\cdot k_i)(\epsilon_i\cdot T'_{mi})+\sum_{i=m+1}^{n-1}(\epsilon_m\cdot k_i) T''_{mi}~,~~~\label{proof-prop1-hnm}
\end{equation}
where $T_{mi}\in \mathcal{V}_{n,m-2}$ and $\epsilon_i\cdot T'_{mi}~,~T''_{mi}\in \mathcal{V}_{n,m-1}$. For $\mathfrak{h}_{n,m}\in \mathrm{Ker}~\mathcal{G}_1\mathcal{G}_2$, by definition we have
\begin{eqnarray}
&&0=\mathfrak{h}_{n,m}|_{\epsilon_1\to k_1\atop \epsilon_2\to k_2}=(\epsilon_m\cdot k_1)T_{m1}^{(2)}+(\epsilon_m\cdot k_2)T_{m2}^{(1)}+\sum_{i=3}^{m-1}(\epsilon_m\cdot \epsilon_i) T_{mi}^{(12)}\label{proof-prop1-recur}\\
&&~~~~~~~~~+(\epsilon_m\cdot k_1)(k_1\cdot T^{'(2)}_{m1})+(\epsilon_m\cdot k_2)(k_2\cdot T^{'(1)}_{m2})+\sum_{i=3}^{m-1}(\epsilon_m\cdot k_i)(\epsilon_i\cdot T^{'(12)}_{mi})+\sum_{i=m+1}^{n-1}(\epsilon_m\cdot k_i)T^{''(12)}_{mi}~,~~~\nonumber
\end{eqnarray}
where the superscript in $T,T'$ and $T''$ denotes the corresponding polarizations to be replaced by their momenta. In (\ref{proof-prop1-recur}), the Lorentz invariants $\epsilon_m\cdot k_i$, $i=1,2,\ldots, m-1,m+1,\ldots, n-1$ and $\epsilon_m\cdot \epsilon_i$, $i=3,4,\ldots, m-1$ are all independent, hence all the coefficients of them should be zero if $\mathfrak{h}_{n,m}|_{\epsilon_1\to k_1\atop \epsilon_2\to k_2}=0$, and we get
\begin{eqnarray}
&& T_{m1}^{(2)}+k_1\cdot T^{'(2)}_{m1}=0~~~,~~~T_{m2}^{(1)}+k_2\cdot T^{'(1)}_{m2}=0~,~~~\label{proof-prop1-coe1}\\
&& T_{mi}^{(12)}=0~~,~~\epsilon_{i}\cdot T_{mi}^{'(12)}=0~~\forall (i=3,\ldots,m-1)~~~,~~~ T^{''(12)}_{mi}=0~~\forall (i=m+1,\ldots,n-1)~.~~~\label{proof-prop1-coe2}
\end{eqnarray}
The result (\ref{proof-prop1-coe2}) tells us that all  $T_{mi}, T'_{mi}, i=3,\ldots,m-1$ and $T''_{mi}, i=m+1,\ldots, n-1$ belong to $\mathrm{Ker}~\mathcal{G}_1\mathcal{G}_2$, and by the induction they belong to $\mathrm{Ker}~\mathcal{G}_1+\mathrm{Ker}~\mathcal{G}_2$. For the remaining terms in (\ref{proof-prop1-hnm}), {\sl i.e.},
\begin{equation}
\mathfrak{h}'_{n,m}=(\epsilon_m\cdot \epsilon_1)T_{m1}+(\epsilon_m\cdot \epsilon_2)T_{m2}+(\epsilon_m\cdot k_1)(\epsilon_1\cdot T'_{m1})+(\epsilon_m\cdot k_2)(\epsilon_2\cdot T'_{m2})~.~~~
\end{equation}
After adding
$
  0=(\epsilon_m\cdot \epsilon_1)(k_1\cdot T'_{m1}) -(\epsilon_m\cdot \epsilon_1)(k_1\cdot T'_{m1})
     +(\epsilon_m\cdot \epsilon_2)(k_2\cdot T'_{m2})
-(\epsilon_m\cdot \epsilon_2)(k_2\cdot T'_{m2})
$
at the RHS of above equation, we can reorganize $\mathfrak{h}'_{n,m}$ to be
\begin{equation}
\mathfrak{h}'_{n,m}=\Big((\epsilon_{m}\cdot \epsilon_2)(T_{m2}+k_2\cdot T'_{m2})+(\epsilon_m\cdot f_1\cdot T'_{m1})\Big)+\Big((\epsilon_{m}\cdot \epsilon_1)(T_{m1}+k_1\cdot T'_{m1})+(\epsilon_m\cdot f_2\cdot T'_{m2})\Big)~.~~~\label{proof-prop1-hnmprime}
\end{equation}
Using the result (\ref{proof-prop1-coe1}) we get
\begin{equation}
\mathcal{G}_i (\epsilon_m\cdot f_i\cdot T'_{mi})=\mathcal{G}_i \big((\epsilon_m\cdot k_i)(\epsilon_i\cdot T'_{mi})-(\epsilon_m\cdot \epsilon_i)(k_i\cdot T'_{mi})\big)=(\epsilon_m\cdot k_i)(k_i\cdot T'_{mi})-(\epsilon_m\cdot k_i)(k_i\cdot T'_{mi})=0~.~~~
\end{equation}
Thus $\mathfrak{h}'_{n,m}$ belongs to $\mathrm{Ker}~\mathcal{G}_1+\mathrm{Ker}~\mathcal{G}_2$. So finally we have proven that $\mathrm{Ker}\ \mathcal{G}_1+\mathrm{Ker}\ \mathcal{G}_2\supseteq\mathrm{Ker}\ \mathcal{G}_1\mathcal{G}_2$ is valid in any vector space $\mathcal{V}_{n,m}$, and the proposition 1 is proven.

~\\~\\
\noindent {\bf Proof of proposition 2}: We want to prove the following distribution formula of linear maps $\mathcal{G}_1,\mathcal{G}_2,\mathcal{G}_3$,
\begin{equation}
(\mathrm{Ker}~\mathcal{G}_1+\mathrm{Ker}~\mathcal{G}_2)\cap \mathrm{Ker}~\mathcal{G}_3=\mathrm{Ker}~\mathcal{G}_1\cap \mathrm{Ker}~\mathcal{G}_3+\mathrm{Ker}~\mathcal{G}_2\cap \mathrm{Ker}~\mathcal{G}_3~.~~~\label{proposition2}
\end{equation}
In order to do so, it is suffice to show
\begin{eqnarray}
(\mathrm{Ker}~\mathcal{G}_1+\mathrm{Ker}~\mathcal{G}_2)\cap \mathrm{Ker}~\mathcal{G}_3 \supseteq\mathrm{Ker}~\mathcal{G}_1\cap \mathrm{Ker}~\mathcal{G}_3+\mathrm{Ker}~\mathcal{G}_2\cap \mathrm{Ker}~\mathcal{G}_3~,~~&&\label{proof-prop2-1}\\
\mbox{and}~ (\mathrm{Ker}~\mathcal{G}_1+\mathrm{Ker}~\mathcal{G}_2)\cap \mathrm{Ker}~\mathcal{G}_3\subseteq\mathrm{Ker}~\mathcal{G}_1\cap \mathrm{Ker}~\mathcal{G}_3+\mathrm{Ker}~\mathcal{G}_2\cap \mathrm{Ker}~\mathcal{G}_3~.~~~\label{proof-prop2-2}&&
\end{eqnarray}

To show (\ref{proof-prop2-1}), noticing that  any vector $v\in \mathrm{Ker}~\mathcal{G}_1\cap \mathrm{Ker}~\mathcal{G}_3+\mathrm{Ker}~\mathcal{G}_2\cap \mathrm{Ker}~\mathcal{G}_3$ can always be written as
\bea v=v_1+v_2~~~,~~~v_1\in \mathrm{Ker}~\mathcal{G}_1\cap \mathrm{Ker}~\mathcal{G}_3~~~,~~~v_2 \in \mathrm{Ker}~\mathcal{G}_2\cap \mathrm{Ker}~\mathcal{G}_3~.~~~\eea
Thus we can check
\bea &&{\cal G}_3  v  =  {\cal G}_3 v_1+{\cal G}_3 v_2=0 ~,~~~\\
&&{\cal G}_1 {\cal G}_2 v =  {\cal G}_2 {\cal G}_1 v_1+
{\cal G}_1 {\cal G}_2 v_2=0~.~~~\eea
Referring to the proposition 1 \eref{App-pro-1}, above result shows that
$v\in (\mathrm{Ker}~\mathcal{G}_1+\mathrm{Ker}~\mathcal{G}_2)\cap \mathrm{Ker}~\mathcal{G}_3 $. Hence  (\ref{proof-prop2-1}) is derived.

To show (\ref{proof-prop2-2}), we again use induction method.  Let us start with the vector space $\mathcal{V}_{n,3}$. A polynomial $\mathfrak{h}_{n,3}\in \mathcal{V}_{n,3}$
has the generic form as
\begin{eqnarray}
\mathfrak{h}_{n,3}&=&\sum_{i,j,l=1\atop i\neq 1,j\neq 2,l\neq 3}^{n-1} \alpha^{ijl}(\epsilon_1\cdot k_i)(\epsilon_2\cdot k_j)(\epsilon_3\cdot k_l)\label{proof-prop2-hn3}\\[-1em]
&&~~~~~~~~~~~~~~~+\sum_{i=1\atop i\neq 1}^{n-1}\beta_{1}^{i}(\epsilon_1\cdot k_i)(\epsilon_2\cdot\epsilon_3)+\sum_{j=1\atop j\neq 2}^{n-1}\beta_{2}^{j}(\epsilon_2\cdot k_j)(\epsilon_1\cdot\epsilon_3)+\sum_{l=1\atop l\neq 3}^{n-1}\beta_{3}^{l}(\epsilon_3\cdot k_l)(\epsilon_1\cdot\epsilon_2)~,~~~\nonumber
\end{eqnarray}
where $k_n$ has been eliminated using the momentum conservation. Now we impose the condition that $\mathfrak{h}_{n,3}\in  (\mathrm{Ker}~\mathcal{G}_1+\mathrm{Ker}~\mathcal{G}_2)\cap \mathrm{Ker}~\mathcal{G}_3$. Imposing  $\mathcal{G}_3 \mathfrak{h}_{n,3}=0$  we get an equation $\mathfrak{h}_{n,3}|_{\epsilon_3\to k_3}=0$. After $\epsilon_3$ is replaced by $k_3$, $\mathfrak{h}_{n,3}$ becomes a polynomial of $\epsilon_1, \epsilon_2$. Since  $(\epsilon_1\cdot \epsilon_2)$ and $(\epsilon_1\cdot k_i)(\epsilon_2\cdot k_j)$'s are all independent in (\ref{proof-prop2-hn3}), their coefficients should be zero by the condition $\mathfrak{h}_{n,3}|_{\epsilon_3\to k_3}=0$. Thus we get
\begin{eqnarray}
&& \sum_{l=1\atop l\neq 3}^{n-1}\alpha^{ijl}(k_3\cdot k_l)=0~\forall (i\neq 3,j\neq 3)~~,~~ \sum_{l=1\atop l\neq 3}^{n-1}\alpha^{i3l}(k_3\cdot k_l)+\beta_1^i=0~\forall (i\neq 3)~,~~~\\
&& \sum_{l=1\atop l\neq 3}^{n-1}\alpha^{3jl}(k_3\cdot k_l)+\beta_2^{j}=0~\forall (j\neq 3)~~,~~\sum_{l=1\atop l\neq 3}^{n-1}\alpha^{33l}(k_3\cdot k_l)+\beta_1^3+\beta_2^3=0~~,~~\sum_{l=1\atop l\neq 3}^{n-1} \beta_3^l(k_3\cdot k_l)=0~.~~~
\end{eqnarray}
From above equations we solve $\beta_3^{n-1}$ and $\alpha^{ij(n-1)}~\forall(i,j)$. Substitute solutions back to (\ref{proof-prop2-hn3}), we get
\begin{eqnarray}
\mathfrak{h}_{n,3}&=&\sum_{i,j=1\atop i\neq 1,j\neq 2}^{n-1} \sum_{l=1\atop l\neq 3}^{n-2} \alpha^{ijl}(\epsilon_1\cdot k_i )(\epsilon_2\cdot k_j)\frac{k_{n-1}\cdot f_3\cdot k_l}{k_3\cdot k_{n-1}}\label{proof-prop2-hn3-2}\\[-1em]
&&~~~~~~~~~~+\sum_{i=1\atop i\neq 1}^{n-1}\beta_1^i(\epsilon_1\cdot k_i)\frac{k_{n-1}\cdot f_3\cdot \epsilon_2}{k_3\cdot k_{n-1}}+\sum_{j=1\atop j\neq 2}^{n-1}\beta_2^j(\epsilon_2\cdot k_j)\frac{k_{n-1}\cdot f_3\cdot \epsilon_1}{k_3\cdot k_{n-1}}+\sum_{l=1\atop l\neq 3}^{n-2}\beta_3^l(\epsilon_1\cdot \epsilon_2)\frac{k_{n-1}\cdot f_3\cdot k_l}{k_3\cdot k_{n-1}}~.~~~\nonumber
\end{eqnarray}
Now we impose the condition $\mathfrak{h}_{n,3}\in \mathrm{Ker}~\mathcal{G}_1+\mathrm{Ker}~\mathcal{G}_2=\mathrm{Ker}~\mathcal{G}_1\mathcal{G}_2$ by acting $\mathcal{G}_1\mathcal{G}_2$ on (\ref{proof-prop2-hn3-2}) to get $\mathfrak{h}_{n,3}|_{\epsilon_1\to k_1\atop \epsilon_2\to k_2}=0$. After $\epsilon_1,\epsilon_2$ being replaced, the remaining polarization $\epsilon_3$ appears as $(k_{n-1}\cdot f_3\cdot k_l)~\forall(l\neq 3, n-1,n)$ in the resulting expression, and all of them are independent. The condition $\mathfrak{h}_{n,3}|_{\epsilon_1\to k_1\atop \epsilon_2\to k_2}=0$ indicates that their coefficients should be zero, resulting to the following equations
\begin{eqnarray}
&&\sum_{i,j=1\atop i\neq 1,j\neq 2}^{n-1} \alpha^{ij1}(k_1\cdot k_i)(k_2\cdot k_j)+\sum_{j=1\atop j\neq 2}^{n-1}\beta_2^j(k_2\cdot k_j)+\beta_3^1(k_1\cdot k_2)=0~,~~~\\
&&\sum_{i,j=1\atop i\neq 1,j\neq 2}^{n-1} \alpha^{ij2}(k_1\cdot k_i)(k_2\cdot k_j)+\sum_{i=1\atop i\neq 1}^{n-1}\beta_1^i(k_1\cdot k_i)+\beta_3^2(k_1\cdot k_2)=0~,~~~\\
&& \sum_{i,j=1\atop i\neq 1,j\neq 2}^{n-1} \alpha^{ijl}(k_1\cdot k_i)(k_2\cdot k_j)+\beta_3^l(k_1\cdot k_2)=0~~\forall(l=4,\ldots,n-2)~.~~~
\end{eqnarray}
From above equations we can solve all $\beta_3^l, l\neq 3, n-1,n$. After substituting  solutions back to (\ref{proof-prop2-hn3-2}) and reorganizing terms, we get
\begin{equation} \mathfrak{h}_{n,3}=v_1+v_2~,~~~\end{equation}
where
\begin{eqnarray}
v_1&=&\sum_{i,j=1\atop i\neq 1,j\neq 2}^{n-1} \sum_{l=1\atop l\neq 3}^{n-2}\alpha^{ijl}\frac{(k_2\cdot k_j)}{(k_3\cdot k_{n-1})(k_1\cdot k_2)}(\epsilon_2\cdot f_1\cdot k_i)(k_{n-1}\cdot f_3\cdot k_l)\\[-1em]
&&~~~~~~~~~~~~+\sum_{i=1\atop i\neq 1}^{n-1}\beta_1^i\frac{(\epsilon_2\cdot f_1\cdot k_i)(k_{n-1}\cdot f_3\cdot k_2)}{(k_3\cdot k_{n-1})(k_1\cdot k_2)}-\sum_{j=1\atop j\neq 2}^{n-1}\beta_2^j\frac{(\epsilon_2\cdot k_j)}{(k_3\cdot k_{n-1})(k_1\cdot k_2)}(k_{n-1}\cdot f_3\cdot f_1\cdot k_2)~,~~~\nonumber
\end{eqnarray}
and
\begin{eqnarray}
v_2&=&\sum_{i,j=1\atop i\neq 1,j\neq 2}^{n-1} \sum_{l=1\atop l\neq 3}^{n-2}\alpha^{ijl}\frac{(\epsilon_1\cdot k_i)}{(k_3\cdot k_{n-1})(k_1\cdot k_2)}(k_1\cdot f_2\cdot k_j)(k_{n-1}\cdot f_3\cdot k_l)\\[-1em]
&&~~~~~~~~~~~~-\sum_{i=1\atop i\neq 1}^{n-1}\beta_1^i\frac{(\epsilon_1\cdot k_i)}{(k_3\cdot k_{n-1})(k_1\cdot k_2)}(k_{n-1}\cdot f_3\cdot f_2\cdot k_1)+\sum_{j=1\atop j\neq 2}^{n-1}\beta_2^j\frac{(\epsilon_1\cdot f_2\cdot k_j)(k_{n-1}\cdot f_3\cdot k_1)}{(k_3\cdot k_{n-1})(k_1\cdot k_2)}~.~~~\nonumber
\end{eqnarray}
Since $f_i$ is gauge invariant under $\mathcal{G}_i$, it is simple to see that $\mathcal{G}_1v_1=\mathcal{G}_3v_1=0$ and $\mathcal{G}_2v_2=\mathcal{G}_3v_2=0$, hence $v_1\in \mathrm{Ker}~\mathcal{G}_1\cap \mathrm{Ker}~\mathcal{G}_3$ and $v_2\in \mathrm{Ker}~\mathcal{G}_2\cap \mathrm{Ker}~\mathcal{G}_3$. Thus for $m=3$ we have shown  the relation (\ref{proof-prop2-2}).

Now let us assume that for all vector spaces $\mathcal{V}_{n,s}, s<m$  relation (\ref{proof-prop2-2}) is always true. For a generic vector $\mathfrak{h}_{n,m}\in \mathcal{V}_{n,m}$ with the form (\ref{proof-prop1-hnm}), we impose the condition
${\cal G}_3 \mathfrak{h}_{n,m}=0$ and ${\cal G}_{12} \mathfrak{h}_{n,m}=0$,
\begin{equation}
\mathfrak{h}_{n,m}|_{\epsilon_3\to k_3}=0~~~,~~~\mathfrak{h}_{n,m}|_{\epsilon_1\to k_1\atop \epsilon_2\to k_2}=0~.~~~
\end{equation}
Considering the independent Lorentz invariant product of polarizations and momenta, we get the following identities
\begin{eqnarray}
&& T_{mi}^{(3)}=0~~~,~~~T_{mi}^{(12)}=0~~\forall(i=4,\ldots,m-1)~,~~~\label{proof-prop2-hnm-part1}\\
&& \epsilon_i\cdot T_{mi}^{'(3)}=0~~~,~~~\epsilon_i\cdot T_{mi}^{'(12)}=0~~\forall(i=4,\ldots,m-1)~,~~~\label{proof-prop2-hnm-part2}\\
&& T_{mi}^{''(3)}=0~~~,~~~T_{mi}^{''(12)}=0~~\forall(i=m+1,\ldots,n-1)~,~~~\label{proof-prop2-hnm-part3}
\end{eqnarray}
as well as
\begin{eqnarray}
&&T_{m1}^{(3)}=0~~~,~~~T_{m2}^{(3)}=0~~~,~~~\epsilon_1\cdot T_{m1}^{'(3)}=0~~~,~~~\epsilon_2\cdot T_{m2}^{'(3)}=0~~~,~~~T_{m3}+k_{3}\cdot T'_{m3}=0~,~~~\label{proof-prop2-hnm-part4}\\
&& T_{m1}^{(2)}+k_1\cdot T_{m1}^{'(2)}=0~~~,~~~T_{m2}^{(1)}+k_2\cdot T_{m2}^{'(1)}=0~~~,~~~T_{m3}^{(12)}=0~~~,~~~\epsilon_3\cdot T_{m3}^{'(12)}=0~.~~~\label{proof-prop2-hnm-part5}
\end{eqnarray}
From results (\ref{proof-prop2-hnm-part1}), (\ref{proof-prop2-hnm-part2}) and (\ref{proof-prop2-hnm-part3}) we immediately know
\begin{equation}
T_{mi}~,~\epsilon_i\cdot T'_{mi}~\forall(i=4,\ldots,m-1)~~,~~T''_{mi}~\forall(i=m+1,\ldots,n-1)~\in  (\mathrm{Ker}~\mathcal{G}_1+\mathrm{Ker}~\mathcal{G}_2)\cap \mathrm{Ker}~\mathcal{G}_3~.~~~\nonumber
\end{equation}
Since $T_{mi} \in \mathcal{V}_{n,m-2}$, $\epsilon_i\cdot T'_{mi}~,~T''_{mi} \in \mathcal{V}_{n,m-1}$, by assumption they satisfy (\ref{proof-prop2-2}). Now we consider the remaining terms in  (\ref{proof-prop1-hnm}), which after reorganization of terms we get\footnote{In the reorganization, there is some freedom to put certain term in either part, so the manifest symmetry between $1\leftrightarrow 2$ is lost.}
\begin{equation}
\mathfrak{h}'_{n,m}=\sum_{i=1,2,3}({\mathfrak{h}}_{n,m}^{1i}+{\mathfrak{h}}_{n,m}^{2i})~~~\mbox{where}~~~{\mathfrak{h}}_{n,m}^{1i}:=(\epsilon_m\cdot \epsilon_i)(T_{mi}+k_i\cdot T'_{mi})~~,~~{\mathfrak{h}}_{n,m}^{2i}:=\epsilon_m\cdot f_i\cdot T'_{mi}~.~~~
\end{equation}
From (\ref{proof-prop2-hnm-part4}) we see that for $i=1,2$ we have $\epsilon_i\cdot T_{mi}^{'(3)}=0$, which means either the Lorentz vector\footnote{The Lorentz index of $(T_{mi}^{'(3)})$ can only be carried by $\eps_i, k_i$ in the construction, especially when the total symmetric tensor $\eps_{\mu_1...\mu_D}$ does not appear.} $(T_{mi}^{'(3)})^\mu=0$ or $(T_{mi}^{'(3)})^\mu\sim k_i^{\mu}$. However in either case we could infer $k_i\cdot T_{mi}^{'(3)}=0$ for $i=1,2$ for massless particles. Similarly, $k_3\cdot T_{m3}^{'(12)}=0$. Combined with results (\ref{proof-prop2-hnm-part4}), (\ref{proof-prop2-hnm-part5}) we can directly check that
\begin{eqnarray}&&\mathcal{G}_2~ \mathfrak{h}_{n,m}^{11}=\mathcal{G}_3~ \mathfrak{h}_{n,m}^{11}=0~~,~~~\mathcal{G}_1~ \mathfrak{h}_{n,m}^{12}=\mathcal{G}_3~\mathfrak{h}_{n,m}^{12}=0~~~,~~~\mathfrak{h}_{n,m}^{13}=0~,~~~\\
&& \mathcal{G}_1~\mathfrak{h}_{n,m}^{21}=\mathcal{G}_3~ \mathfrak{h}_{n,m}^{21}=0~~~,~~~\mathcal{G}_2~ \mathfrak{h}_{n,m}^{22}=\mathcal{G}_3 ~\mathfrak{h}_{n,m}^{22}=0~~~,~~~\mathcal{G}_1~ \mathfrak{h}_{n,m}^{23}=\mathcal{G}_2 ~\mathfrak{h}_{n,m}^{23}=\mathcal{G}_3 ~ \mathfrak{h}_{n,m}^{23}=0~.~~~
\end{eqnarray}
Hence if  we reorganize $\mathfrak{h}_{n,m}'$ as
\begin{equation}
\mathfrak{h}'_{n,m}=\Big(\mathfrak{h}_{n,m}^{12}+\mathfrak{h}_{n,m}^{21}+\frac{1}{2}\mathfrak{h}_{n,m}^{23}\Big)+\Big(\mathfrak{h}_{n,m}^{11}+\mathfrak{h}_{n,m}^{22}+\frac{1}{2}\mathfrak{h}_{n,m}^{23}\Big)~,~~~
\end{equation}
expression in the first bracket belongs to $\mathrm{Ker}~\mathcal{G}_1\cap \mathrm{Ker}~\mathcal{G}_3$ and that in the second bracket belongs to $\mathrm{Ker}~\mathcal{G}_2\cap \mathrm{Ker}~\mathcal{G}_3$. Thus we have successfully separated $\mathfrak{h}_{n,m}$ to two parts satisfying (\ref{proof-prop2-2}) in general vector space $\mathcal{V}_{n,m}$, and proposition 2 is proven.

For completeness let us present the proof of  \eref{gen-m-I} and \eref{gen-m-II} as follows,
{\small
\begin{eqnarray}
 &&\text{dim}(U_1+\cdots+U_m)
 = \text{dim}(U_1+\cdots+U_{m-1})+\text{dim}U_m-\text{dim}((U_1+\cdots+U_{m-1})\cap U_m) \nonumber\\
 &&= \sum_{s=1}^{m-1}\sum_{i_1<\cdots<i_{s}}(-1)^{s-1} \text{dim}(U_{i_1}\cap \cdots\cap U_{i_s})+\text{dim}U_m-\text{dim}(U_1\cap U_m+\cdots+U_{m-1}\cap U_m) \nonumber\\
 &&= \sum_{s=1}^{m-1}\sum_{i_1<\cdots<i_{s}}(-1)^{s-1} \text{dim}(U_{i_1}\cap \cdots\cap U_{i_s})+\text{dim}U_m-\sum_{s=1}^{m-1}\sum_{i_1<\cdots<i_{s}}(-1)^{s-1} \text{dim}((U_{i_1}\cap U_m)\cap \cdots\cap (U_{i_s}\cap U_m)) \nonumber\\
 &&= \sum_{s=1}^{m-1}\sum_{i_1<\cdots<i_{s}}(-1)^{s-1} \text{dim}(U_{i_1}\cap \cdots\cap U_{i_s})+\text{dim}U_m  +\sum_{s=1}^{m-1}\sum_{i_1<\cdots<i_{s}}(-1)^{s} \text{dim}(U_{i_1}\cap \cdots\cap U_{i_s}\cap U_m) \nonumber\\
 &&= \sum_{s=1}^{m}\sum_{i_1<\cdots<i_{s}}(-1)^{s-1} \text{dim}(U_{i_1}\cap \cdots\cap U_{i_s})~,~~~\nonumber
\end{eqnarray}}
and
{\small
\begin{eqnarray}
 &&\text{dim}(U_1\cap\cdots \cap U_{m})
 = \text{dim}(U_1\cap\cdots \cap U_{m-1})+\text{dim}U_{m}-\text{dim}(U_1\cap\cdots \cap U_{m-1}+ U_{m}) \nonumber\\
 && = \sum_{s=1}^{m-1}\sum_{i_1<\cdots<i_{s}}(-1)^{s-1} \text{dim}(U_{i_1}+\cdots+U_{i_s})+\text{dim}U_{m} -\text{dim}((U_1+ U_{m})\cap\cdots \cap (U_{m-1}+ U_{m})) \nonumber\\
 &&= \sum_{s=1}^{m-1}\sum_{i_1<\cdots<i_{s}}(-1)^{s-1} \text{dim}(U_{i_1}+\cdots+U_{i_s})+\text{dim}U_{m}  +\sum_{s=1}^{m-1}\sum_{i_1<\cdots<i_{h}}(-1)^{s} \text{dim}(U_{i_1}+\cdots+U_{i_h}+U_{m}) \nonumber\\
 &&=\sum_{s=1}^{m}\sum_{i_1<\cdots<i_{s}}(-1)^{s-1} \text{dim}(U_{i_1}+\cdots+U_{i_s})~.~~~\nonumber %\label{inverse-m}
\end{eqnarray}}
%
%In the above deduction, we have used the distribution formula.

%%%%%%%%%%%%%%%%%%%%%%
\section{Explicit BCJ coefficients}
\label{sec-appen-BCJ}
%%%%%%%%%%%%%%%%%%%%%%

In this appendix, we provide some explanation for notations in (\ref{BCJ-exp-2}). For convenience we also collect some explicit BCJ coefficients which are used in the computation. In formula (\ref{BCJ-exp-2}), we have
\bea {\cal F}_{\b_k}(\{\a\},\{\b\};\{\xi\})& = &
\theta(\xi_{\b_{k}}-\xi_{k-1})\left\{k_{\b_k}\cdot
W^{(R,R)}_{\b_k}+\theta(\xi_{\b_{k+1}}-\xi_{\b_k}) {\cal
K}_{1\b_1...\b_k} \right\}\nn & & +
\theta(\xi_{\b_{k-1}}-\xi_{k})\left\{-k_{\b_k}\cdot
(W^{(L,R)}_{\b_k}-k_1)-\theta(\xi_{\b_{k}}-\xi_{\b_{k+1}}) {\cal
K}_{1\b_1...\b_k}\right\} ~,~~~\label{BCJ-exp-3}\eea
with $\theta(x)=1$ when $x>0$ and otherwise $\theta(x)=0$. Some notations are defined as follows. The shuffle permutation $\shuffle$ of two lists is a summation of lists, which can be obtained recursively as
\bea & & \a\shuffle \emptyset =\a~~~,~~~\emptyset \shuffle \b=\b~,~~~\nn
& &  \{\a_1,...,\a_m\}\shuffle\{\b_1,...,\b_k\} =\{\a_1,
\{\a_2,...,\a_m\}\shuffle \b\}+\{\b_1,
\a\shuffle\{\b_2,...,\b_k\}\}~.~~~\label{shuffle}\eea
The  ${\cal K}$ is defined as
\bea {\cal K}_{\a}=\sum_{i< j;i,j\in \a} k_i\cdot
k_j~.~~\label{K-contract}\eea
Definition of $W$ needs further explanations. Given two ordered sets
$\Xi=\{\xi_1,\xi_2,...,\xi_n\}$ and  $\b=\{\b_1,...,\b_r\}$ where
set $\b$ is a subset of $\Xi$, for a given element $p\in \Xi$
with its position $K$ in $\Xi$, {\sl i.e.}, $\xi_K=p$, we define
\bea X_p= \sum_{i=1}^{K-1} k_{\xi_i}~~~,~~~
Y_p=\sum_{i=1, \xi_i\not\in \b}^{K-1}
k_{\xi_i}~.~~~\label{XpYp-def}\eea
Furthermore, since $p$ has split
set $\b$ into two subsets $\b_{p}^{L}$ and $\b_{p}^{R}$, {\sl i.e.}, the
collections of elements on the LHS and RHS of $p$ respectively,  we can define
\begin{eqnarray}
\begin{array}{l}
W^{(L,L)}_{p}  =  \sum_{i=1, \xi\not\in \b_{p}^{R}}^{K-1}
k_{\xi_i}~~~,~~~
 W^{(L,R)}_{p}  =  \sum_{i=1, \xi\not\in \b_{p}^{L}}^{K-1}
k_{\xi_i}~,~~~\\
 W^{(R,L)}_{p} = \sum_{i=K+1, \xi\not\in \b_{p}^{R}}^{n}
k_{\xi_i}~~~,~~~
 W^{(R,R)}_{p}  =  \sum_{i=K+1, \xi\not\in \b_{p}^{L}}^{n}
k_{\xi_i}~.~~~
\end{array}~~~~\label{W-def}
\end{eqnarray}
%
%
%\bea W^{(L,L)}_{p}  =  \sum_{i=1, \xi\not\in \b_{p}^{R}}^{K-1}
%k_{\xi_i}~,~
% W^{(L,R)}_{p}  =  \sum_{i=1, \xi\not\in \b_{p}^{L}}^{K-1}
%k_{\xi_i},~
% W^{(R,L)}_{p} = \sum_{i=K+1, \xi\not\in \b_{p}^{R}}^{n}
%k_{\xi_i},~
% W^{(R,R)}_{p} & = & \sum_{i=K+1, \xi\not\in \b_{p}^{L}}^{n}
%k_{\xi_i}.\Label{W-def}
%\eea
%

Next we provide some examples. We consider the BCJ basis with legs $1, 2$ being fixed in the first two positions and leg $n$ in the last position in the color-ordering. For an arbitrary amplitude with one or two gluons inserted between legs $1,2$, we have the BCJ relations
\bea A_{n+1}^{\rm YM}(1,p,\{2,\ldots,n-1\},n)={-(k_p\cdot X_p)\over
(k_p\cdot k_1)}~ A_{n+1}^{\rm
YM}(1,2,\{3,\ldots,n-1\}\shuffle\{p\},n)~,~~~\label{Fund-BCJ-deform-1}\eea
and
\bea A^{\rm YM}_{n+2}(1,p,q, \{2,...,n-1\},n)
& = &  {(k_p\cdot k_1+k_q\cdot (Y_q+k_p))(k_p\cdot (Y_p+k_q))\over
{\cal K}_{1pq} {\cal K}_{1p}}  ~A_{n+1}^{\rm
YM}(1,2,\{3,\ldots,n-1\}\shuffle\{q,p\},n)\nn
& & +{(k_p\cdot (Y_p- k_1))(k_q\cdot (Y_q+k_p))\over {\cal K}_{1pq}
{\cal K}_{1p}}~A^{\rm YM}_{n+2}(1,2,
\{3,...,n-1\}\shuffle\{p,q\},n)~.~~~\label{G-BCJ-b=2-3}\eea
For amplitude with three gluons between legs $1,2$ we have
\bea
A^{\rm YM}_{n+2}(1,p,q,r, \{2,...,n-1\},n)= \sum_{\rho\in S_3} {\cal C}[\{p,q,r\}; \rho\{p,q,r\}] A^{\rm
YM}_{n+3}(1,
2,\{3,...,n-1\}\shuffle\{\rho\{p,q,r\}\},n)~,~~~\label{3legs-BCJ}\eea
where
\begin{eqnarray}
\begin{array}{l}
{\cal C}[\{p,q,r\};\{p,q,r\}] =   {  -(k_p\cdot ( Y_p-
k_1))\over {\cal K}_{1p}}\times {({\cal K}_{1pq}-k_q\cdot X_q)\over
{\cal K}_{1pq}}\times {-(k_r\cdot X_r)\over {\cal K}_{1pqr}}~,\\
{\cal C}[\{p,q,r\};\{p,r,q\} ]  =   {- (k_p\cdot ( Y_p-
k_1))\over {\cal K}_{1p}} \times { -(k_q\cdot X_q)\over {\cal
K}_{1pq}} \times{ -k_r\cdot (Y_r-k_1)-{\cal K}_{1pqr} \over {\cal
K}_{1pqr}}~, \\
{\cal C}[\{p,q,r\};\{q,p,r\} ]  =  { -k_p\cdot X_p\over {\cal
K}_{1p}} \times { -(k_q\cdot (Y_q-k_1))\over {\cal K}_{1pq}} \times{
-k_r\cdot X_r\over {\cal K}_{1pqr}}~, \\
{\cal C}[\{p,q,r\};\{q,r,p\} ] =   { -k_p\cdot X_p\over {\cal
K}_{1p}} \times { -(k_q\cdot (Y_q-k_1))\over {\cal K}_{1pq}} \times{
-k_r\cdot (X_r+k_p)\over {\cal K}_{1pqr}}~, \\
{\cal C}[\{p,q,r\};\{r,p,q\} ] =   { -(k_p\cdot (X_p-k_1))\over
{\cal K}_{1p}} \times {-k_q\cdot X_q \over {\cal K}_{1pq}} \times{
-k_r\cdot (Y_r-k_1)-{\cal K}_{1pqr}\over {\cal K}_{1pqr}} ~,\\
{\cal C}[\{p,q,r\};\{r,q,p\} ] =   { -k_p\cdot X_p\over {\cal
K}_{1p}} \times { -k_q\cdot (X_q-k_1)-{\cal K}_{1pq} \over {\cal
K}_{1pq}} \times{ -k_r\cdot (Y_r-k_1)-{\cal K}_{1pqr}\over {\cal
K}_{1pqr}} ~.
\end{array}\label{BCJ-pqr-collect}
\end{eqnarray}
%
\iffalse
where
%
\bea {\cal C}[\{p,q,r\};\{p,q,r\}] &= &  {  -(k_p\cdot ( Y_p-
k_1))\over {\cal K}_{1p}}\times {({\cal K}_{1pq}-k_q\cdot X_q)\over
{\cal K}_{1pq}}\times {-(k_r\cdot X_r)\over {\cal K}_{1pqr}}\nn
%
{\cal C}[\{p,q,r\};\{p,r,q\} ] & = &   {- (k_p\cdot ( Y_p-
k_1))\over {\cal K}_{1p}} \times { -(k_q\cdot X_q)\over {\cal
K}_{1pq}} \times{ -k_r\cdot (Y_r-k_1)-{\cal K}_{1pqr} \over {\cal
K}_{1pqr}} \nn
%
{\cal C}[\{p,q,r\};\{q,p,r\} ] & = &  { -k_p\cdot X_p\over {\cal
K}_{1p}} \times { -(k_q\cdot (Y_q-k_1))\over {\cal K}_{1pq}} \times{
-k_r\cdot X_r\over {\cal K}_{1pqr}} \nn
%
{\cal C}[\{p,q,r\};\{q,r,p\} ]& = &  { -k_p\cdot X_p\over {\cal
K}_{1p}} \times { -(k_q\cdot (Y_q-k_1))\over {\cal K}_{1pq}} \times{
-k_r\cdot (X_r+k_p)\over {\cal K}_{1pqr}} \nn
%
{\cal C}[\{p,q,r\};\{r,p,q\} ]& = &  { -(k_p\cdot (X_p-k_1))\over
{\cal K}_{1p}} \times {-k_q\cdot X_q \over {\cal K}_{1pq}} \times{
-k_r\cdot (Y_r-k_1)-{\cal K}_{1pqr}\over {\cal K}_{1pqr}} \nn
%
{\cal C}[\{p,q,r\};\{r,q,p\} ]& = &  { -k_p\cdot X_p\over {\cal
K}_{1p}} \times { -k_q\cdot (X_q-k_1)-{\cal K}_{1pq} \over {\cal
K}_{1pq}} \times{ -k_r\cdot (Y_r-k_1)-{\cal K}_{1pqr}\over {\cal
K}_{1pqr}} ~~~~~~~\label{BCJ-pqr-collect}\eea
%
\fi
For amplitude with four gluons between legs $1,2$ we have
\begin{eqnarray}
&&A_{n+4}(1,h_1,h_2,h_3,h_4,2,\cdots,n) \notag\\
&&= \sum_{\shuffle}\sum_{\mathcal{P}} \mathcal{C}_\shuffle(\{h_1,\cdots,h_4\}|\mathcal{P}\{h_1,\cdots,h_4\}) A_n(1,2,\{3,\cdots,n-1\}\shuffle \mathcal{P}\{h_1,\cdots,h_4\},n)~.~~~\label{BCJ-4g}
\end{eqnarray}
with coefficients(For simplicity we ignored the first list  $\{h_1,\cdots,h_4\}$ and $\shuffle$)
\begin{eqnarray}
\begin{array}{l}
\mathcal{C}(\{h_1,h_2,h_3,h_4\})
= \frac{[k_{h_1}\cdot (X_{h_1}-k_1)]}{K_{1h_1}} \frac{[(k_{h_2}\cdot X_{h_2})-K_{1h_1h_2}]}{K_{1h_1h_2}} \frac{[(k_{h_3}\cdot X_{h_3})-K_{1h_1h_2h_3}]}{K_{1h_1h_2h_3}} \frac{(k_{h_4}\cdot X_{h_4})}{K_{1h_1h_2h_3h_4}}~,\\
\mathcal{C}(\{h_1,h_2,h_4,h_3\})
= \frac{[k_{h_1}\cdot(X_{h_1}-k_1)]}{K_{1h_1}} \frac{[(k_{h_2}\cdot X_{h_2})-K_{1h_1h_2}]}{K_{1h_1h_2}} \frac{(k_{h_3}\cdot X_{h_3})}{K_{1h_1h_2h_3}} \frac{[k_{h_4}\cdot (Y_{h_4}-k_1)]+K_{1h_1h_2h_3h_4}}{K_{1h_1h_2h_3h_4}}~,\\
\mathcal{C}(\{h_1,h_3,h_2,h_4\})
= \frac{[k_{h_1}\cdot(X_{h_1}-k_1)]}{K_{1h_1}} \frac{(k_{h_2}\cdot X_{h_2})}{K_{1h_1h_2}} \frac{[k_{h_3}\cdot (Y_{h_3}-k_1)]}{K_{1h_1h_2h_3}} \frac{(k_{h_4}\cdot X_{h_4})}{K_{1h_1h_2h_3h_4}}~, \\
\mathcal{C}(\{h_1,h_3,h_4,h_2\})
= \frac{[k_{h_1}\cdot(X_{h_1}-k_1)]}{K_{1h_1}} \frac{(k_{h_2}\cdot X_{h_2})}{K_{1h_1h_2}} \frac{[k_{h_3}\cdot( Y_{h_3}-k_1)}{K_{1h_1h_2h_3}} \frac{[k_{h_4}\cdot (X_{h_4}+k_{h_2})]}{K_{1h_1h_2h_3h_4}}~,\\
\mathcal{C}(\{h_1,h_4,h_2,h_3\})
= \frac{[k_{h_1}\cdot(X_{h_1}-k_1)]}{K_{1h_1}} \frac{[(k_{h_2}\cdot X_{h_2})-K_{1h_1h_2}]}{K_{1h_1h_2}} \frac{(k_{h_3}\cdot X_{h_3})}{K_{1h_1h_2h_3}} \frac{[k_{h_4}\cdot (Y_{h_4}-k_1)]+K_{1h_1h_2h_3h_4}}{K_{1h_1h_2h_3h_4}}~,\\
\mathcal{C}(\{h_1,h_4,h_3,h_2\})
= \frac{[k_{h_1}\cdot(X_{h_1}-k_1)]}{K_{1h_1}} \frac{(k_{h_2}\cdot X_{h_2})}{K_{1h_1h_2}} \frac{[k_{h_3}\cdot (X_{h_3}-k_{h_1}-k_1)]+K_{1h_1h_2h_3}}{K_{1h_1h_2h_3}}\frac{[k_{h_4}\cdot (Y_{h_4}-k_1)]+K_{1h_1h_2h_3h_4}}{K_{1h_1h_2h_3h_4}}~,
\end{array}
\end{eqnarray}
%
\iffalse
%
\begin{align}
\mathcal{C}(\{h_1,h_2,h_3,h_4\})
=& \frac{[k_{h_1}\cdot (X_{h_1}-k_1)]}{K_{1h_1}} \frac{[(k_{h_2}\cdot X_{h_2})-K_{1h_1h_2}]}{K_{1h_1h_2}} \frac{[(k_{h_3}\cdot X_{h_3})-K_{1h_1h_2h_3}]}{K_{1h_1h_2h_3}} \frac{(k_{h_4}\cdot X_{h_4})}{K_{1h_1h_2h_3h_4}}, \notag\\
\mathcal{C}(\{h_1,h_2,h_4,h_3\})
=& \frac{[k_{h_1}\cdot(X_{h_1}-k_1)]}{K_{1h_1}} \frac{[(k_{h_2}\cdot X_{h_2})-K_{1h_1h_2}]}{K_{1h_1h_2}} \frac{(k_{h_3}\cdot X_{h_3})}{K_{1h_1h_2h_3}} \frac{[k_{h_4}\cdot (Y_{h_4}-k_1)]+K_{1h_1h_2h_3h_4}}{K_{1h_1h_2h_3h_4}}, \notag\\
\mathcal{C}(\{h_1,h_3,h_2,h_4\})
=& \frac{[k_{h_1}\cdot(X_{h_1}-k_1)]}{K_{1h_1}} \frac{(k_{h_2}\cdot X_{h_2})}{K_{1h_1h_2}} \frac{[k_{h_3}\cdot (Y_{h_3}-k_1)]}{K_{1h_1h_2h_3}} \frac{(k_{h_4}\cdot X_{h_4})}{K_{1h_1h_2h_3h_4}}, \notag\\
\mathcal{C}(\{h_1,h_3,h_4,h_2\})
=& \frac{[k_{h_1}\cdot(X_{h_1}-k_1)]}{K_{1h_1}} \frac{(k_{h_2}\cdot X_{h_2})}{K_{1h_1h_2}} \frac{[k_{h_3}\cdot( Y_{h_3}-k_1)}{K_{1h_1h_2h_3}} \frac{[k_{h_4}\cdot (X_{h_4}+k_{h_2})]}{K_{1h_1h_2h_3h_4}}, \notag\\
\mathcal{C}(\{h_1,h_4,h_2,h_3\})
=& \frac{[k_{h_1}\cdot(X_{h_1}-k_1)]}{K_{1h_1}} \frac{[(k_{h_2}\cdot X_{h_2})-K_{1h_1h_2}]}{K_{1h_1h_2}} \frac{(k_{h_3}\cdot X_{h_3})}{K_{1h_1h_2h_3}} \frac{[k_{h_4}\cdot (Y_{h_4}-k_1)]+K_{1h_1h_2h_3h_4}}{K_{1h_1h_2h_3h_4}}, \notag\\
\mathcal{C}(\{h_1,h_4,h_3,h_2\})
=& \frac{[k_{h_1}\cdot(X_{h_1}-k_1)]}{K_{1h_1}} \frac{(k_{h_2}\cdot X_{h_2})}{K_{1h_1h_2}} \frac{[k_{h_3}\cdot (X_{h_3}-k_{h_1}-k_1)]+K_{1h_1h_2h_3}}{K_{1h_1h_2h_3}}\notag\\
& \frac{[k_{h_4}\cdot (Y_{h_4}-k_1)]+K_{1h_1h_2h_3h_4}}{K_{1h_1h_2h_3h_4}},
\end{align}
\fi
%
\begin{eqnarray}
\begin{array}{l}
\mathcal{C}(\{h_2,h_1,h_3,h_4\})
= \frac{(k_{h_1}\cdot X_{h_1})}{K_{1h_1}} \frac{[k_{h_2}\cdot (X_{h_2}-k_1)]}{K_{1h_1h_2}} \frac{(k_{h_3}\cdot X_{h_3})-K_{1h_1h_2h_3}}{K_{1h_1h_2h_3}} \frac{(k_{h_4}\cdot X_{h_4})}{K_{1h_1h_2h_3h_4}}~,\\
\mathcal{C}(\{h_2,h_1,h_4,h_3\})
= \frac{(k_{h_1}\cdot X_{h_1})}{K_{1h_1}} \frac{[k_{h_2}\cdot (X_{h_2}-k_1)]}{K_{1h_1h_2}} \frac{(k_{h_3}\cdot X_{h_3})}{K_{1h_1h_2h_3}} \frac{[k_{h_4}\cdot (Y_{h_4}-k_1)]+K_{1h_1h_2h_3h_4}}{K_{1h_1h_2h_3h_4}}~,\\
\mathcal{C}(\{h_2,h_3,h_1,h_4\})
= \frac{(k_{h_1}\cdot X_{h_1})}{K_{1h_1}} \frac{[k_{h_2}\cdot (X_{h_2}-k_1)]}{K_{1h_1h_2}} \frac{[k_{h_3}\cdot (X_{h_3}+k_{h_1})]-K_{1h_1h_2h_3}}{K_{1h_1h_2h_3}} \frac{(k_{h_4}\cdot X_{h_4})}{K_{1h_1h_2h_3h_4}}~,\\
\mathcal{C}(\{h_2,h_3,h_4,h_1\})
= \frac{(k_{h_1}\cdot X_{h_1})}{K_{1h_1}} \frac{[k_{h_2}\cdot (X_{h_2}-k_1)]}{K_{1h_1h_2}} \frac{[k_{h_3}\cdot (X_{h_3}+k_{h_1})]-K_{1h_1h_2h_3}}{K_{1h_1h_2h_3}} \frac{[k_{h_4}\cdot (X_{h_4}+k_{h_1})]}{K_{1h_1h_2h_3h_4}}~, \\
\mathcal{C}(\{h_2,h_4,h_3,h_1\})
=\frac{(k_{h_1}\cdot X_{h_1})}{K_{1h_1}} \frac{[k_{h_2}\cdot (X_{h_2}-k_1)]}{K_{1h_1h_2}} \frac{[k_{h_3}\cdot (X_{h_3}+k_{h_1})]}{K_{1h_1h_2h_3}} \frac{[k_{h_4}\cdot (Y_{h_4}-k_1)]+K_{1h_1h_2h_3h_4}}{K_{1h_1h_2h_3h_4}}~,\\
\mathcal{C}(\{h_2,h_4,h_1,h_3\})
= \frac{(k_{h_1}\cdot X_{h_1})}{K_{1h_1}} \frac{[k_{h_2}\cdot (X_{h_2}-k_1)]}{K_{1h_1h_2}} \frac{(k_{h_3}\cdot X_{h_3})}{K_{1h_1h_2h_3}} \frac{[k_{h_4}\cdot (Y_{h_4}-k_1)]+K_{1h_1h_2h_3h_4}}{K_{1h_1h_2h_3h_4}}~,
\end{array}
\end{eqnarray}
%
\iffalse
\begin{align}
\mathcal{C}(\{h_2,h_1,h_3,h_4\})
=& \frac{(k_{h_1}\cdot X_{h_1})}{K_{1h_1}} \frac{[k_{h_2}\cdot (X_{h_2}-k_1)]}{K_{1h_1h_2}} \frac{(k_{h_3}\cdot X_{h_3})-K_{1h_1h_2h_3}}{K_{1h_1h_2h_3}} \frac{(k_{h_4}\cdot X_{h_4})}{K_{1h_1h_2h_3h_4}}, \notag\\
\mathcal{C}(\{h_2,h_1,h_4,h_3\})
=& \frac{(k_{h_1}\cdot X_{h_1})}{K_{1h_1}} \frac{[k_{h_2}\cdot (X_{h_2}-k_1)]}{K_{1h_1h_2}} \frac{(k_{h_3}\cdot X_{h_3})}{K_{1h_1h_2h_3}} \frac{[k_{h_4}\cdot (Y_{h_4}-k_1)]+K_{1h_1h_2h_3h_4}}{K_{1h_1h_2h_3h_4}}, \notag\\
\mathcal{C}(\{h_2,h_3,h_1,h_4\})
=& \frac{(k_{h_1}\cdot X_{h_1})}{K_{1h_1}} \frac{[k_{h_2}\cdot (X_{h_2}-k_1)]}{K_{1h_1h_2}} \frac{[k_{h_3}\cdot (X_{h_3}+k_{h_1})]-K_{1h_1h_2h_3}}{K_{1h_1h_2h_3}} \frac{(k_{h_4}\cdot X_{h_4})}{K_{1h_1h_2h_3h_4}}, \notag\\
\mathcal{C}(\{h_2,h_3,h_4,h_1\})
=& \frac{(k_{h_1}\cdot X_{h_1})}{K_{1h_1}} \frac{[k_{h_2}\cdot (X_{h_2}-k_1)]}{K_{1h_1h_2}} \frac{[k_{h_3}\cdot (X_{h_3}+k_{h_1})]-K_{1h_1h_2h_3}}{K_{1h_1h_2h_3}} \frac{[k_{h_4}\cdot (X_{h_4}+k_{h_1})]}{K_{1h_1h_2h_3h_4}}, \notag\\
\mathcal{C}(\{h_2,h_4,h_3,h_1\})
=& \frac{(k_{h_1}\cdot X_{h_1})}{K_{1h_1}} \frac{[k_{h_2}\cdot (X_{h_2}-k_1)]}{K_{1h_1h_2}} \frac{[k_{h_3}\cdot (X_{h_3}+k_{h_1})]}{K_{1h_1h_2h_3}} \frac{[k_{h_4}\cdot (Y_{h_4}-k_1)]+K_{1h_1h_2h_3h_4}}{K_{1h_1h_2h_3h_4}}, \notag\\
\mathcal{C}(\{h_2,h_4,h_1,h_3\})
=& \frac{(k_{h_1}\cdot X_{h_1})}{K_{1h_1}} \frac{[k_{h_2}\cdot (X_{h_2}-k_1)]}{K_{1h_1h_2}} \frac{(k_{h_3}\cdot X_{h_3})}{K_{1h_1h_2h_3}} \frac{[k_{h_4}\cdot (Y_{h_4}-k_1)]+K_{1h_1h_2h_3h_4}}{K_{1h_1h_2h_3h_4}},
\end{align}
\fi
%
\begin{eqnarray}
\begin{array}{l}
\mathcal{C}(\{h_3,h_1,h_2,h_4\})
= \frac{[k_{h_1}\cdot(X_{h_1}-k_1)]}{K_{1h_1}} \frac{(k_{h_2}\cdot X_{h_2})}{K_{1h_1h_2}} \frac{[k_{h_3}\cdot (X_{h_3}-k_1)]}{K_{1h_1h_2h_3}} \frac{(k_{h_4}\cdot X_{h_4})}{K_{1h_1h_2h_3h_4}}~,\\
\mathcal{C}(\{h_3,h_1,h_4,h_2\})
= \frac{[k_{h_1}\cdot(X_{h_1}-k_1)]}{K_{1h_1}} \frac{(k_{h_2}\cdot X_{h_2})}{K_{1h_1h_2}} \frac{[k_{h_3}\cdot (X_{h_3}-k_1)]}{K_{1h_1h_2h_3}} \frac{[k_{h_4}\cdot (X_{h_4}+k_{h_2})]}{K_{1h_1h_2h_3h_4}}~,\\
\mathcal{C}(\{h_3,h_2,h_1,h_4\})
= \frac{(k_{h_1}\cdot X_{h_1})}{K_{1h_1}} \frac{[k_{h_2}\cdot (X_{h_2}-k_1)]+K_{1h_1h_2}}{K_{1h_1h_2}} \frac{[k_{h_3}\cdot (X_{h_3}-k_1)]}{K_{1h_1h_2h_3}} \frac{(k_{h_4}\cdot X_{h_4})}{K_{1h_1h_2h_3h_4}}~,\\
\mathcal{C}(\{h_3,h_2,h_4,h_1\})
= \frac{(k_{h_1}\cdot X_{h_1})}{K_{1h_1}} \frac{[k_{h_2}\cdot (X_{h_2}-k_1)]+K_{1h_1h_2}}{K_{1h_1h_2}} \frac{[k_{h_3}\cdot (X_{h_3}-k_1)]}{K_{1h_1h_2h_3}} \frac{[k_{h_4}\cdot (X_{h_4}+k_{h_1})]}{K_{1h_1h_2h_3h_4}}~,\\
\mathcal{C}(\{h_3,h_4,h_1,h_2\})
= \frac{[k_{h_1}\cdot(X_{h_1}-k_1)]}{K_{1h_1}} \frac{(k_{h_2}\cdot X_{h_2})}{K_{1h_1h_2}} \frac{[k_{h_3}\cdot (X_{h_3}-k_1)]}{K_{1h_1h_2h_3}} \frac{[k_{h_4}\cdot (X_{h_4}+k_{h_1}+k_{h_2})]}{K_{1h_1h_2h_3h_4}}~,\\
\mathcal{C}(\{h_3,h_4,h_2,h_1\})
= \frac{(k_{h_1}\cdot X_{h_1})}{K_{1h_1}} \frac{[k_{h_2}\cdot (X_{h_2}-k_1)]+K_{1h_1h_2}}{K_{1h_1h_2}} \frac{[k_{h_3}\cdot (X_{h_3}-k_1)]}{K_{1h_1h_2h_3}} \frac{[k_{h_4}\cdot (X_{h_4}+k_{h_1}+k_{h_2})]}{K_{1h_1h_2h_3h_4}}~,
\end{array}
\end{eqnarray}
%
\iffalse
\begin{align}
\mathcal{C}(\{h_3,h_1,h_2,h_4\})
=& \frac{[k_{h_1}\cdot(X_{h_1}-k_1)]}{K_{1h_1}} \frac{(k_{h_2}\cdot X_{h_2})}{K_{1h_1h_2}} \frac{[k_{h_3}\cdot (X_{h_3}-k_1)]}{K_{1h_1h_2h_3}} \frac{(k_{h_4}\cdot X_{h_4})}{K_{1h_1h_2h_3h_4}}, \notag\\
\mathcal{C}(\{h_3,h_1,h_4,h_2\})
=& \frac{[k_{h_1}\cdot(X_{h_1}-k_1)]}{K_{1h_1}} \frac{(k_{h_2}\cdot X_{h_2})}{K_{1h_1h_2}} \frac{[k_{h_3}\cdot (X_{h_3}-k_1)]}{K_{1h_1h_2h_3}} \frac{[k_{h_4}\cdot (X_{h_4}+k_{h_2})]}{K_{1h_1h_2h_3h_4}}, \notag\\
\mathcal{C}(\{h_3,h_2,h_1,h_4\})
=& \frac{(k_{h_1}\cdot X_{h_1})}{K_{1h_1}} \frac{[k_{h_2}\cdot (X_{h_2}-k_1)]+K_{1h_1h_2}}{K_{1h_1h_2}} \frac{[k_{h_3}\cdot (X_{h_3}-k_1)]}{K_{1h_1h_2h_3}} \frac{(k_{h_4}\cdot X_{h_4})}{K_{1h_1h_2h_3h_4}}, \notag\\
\mathcal{C}(\{h_3,h_2,h_4,h_1\})
=& \frac{(k_{h_1}\cdot X_{h_1})}{K_{1h_1}} \frac{[k_{h_2}\cdot (X_{h_2}-k_1)]+K_{1h_1h_2}}{K_{1h_1h_2}} \frac{[k_{h_3}\cdot (X_{h_3}-k_1)]}{K_{1h_1h_2h_3}} \frac{[k_{h_4}\cdot (X_{h_4}+k_{h_1})]}{K_{1h_1h_2h_3h_4}}, \notag\\
\mathcal{C}(\{h_3,h_4,h_1,h_2\})
=& \frac{[k_{h_1}\cdot(X_{h_1}-k_1)]}{K_{1h_1}} \frac{(k_{h_2}\cdot X_{h_2})}{K_{1h_1h_2}} \frac{[k_{h_3}\cdot (X_{h_3}-k_1)]}{K_{1h_1h_2h_3}} \frac{[k_{h_4}\cdot (X_{h_4}+k_{h_1}+k_{h_2})]}{K_{1h_1h_2h_3h_4}}, \notag\\
\mathcal{C}(\{h_3,h_4,h_2,h_1\})
=& \frac{(k_{h_1}\cdot X_{h_1})}{K_{1h_1}} \frac{[k_{h_2}\cdot (X_{h_2}-k_1)]+K_{1h_1h_2}}{K_{1h_1h_2}} \frac{[k_{h_3}\cdot (X_{h_3}-k_1)]}{K_{1h_1h_2h_3}} \frac{[k_{h_4}\cdot (X_{h_4}+k_{h_1}+k_{h_2})]}{K_{1h_1h_2h_3h_4}}, \notag\\
\end{align}
\fi
%
\begin{eqnarray}
\begin{array}{l}
 \mathcal{C}(\{h_4,h_1,h_2,h_3\})
= \frac{[k_{h_1}\cdot(X_{h_1}-k_1)]}{K_{1h_1}} \frac{(k_{h_2}\cdot X_{h_2})-K_{1h_1h_2}}{K_{1h_1h_2}} \frac{(k_{h_3}\cdot X_{h_3})}{K_{1h_1h_2h_3}} \frac{[k_{h_4}\cdot (Y_{h_4}-k_1)]+K_{1h_1h_2h_3h_4}}{K_{1h_1h_2h_3h_4}}~,\\
\mathcal{C}(\{h_4,h_1,h_3,h_2\})
= \frac{[k_{h_1}\cdot(X_{h_1}-k_1)]}{K_{1h_1}} \frac{(k_{h_2}\cdot X_{h_2})}{K_{1h_1h_2}} \frac{[k_{h_3}\cdot (X_{h_3}-k_{h_1}-k_1)]+K_{1h_1h_2h_3}}{K_{1h_1h_2h_3}}\frac{[k_{h_4}\cdot (Y_{h_4}-k_1)]+K_{1h_1h_2h_3h_4}}{K_{1h_1h_2h_3h_4}}~,\\
\mathcal{C}(\{h_4,h_2,h_1,h_3\})
= \frac{(k_{h_1}\cdot X_{h_1})}{K_{1h_1}} \frac{[k_{h_2}\cdot (X_{h_2}-k_1)]}{K_{1h_1h_2}} \frac{(k_{h_3}\cdot X_{h_3})}{K_{1h_1h_2h_3}} \frac{[k_{h_4}\cdot (Y_{h_4}-k_1)]+K_{1h_1h_2h_3h_4}}{K_{1h_1h_2h_3h_4}}~,\\
\mathcal{C}(\{h_4,h_2,h_3,h_1\})
= \frac{(k_{h_1}\cdot X_{h_1})}{K_{1h_1}} \frac{[k_{h_2}\cdot (X_{h_2}-k_1)]}{K_{1h_1h_2}} \frac{[k_{h_3}\cdot (X_{h_3}+k_{h_1})]}{K_{1h_1h_2h_3}} \frac{[k_{h_4}\cdot (Y_{h_4}-k_1)]+K_{1h_1h_2h_3h_4}}{K_{1h_1h_2h_3h_4}}~,\\
\mathcal{C}(\{h_4,h_3,h_1,h_2\})
=\frac{[k_{h_1}\cdot(X_{h_1}-k_1)]}{K_{1h_1}} \frac{(k_{h_2}\cdot X_{h_2})}{K_{1h_1h_2}} \frac{[k_{h_3}\cdot (X_{h_3}-k_{1})]+K_{1h_1h_2h_3}}{K_{1h_1h_2h_3}} \frac{[k_{h_4}\cdot (Y_{h_4}-k_1)]+K_{1h_1h_2h_3h_4}}{K_{1h_1h_2h_3h_4}}~,\\
\mathcal{C}(\{h_4,h_3,h_2,h_1\})
= \frac{(k_{h_1}\cdot X_{h_1})}{K_{1h_1}} \frac{[k_{h_2}\cdot (X_{h_2}-k_1)]+K_{1h_1h_2}}{K_{1h_1h_2}} \frac{[k_{h_3}\cdot (X_{h_3}-k_{h_1})]+K_{1h_1h_2h_3}}{K_{1h_1h_2h_3}} \frac{[k_{h_4}\cdot (Y_{h_4}-k_1)]+K_{1h_1h_2h_3h_4}}{K_{1h_1h_2h_3h_4}}~.
\end{array}
\end{eqnarray}
In above expressions for simplicity we have used $Y_{h_i}$ to replace $X_{h_i}$ in some terms.

%%%%%%%%%%%%%%%%%%%
%\section{Overview of the BCJ relations}
%%%%%%%%%%%%%%%%%%%%%

%%%%%%%%%%%%%%%%%%%%%%%%%%%%%%%%%%%% %%%%%%%

\appendix

\newpage
\bibliographystyle{JHEP}
\bibliography{reference2}

\end{document}